\documentclass[12pt]{article}
\usepackage[top=2cm,bottom=3cm,left=2cm,right=2cm]{geometry}
\usepackage{amsmath}
\usepackage{amssymb} 
\usepackage{amsthm}
\usepackage{stmaryrd}
\usepackage{graphicx}
\usepackage{hyperref}
\usepackage{wrapfig}
\usepackage{array}
\usepackage{mathtools}
\usepackage[mathscr]{euscript}
\usepackage{bm}

\numberwithin{equation}{section}
\numberwithin{figure}{section}

\def\eq#1{(\ref{eq:#1})}
\def\lineup{\!\!\!\!\!\!\!\! &&}

\def\d{\partial}
\def\eps{\epsilon}

\def\zbar{\overline{z}}
\def\H{\mathcal{H}}
\def\Hhat{\widehat{\mathcal{H}}}
\def\P{\mathcal{P}}
\def\Phat{\widehat{\mathcal{P}}}
\def\NS{{\mathrm{NS}}}
\def\R{{\mathrm{R}}}

\newtheorem{exercise}{Exercise}

\begin{document}
\begin{titlepage}
\rightline\today

\begin{center}
\vskip 3.5cm

{\large \bf{Four Lectures on Closed String Field Theory}}

\vskip 1.0cm

{\large {Theodore Erler\footnote{tchovi@gmail.com}}}

\vskip 1.0cm

{\it Institute of Physics of the ASCR, v.v.i.}\\
{\it Na Slovance 2, 182 21 Prague 8, Czech Republic}\\
\vskip .5cm
{\it Institute of Mathematics, the Czech Academy of Sciences}\\ {\it Zinta 25, 11567 Prague 1, Czech Republic}\\

\vskip 2.0cm

{\bf Abstract} 

\end{center}

The following notes derive from review lectures on closed string field theory given at the Galileo Galilei Institute for Theoretical Physics in March 2019.

\end{titlepage}

\tableofcontents

\section{Preface}

These are notes for four lectures on the topic of closed string field theory (closed SFT). Videos of the lectures can be found by clicking \href{https://www.youtube.com/watch?v=ZtN3BUBi1TE&list=PL1CFLtxeIrQp8mo74ExxEDIYnjmXfPv5Y}{here}, though in four allotted slots I covered only the first two lectures. Even with all four lectures, the scope of what is covered is limited. The definitive reference is Zwiebach's 1992 paper~\cite{Zwiebach}, which however only discusses the bosonic string. The superstring requires a few additional considerations which have only been taken up in recent years. A review of these developments can be found in~\cite{SenErbin}, which perhaps more importantly discusses physical applications which have been the focus of recent interest. However, we do not discuss applications here. Our aim is only to describe the formalism. 

We assume familiarity with two dimensional conformal field theory, worldsheet BRST symmetry, and the basics of on-shell string amplitudes, and, for the superstring, the RNS formalism. An elementary course on string theory, or the relevant chapters in Polchinski \cite{Polchinski} should be more than sufficient to follow. Also included are a few simple exercises. 

\section{Intro}

Closed string field theory is a quantum field theory constructed in such a way that its Feynman diagram expansion computes string S-matrix elements. The utility of this formalism is that it seems necessary to give a fully complete and consistent definition of string perturbation theory. Applications have been especially highlighted in recent work by Sen and others \cite{SenErbin}, and include
\begin{itemize}
\item {\it Consistent treatment of divergences.} Several kinds of divergences can appear in the standard worldsheet approach to string perturbation theory. Unphysical divergences can appear from using a Euclidean worldsheet description towards the boundary of moduli space, or from spurious singularities in correlation functions of the $\beta\gamma$ superconformal ghosts. Physical infrared divergences can appear in compactifications down to four dimensions or less, or if the tree level vacuum and/or physical state condition receive quantum corrections (vacuum shift \cite{susy} and mass renormalization \cite{mass_ren}). Closed SFT tells you how to avoid or cancel such divergences.
\item {\it Fundamental formal properties of string perturbation theory.} Perturbative string theory is widely assumed to have a number of rudimentary physical properties that are nevertheless difficult to establish in the standard worldsheet approach. These include unitarity \cite{unitarity} and crossing symmetry \cite{crossing} of amplitudes, and independence of the theory under continuous deformation of the background where we quantize the string \cite{BI,BIsuper}. These questions can be addressed in closed~SFT.
\item {\it Access to new observables.} In closed SFT we can in principle compute physical quantities in situations where the standard worldsheet formulation breaks down. Examples include strings in Ramond-Ramond backgrounds in the RNS formalism \cite{Yin}, or amplitudes when the vacuum and physical state condition receive quantum correction.
\end{itemize}
Perhaps it is surprising that these results do not follow from some incremental improvement to the sum over worldsheets approach, which otherwise seems so effective. Instead, they require a fundamentally different way of thinking about string perturbation theory. Closed SFT offers the rigor and conceptual clarity of perturbative quantum field theory, and provides an exact spacetime action for string theory whose gauge symmetry---it can be argued---takes the most elegant possible form. But computations with the theory are difficult. Its physical content is buried underneath mountains of unphysical and computationally inaccessible data. With more work we can understand how to calculate with the theory more effectively, or, perhaps, replace the formalism with something better.

\subsection*{First Look}

Closed SFT is a field theory of fluctuations of a closed string (or spacetime) background in string theory. A closed string background is specified by a matter+ghost worldsheet conformal field theory with vanishing central charge. Fluctuations of the background are therefore equivalent to deformations of this conformal field theory. The deformations can be characterized by the vector space of local operators in the conformal field theory; given such an operator $\mathcal{O}(z,\zbar)$, we can deform the worldsheet action by
\begin{equation}S'=S+\int d^2 z\,\mathcal{O}(z,\zbar).\end{equation}
The new action describes a closed string moving in a deformed background. This deformation does not necessarily preserve conformal invariance, but this is okay: An arbitrary configuration of a fluctuation field is not necessarily meaningful, since it is not guaranteed to satisfy the equations of motion. Presently we are concerned with the nature of the fluctuation field for closed strings, not whether the fluctuation is on-shell. The equations of motion of closed SFT are believed to be equivalent to the requirement of conformal invariance of the worldsheet theory.

An operator of the worldsheet theory defines a state through the state-operator mapping. We therefore claim the following:
\begin{quote}{\it A closed string field is an element of the vector space $\mathcal{H}$ of states of the worldsheet conformal field theory defining the reference closed string background.}\end{quote}
The state space $\mathcal{H}$ has two important gradings: a $\mathbb{Z}$ grading called {\it ghost number} and a $\mathbb{Z}_2$ grading called {\it Grassmann parity}. For the bosonic string, ghost number counts the number of $c$ ghost minus the number of $b$ ghost insertions in the vertex operator defining the state (for the superstring in the RNS formalism, we must add to this the number of $\gamma$ minus $\beta$ ghosts). Grassmann parity tells us whether the vertex operator defining the state is commuting or anticommuting. This is determined both by whether the worldsheet fields entering the vertex operator are commuting or anticommuting, and by whether the vertex operator comes multiplied by a commuting or anticommuting coefficient. The Grassmann parity of a state $|A\rangle$ will be denoted $|A|$. Depending on the string theory we consider, there may be other gradings (such as picture number). From the point of view of the 2 dimensional field theory of the string worldsheet, a vector in $\H$ is a quantum state which would conventionally be written as a ket $|A\rangle$. From the point of view of string field theory, a vector in $\H$ is a spacetime field analogous to (and generalizing) the metric of general relativity. In this interpretation the ket notation seems less appropriate. Usually we write string fields without the ket. 

As it turns out, the dynamical variable of closed SFT is not a generic closed string field, but is a state of a particular kind:
\begin{quote}{\it The dynamical string field $\Phi$---the dynamical variable which appears in the action---is a Grassmann even state in a linear subspace $\Hhat\subset\H$ defined by certain constraints, notably the level matching condition. Classically the dynamical string field has ghost number~2, but in the quantum theory it contains components at all ghost numbers which play a role similar to Faddeev-Popov ghosts.}
\end{quote}
The necessity of restricting the dynamical field to a linear subspace $\Hhat$ will be explained in due course. The condition on Grassmann parity and ghost number comes from the fact that ``physical" vertex operators take the form (in bosonic string theory)
\begin{equation}c\overline{c}\mathcal{V}(z,\zbar),\end{equation}
where $\mathcal{V}$ is a weight $(1,1)$ matter primary. In closed SFT, such vertex operators will be understood as solutions to the linearized equations of motion.

The action of closed SFT takes the form\footnote{The path integral is defined with the weight factor $e^S$. We set the closed string coupling constant~to~$1$.} 
\begin{eqnarray}
S \lineup= \frac{1}{2!}\omega(\Phi,Q\Phi)+\frac{1}{3!}\omega(\Phi,L_{0,2}(\Phi,\Phi))+\frac{1}{4!}\omega(\Phi,L_{0,3}(\Phi,\Phi,\Phi))+\frac{1}{5!}\omega(\Phi,L_{0,4}(\Phi,\Phi,\Phi,\Phi))+...\nonumber\\
\lineup \ \ \ \ \ \ \ \ \ \ \ \ \ \ \ \ \ \ \ +\ \  \ \ \ \frac{1}{1!}\omega(\Phi,L_{1,0})\ \ \ \ +\ \ \ \ \frac{1}{2!}\omega(\Phi,L_{1,1}(\Phi))\ \ \  + \ \ \ \frac{1}{3!}\omega(\Phi,L_{1,2}(\Phi,\Phi))\ \ \ \ +...\nonumber\\
\lineup \ \ \ \ \ \ \ \ \ \ \ \ \ \ \ \ \ \ \ \ \ \ \ \ \ \ \ \ \ \ \ \ \ \ \ \ \ \ \ \ \ \ \ \ \ \ \ \ \ \ \ \ \ \ \ \ \ \ \ \ \ \ \ \ \ \ \ \ \ \ \ \ \ \ \ \ \ \, +\, \ \ \ \ \ \ \ \ \frac{1}{1!}\omega(\Phi,L_{2,0})\ \ \ \ \ \ \  +...\nonumber\\ \lineup \ \ \ \ \ \ \ \ \ \ \ \ \ \ \ \ \ \ \ \ \ \ \ \ \ \ \ \ \ \ \ \ \ \ \ \ \ \ \ \ \ \ \ \ \ \ \ \ \ \ \ \ \ \ \ \ \ \ \ \ \ \ \ \ \ \ \ \ \ \ \ \ \ \ \ \ \ \ \ \ \ \ \ \ \ \ \ \ \ \ \ \ \ \ \ \ddots\ \ .
\label{eq:action}\end{eqnarray}
Technically, this is a {\it quantum master action} within the framework of the Batalin-Vilkovisky formalism. The ingredients are as follows:

\begin{itemize}
\item $\omega(\cdot,\cdot)$ is a symplectic form on the state space. It is graded antisymmetric, nondegenerate, and is nonvanishing only between states whose ghost number adds up to $5$. The symplectic form can be viewed as a linear map from the tensor product of two copies of the state space $\Hhat\otimes\Hhat =\Hhat^{\otimes 2}$ into numbers, which we identify as ``zero" copies of the state space $\Hhat^{\otimes 0}$. We will sometimes write the symplectic form as a ``double bra" state $\langle \omega|$:
\begin{equation}\langle \omega|:\Hhat^{\otimes 2}\to \Hhat^{\otimes 0}.\end{equation}
The double bra $\langle \omega|$ is Grassmann odd and carries ghost number $-5$. The inverse of the symplectic form is a ``double ket'' state  $|\omega^{-1}\rangle\in \Hhat^{\otimes 2}$ which may be called the Poisson bivector. By definition it satisfies
\begin{equation}(\langle \omega|\otimes\mathbb{I})(\mathbb{I}\otimes |\omega^{-1}\rangle) = \mathbb{I},\label{eq:wwinv}\end{equation}
where $\mathbb{I}$ is the identity operator on $\Hhat$. The tensor product notation in this equation will be used extensively in these notes, and is explained in appendix \ref{app:tensor}. The Poisson bivector is Grassmann odd and carries ghost number $+5$. Many discussions of closed SFT do not mention the symplectic form, but rather express the action in terms of the conformal field theory inner product. This is related to the symplectic form through a $c$ ghost and possibly other operator insertions.
\item $Q:\Hhat\to \Hhat$ is the BRST operator: 
\begin{equation}Q = \oint \frac{dz}{2\pi i}j_B(z) - \oint \frac{d\overline{z}}{2\pi i}\overline{j}_B(\overline{z}),\end{equation}
where $j_B(z)$ is the holomorphic BRST current and $\overline{j}_B(\overline{z})$ its antiholomorphic counterpart. The BRST operator is Grassmann odd and carries ghost number $1$. Its most important property is that it is nilpotent:
\begin{equation}Q^2=0.\end{equation}
Another important fact is that the BRST variation of the $b$-ghost gives the energy-momentum tensor of the total matter+ghost worldsheet conformal field theory:
\begin{equation}Q\cdot b(z) = T(z),\ \ \ Q\cdot \overline{b}(\overline{z}) = \overline{T}(\overline{z}).\end{equation}
Since the total worldsheet conformal field theory has vanishing central charge, $T(z)$ is a primary operator of weight $(2,0)$ and $\overline{T}(\overline{z})$ is a primary operator of weight $(0,2)$.
\end{itemize}

\begin{wrapfigure}{l}{.37\linewidth}
\centering
\resizebox{2.8in}{1.4in}{\includegraphics{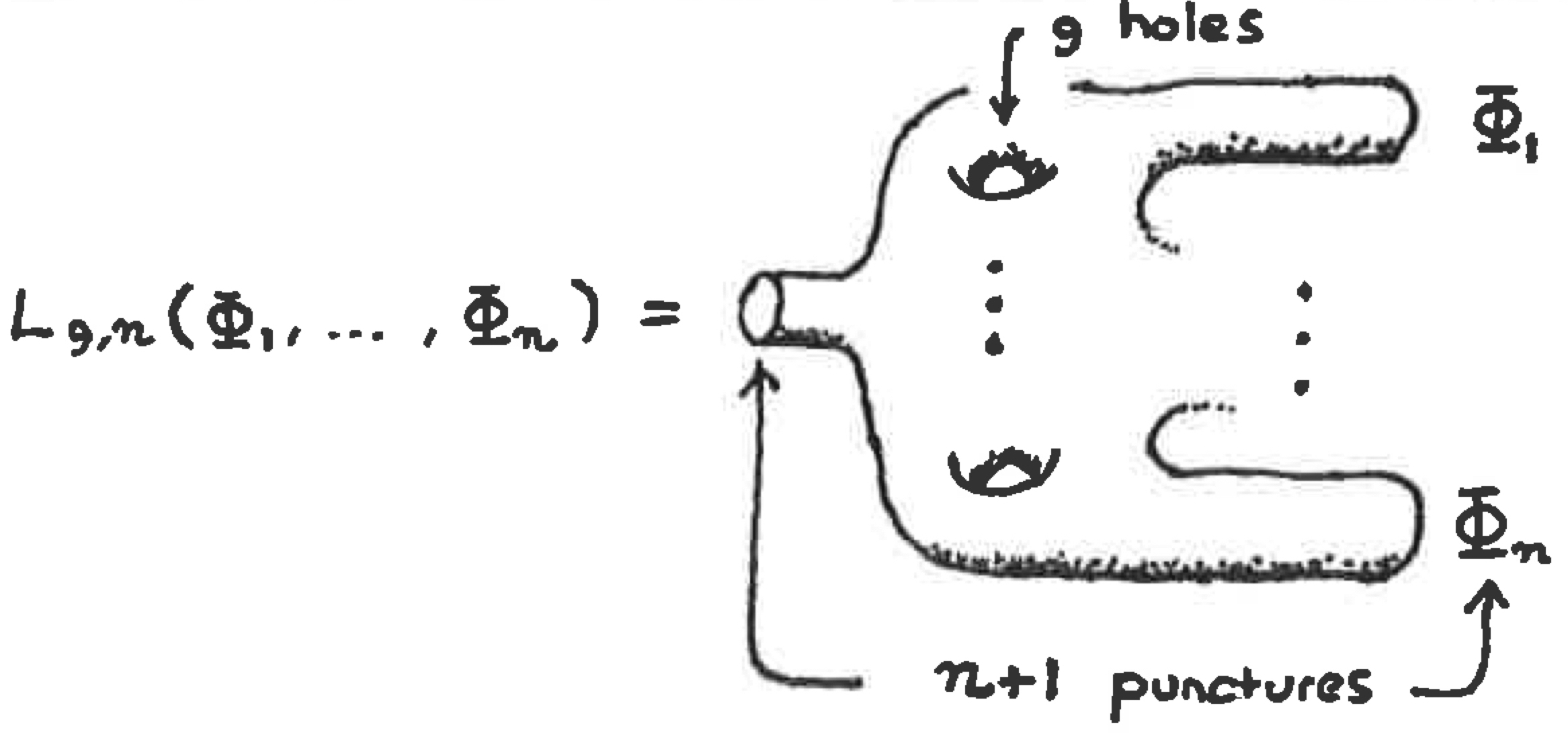}}
\end{wrapfigure}\ \ 

\vspace{-1cm}

\begin{itemize}
\item\ $L_{g,n}$ are linear maps $\Hhat^{\otimes n} \to \Hhat$ called {\it multi-string products}. The index $n$ refers to the number of states being multiplied, and $g$ refers to the ``genus" of the product. They are Grassmann odd and carry ghost number $3-2n$. The multi-string products  are the most nontrivial ingredient in the construction of closed SFT. Roughly, $L_{g,n}$ is defined by a worldsheet path integral 
over a genus $g$ Riemann surface with $n+1$ punctures, with $n$ of those punctures representing the states being multiplied and the final puncture representing their product. The products satisfy a hierarchy of algebraic relations which generalize the statement that $Q$ is nilpotent.  As we will describe later, there is a canonical way to promote multi-string products into operators which act within the symmetrized tensor algebra $L_{g,n}:S\Hhat \to S\Hhat$. In this context is is meaningful to add products which multiply different numbers of states, and we can form a composite object
\begin{equation}L =\sum_{g,n>0}L_{g,n}+|\omega^{-1}\rangle.\end{equation}
Note that $L_{0,0}=0$ and $L_{0,1}=Q$. The algebraic relations satisfied by the multi-string products are encapsulated in the statement that this object is nilpotent,
\begin{equation}L^2 =0,\end{equation}
in exact analogy to the BRST operator. This defines what is called a {\it loop homotopy algebra} \cite{Markl} or {\it quantum $L_\infty$ algebra}. If we drop all of the higher genus products, we have a purely classical closed SFT. The sum of all genus zero products 
\begin{equation}L_{g = 0}  = \sum_{n\geq 1}L_{0,n}\end{equation}
is also nilpotent. This defines an (ordinary) {\it $L_\infty$ algebra}. In addition, the products are required to be graded commutative, and satisfy a compatibility condition with respect to the symplectic form which ensures that the resulting vertices in the action are totally symmetric upon interchange of states.
\end{itemize}

\noindent We make the following comments:

\begin{itemize}
\item The closed SFT action takes this form regardless of what kind of string theory you are considering (bosonic, heterotic, type II) or what background within that string theory whose fluctuations interest you. The differences between these closed SFTs mostly concern the worldsheet conformal field theory you consider, and some technical differences in how the products and symplectic form are realized. In these lectures we spend the bulk of our time discussing closed bosonic SFT. Closed super SFTs require a little extra dressing which we discuss at the end.
\item Even if you have fixed your choice of string theory and background within that string theory, the closed SFT action is not unique. There is some freedom in the choice of vertices consistent with (quantum) $L_\infty$ relations, corresponding to the freedom of field redefinition. It is nevertheless an interesting question whether there is a systematic and unique definition of the vertices which defines closed SFT in the best or most convenient possible form. This question leads to the most nontrivial mathematics of the subject: minimal area metrics and systolic geometry \cite{min_area,systolic}, homotopy algebra techniques \cite{WittenSS}, hyperbolic geometry \cite{Pius1}, and super-Rieman surface theory \cite{revisited,Jurco,Ohmori}. None of these ideas have played an important role in recent physical applications, but this could change in the future.
\item The closed SFT action is nonpolynomial at the classical level and more so at the quantum level. This is reminiscent of an expansion of the Einstein-Hilbert action around flat space in powers of the metric perturbation. In fact, the higher genus vertices of closed SFT can be seen as analogous to the infinite set of counterterms needed to compute loop amplitudes in perturbative general relativity. An important difference is that the ``counterterms" of closed SFT are actually finite, and are likely uniquely determined (up to field redefinition) by quantum gauge invariance.\footnote{In fact, quantum gauge invariance probably fixes all vertices up to field redefinition. If this were not the case, closed strings could couple to each other in a different way from that implied by the usual genus expansion of perturbative string theory. In principle this is something that could be tested in the formalism of closed SFT.} Another difference in closed SFT is that we do not have anything like the Minkowski metric that could be added to $\Phi$ to form a background independent field variable. 
\item We have ordered the terms in the action in \eq{action} in such a way as to highlight the relative complexity of vertices. Starting from the cubic vertex at genus zero, each step to the right and each step downwards increases the real dimension of the moduli space integral appearing in the associated vertex by two. Moreover, a consistent definition of the product $L_{g,n}$ requires data specifying all lower order vertices appearing above and to the left. This is to ensure that $L_{g,n}$ correctly compensates for the Feynman diagrams of lower order vertices to reproduce the correct $n+1$ string amplitude at genus~$g$. 
\end{itemize}

\noindent To illustrate the last point, consider the 1-loop 2-point amplitude. This involves the following diagrams:
\begin{wrapfigure}{l}{1\linewidth}
\centering
\resizebox{5in}{2in}{\includegraphics{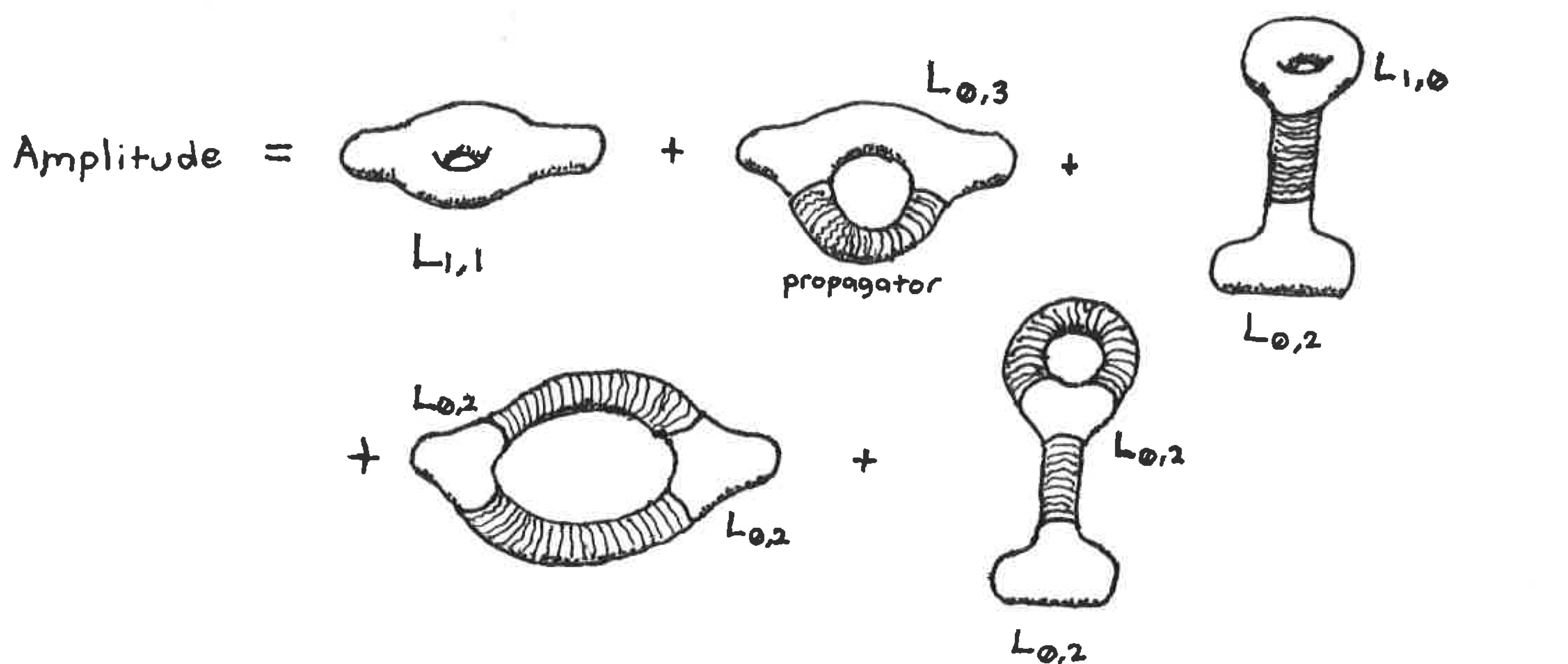}}
\end{wrapfigure}\\ \\ \\ \\ \\ \\ \\ \\ \\ \\ \\ \\
\noindent The product $L_{1,1}$ is chosen so that this sum of Feynman graphs gives the correct amplitude.
\begin{exercise}
Draw all Feynman diagrams necessary for the 2-loop tadpole amplitude and label the vertices according to the corresponding products.
\end{exercise}

\pagebreak

\section{Lecture 1: Off-Shell Amplitudes}

We begin by describing off-shell amplitudes in bosonic string theory. By this we have in mind, at the very least, some kind of continuation of physical amplitudes to generic momenta which are not constrained to lie on the mass shell. More precisely, we are looking for a multilinear map
\begin{equation}\langle \mathcal{A}_{g,n}|:\Hhat^{\otimes n}\to \Hhat^{\otimes 0}\end{equation}
subject to the following conditions:
\begin{description}
\item{(1)} The map is defined on a vector space $\Hhat$ satisfying
\begin{equation}\mathcal{H}_Q\subset\Hhat\subseteq\H,\end{equation}
where $\H$ is the full conformal field theory state space and $\mathcal{H}_Q$ is the vector space of BRST invariant conformal vertex operators
\begin{equation}c\overline{c}\mathcal{V}(0,0)\in \mathcal{H}_Q,\end{equation}
where $\mathcal{V}(0,0)$ is a weight (1,1) primary operator in the $c=26$ ``matter" factor of the conformal field theory. Together with the $c$ ghost factors, the total vertex operator is a primary of weight (0,0). (This is why the vertex operator is called ``conformal").  We would like to choose $\Hhat$ to be as large as conveniently possible---possibly as large as $\H$ itself, but this would encounter complications as we will see later.
\item{(2)} Acting on states in $\mathcal{H}_Q$, $\langle\mathcal{A}_{g,n}|$ gives the physical $n$ point amplitude at genus $g$ of the corresponding states when this amplitude is finite.
\item{(3)} The off-shell amplitude is BRST invariant
\begin{equation}\langle \mathcal{A}_{g,n}|Q^{(n)}=0,\end{equation}
assuming on the left hand side we ignore contributions from the boundary of moduli space. The operator $Q^{(n)}:\H^{\otimes n}\to\H^{\otimes n}$ is defined by a sum of BRST operators acting on each of the $n$ states in the amplitude:
\begin{eqnarray}Q^{(n)}\lineup = Q\otimes(\underbrace{\mathbb{I}\otimes ...\otimes \mathbb{I}}_{n-1\text{ times}}) \, + \, \mathbb{I}\otimes Q\otimes (\underbrace{\mathbb{I}\otimes ...\otimes \mathbb{I}}_{n-2\text{ times}}) \,+\,  ...\, +\, (\underbrace{\mathbb{I}\otimes ...\otimes \mathbb{I}}_{n-1\text{ times}})\otimes Q \nonumber\\
\lineup = \sum_{A=1}^{n}\mathbb{I}^{\otimes A-1}\otimes Q\otimes\mathbb{I}^{\otimes n-A},\ \ \ \ \ \ \ \ \ \ \label{eq:Qcod}\end{eqnarray}
\end{description}
Off shell amplitudes in string theory are not observable, and without additional structure are not of much physical interest. We discuss them since they are very closely related to the vertices of the closed SFT action; roughly speaking, SFT vertices are off-shell amplitudes with integration near the boundary of moduli space excluded.

To start, let us consider the 4-point amplitude on the sphere, and understand what is involved in extending this amplitude off shell. For states $\Phi_1,...,\Phi_4\in\mathcal{H}_Q$, the amplitude is given by
\begin{equation}
\langle \mathcal{A}_{0,4}|\Phi_1\otimes\Phi_2\otimes\Phi_3\otimes\Phi_4 = \int_\mathbb{C}dz\wedge d\zbar\,\Big\langle \Phi_1(0,0)\Big(b_{-1}\overline{b}_{-1}\cdot\Phi_2(z,\zbar)\Big)\Phi_3(1,1)\Phi_4(\infty,\infty)\Big\rangle_\mathbb{C}.
\end{equation}
On the right hand side is a correlation function on the complex plane (Riemann sphere) of vertex operators corresponding to the states. Using $SL(2,\mathbb{C})$ invariance we fix the position of the first, third, and fourth vertex operator to $0,1$, and $\infty$ respectively, The position of the second vertex operator is integrated over the complex plane. This is the integration over the moduli space $\mathcal{M}_{0,4}$ of the 4-punctured sphere. To give the correct integration measure, it is necessary to act $b$-ghosts on $\Phi_2(z,\zbar)$ given by 
\begin{equation}b_{-1}=\oint \frac{dz}{2\pi i} b(z),\ \ \ \ \ \overline{b}_{-1} = \oint \frac{d\zbar}{2\pi i} \overline{b}(\zbar).\end{equation}
The $b$-ghosts have the effect of removing the $c\overline{c}$ factor from $\Phi_2$, producing the so-called ``integrated" vertex operator. The notion of integrated vertex operator does not generalize very cleanly off-shell, so we express the amplitude using unintegrated vertex operators and $b$-ghosts. 

At the level of states in $\mathcal{H}_Q$, the condition of BRST invariance (3) is equivalent to the statement that BRST trivial states produce a vanishing amplitude. Let us assume $\Phi_1=Q\Lambda$. Pull the BRST contour off $\Lambda$ to surround the remaining operators in the correlator. Since the vertex operators are BRST invariant, the only contribution comes from $Q$ acting on $b_{-1}\overline{b}_{-1}$:
\begin{eqnarray}
Q\cdot(b_{-1}\overline{b}_{-1}\cdot\Phi_2(z,\zbar)) \lineup = L_{-1}\overline{b}_{-1}\cdot\Phi_2(z,\overline{z})-b_{-1}\overline{L}_{-1}\cdot\Phi_2(z,\zbar)\nonumber\\
\lineup = \overline{b}_{-1}\d \Phi_2(z,\zbar) - b_{-1}\overline{\d}\Phi_2(z,\zbar),
\end{eqnarray}
where in the last step we turned $L_{-1}$ and $\overline{L}_{-1}$ into derivatives with respect to the modulus using the OPE of $\Phi_2$ with the energy momentum tensor. In this way, the amplitude with $\Phi_1= Q\Lambda$ can be written as the integral of a total derivative on the moduli space:
\begin{equation}
\langle \mathcal{A}_{0,4}|Q\Lambda\otimes\Phi_2\otimes\Phi_3\otimes\Phi_4 = \int_\mathbb{C}d \Big\langle \Lambda(0,0)\Big[(dz b_{-1}+d\zbar\overline{b}_{-1})\cdot\Phi_2(z,\zbar)\Big]\Phi_3(1,1)\Phi_4(\infty,\infty)\Big\rangle_\mathbb{C}.
\end{equation}
Ignoring contributions from the boundary of moduli space, this vanishes as expected. Now we can ask whether we are justified in ignoring  boundary contributions. These correspond to integrals of $\Phi_2$ around small circles at $0$ and $1$ and a very large circle around $\infty$. This might look problematic, since the OPE of $\Phi_2$ and the operators at the punctures can be singular. The standard argument is that there is a kinematic region in momentum space where the OPEs do not lead to divergence, and the boundary contributions vanish. We can then argue that the boundary terms vanish for generic momenta by analytic continuation. In loop amplitudes, divergences from the boundary of moduli space can appear which cannot be removed by passing to a nice kinematic region. These are physical infrared divergences, for example associated to mass renormalization and vacuum shift. The proper treatment in this case requires closed SFT. For present purposes, we simply ignore boundary contributions.

Now let us continue the amplitude off-shell. For example, we can take $\Phi_1,...,\Phi_4$ to be (non-BRST invariant) primary operators with nonvanishing conformal weight. Conformal transformation of a primary of weight $(h,\overline{h})$ by a holomorphic function $f(z)$ is given by
\begin{equation}f\circ \Phi(z,\zbar) = \left(\frac{\d f(z)}{\d z}\right)^h \left(\frac{\d\overline{f}(\zbar)}{\d\zbar}\right)^{\overline{h}}\Phi(f(z),\overline{f}(\zbar)). \end{equation}
This generalization immediately runs into problems with $\Phi_4$, since the amplitude will contain a singular factor 
\begin{equation}\lim_{z\to\infty}z^{-2h}\overline{z}^{-2\overline{h}}\end{equation}
as $\Phi_4$ is pushed to infinity. This can be remedied by fixing $\Phi_1,\Phi_3,\Phi_4$ to some finite positions $z_1,z_3,z_4$ before going off-shell. Any two choices of $z_1,z_3,z_4$ can be related by $SL(2,\mathbb{C})$ transformation, but since the vertex operators pick up a nontrivial factor under conformal transformation, the resulting off-shell continuations of the amplitude are nevertheless different. Moreover, though the complex plane is the simplest coordinate system on the Riemann sphere, there are an infinite number of other possibilities, and they all lead to different off-shell continuations.

The fact that the off-shell continuation of the amplitude is ambiguous is not an immediate source of concern. In closed SFT, this ambiguity corresponds to our freedom to redefine the closed string field
\begin{equation}\Phi' = F[\Phi].\end{equation}
where $F$ is an invertible and potentially nonlinear map from $\Hhat$ to itself that preserves Grassmann parity and ghost number. Off-shell amplitudes in a standard QFT are ambiguous for the same reason, though this is seldom emphasized since typically we are given a canonical choice of field variable. If there is such a canonical choice for closed SFT, we do not know it yet, so the off-shell continuation is not {\it a priori} determined. Still, we need to find a convenient characterization of the extra data that goes into specifying an off-shell extension.

The right way to think about this is that we are looking for off-shell amplitudes of closed string {\it states}, not closed string vertex operators. The concepts are closely related, but there are 
\begin{wrapfigure}{l}{.45\linewidth}
\centering
\resizebox{3in}{1.5in}{\includegraphics{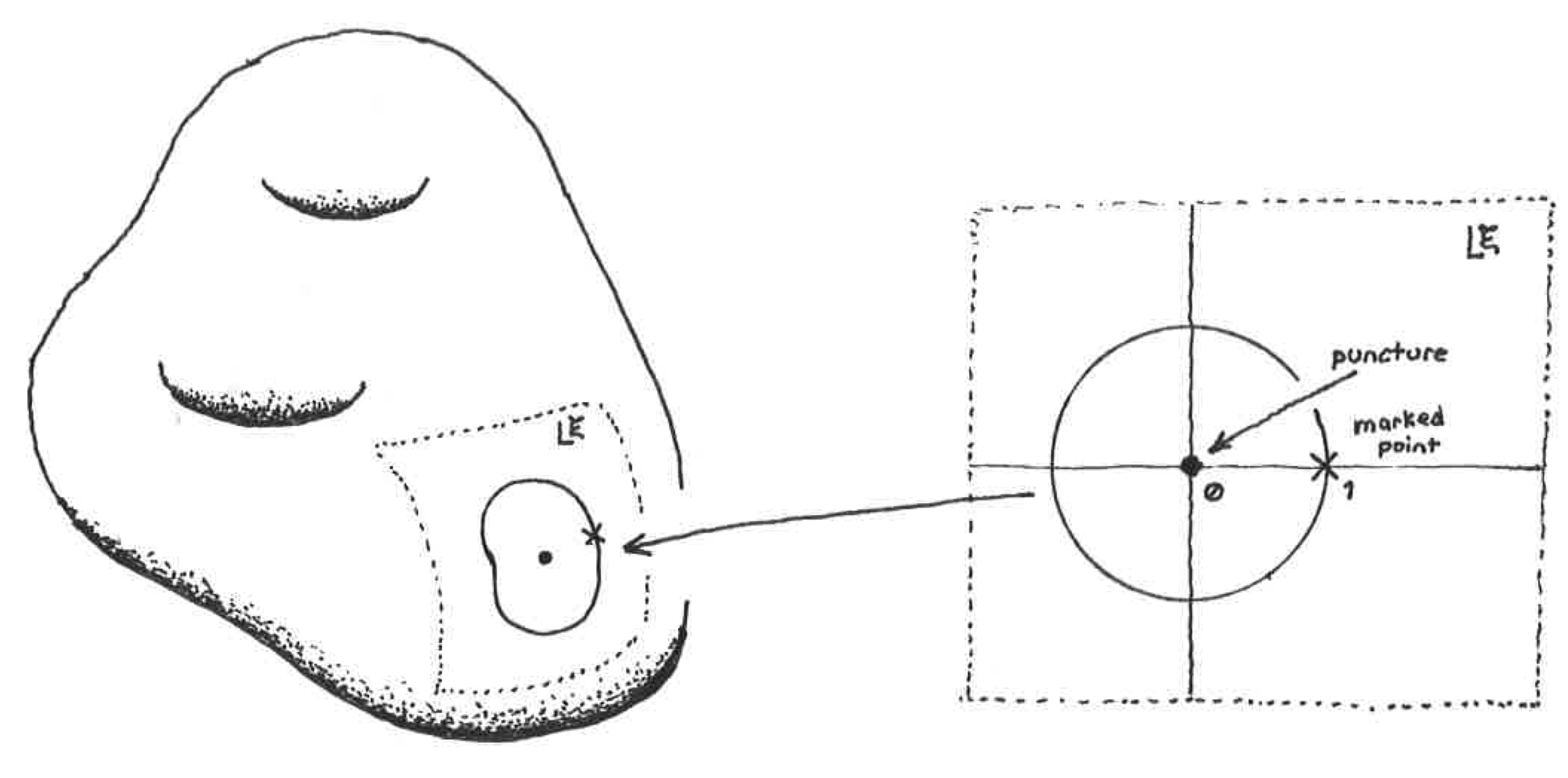}}
\end{wrapfigure}
subtle differences which in the present context are important. A closed string state is defined by a boundary condition for the worldsheet path integral along a contractible, simple closed curve on the Riemann surface. To fix the boundary condition it is necessary to parameterize the curve.  Conventionally, this is done in a specific way by specifying a point in the interior region of the closed curve, called a {\it puncture}, together with a marked point on the curve.\footnote{We could up front consider an arbitrary parameterization, but by deforming the surface this can be made equivalent to the kind of parameterization considered here.} There is a unique local coordinate system on the Riemann surface (denoted $\xi$) where the closed curve appears as a unit circle centered on the puncture at the origin, and the marked point sits on the real axis at $\xi=1$. This defines a parameterization $\sigma\in [0,2\pi]$ of the closed curve as the angle $e^{i\sigma}$ along the unit circle.  A local vertex operator defines a closed string state by inserting it on the puncture in this coordinate system,~and 

\begin{wrapfigure}{l}{.3\linewidth}
\centering
\resizebox{1.9in}{1.4in}{\includegraphics{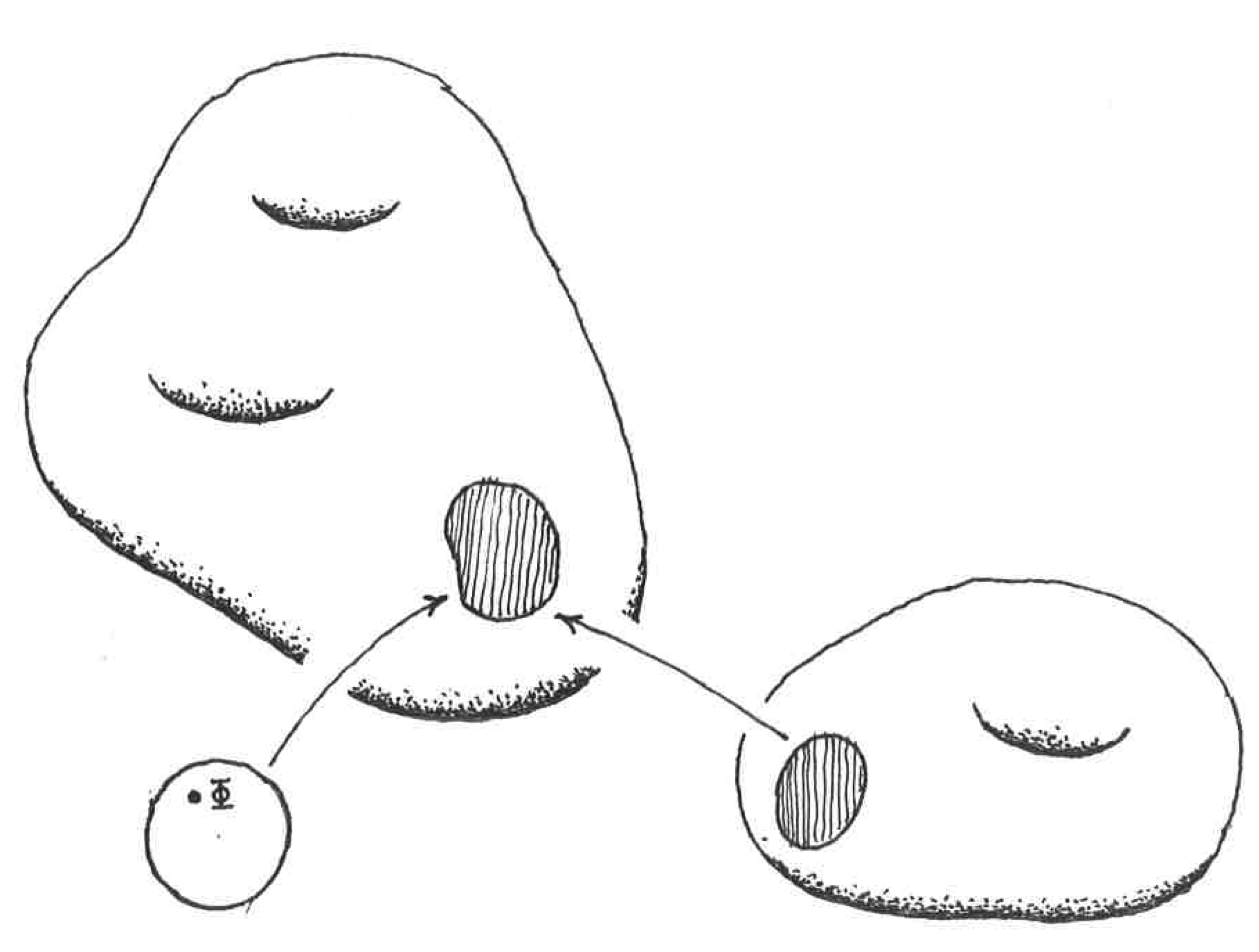}}
\end{wrapfigure}
\noindent evaluating the path integral on the surface completed with the interior of the closed curve and
vertex  operator at the puncture. This prescription defines what we may call a {\it Fock space state}. Representative examples are given by acting a plane wave vertex operator and a finite number of mode oscillators on the $SL(2,\mathbb{C})$ vacuum state of the CFT. However, there are other ways to prepare boundary conditions on the curve. We could, for example, insert a local vertex operator inside the unit disk but away from the puncture. More drastically, we can give a boundary condition by fusing two nontrivial Riemann surfaces together along the closed curve in a prescribed way. These examples do not correspond to Fock space states. However, Fock space states can be used to form a basis for $\H$, so that any state can be written as an infinite linear combination of them. The infinite sum, however, may not be describable as a local vertex operator inserted at the puncture.

This implies that the relevant question for defining an off-shell 4-point amplitude is not where to insert vertex operators, but how to patch unit disks onto the Riemann sphere. If $\xi_1,...,\xi_4$ represent coordinates on~the unit disks of $\Phi_1,...,\Phi_4$, we need to specify four one-to-one  holo-
\begin{wrapfigure}{l}{.45\linewidth}
\centering
\resizebox{3in}{2.1in}{\includegraphics{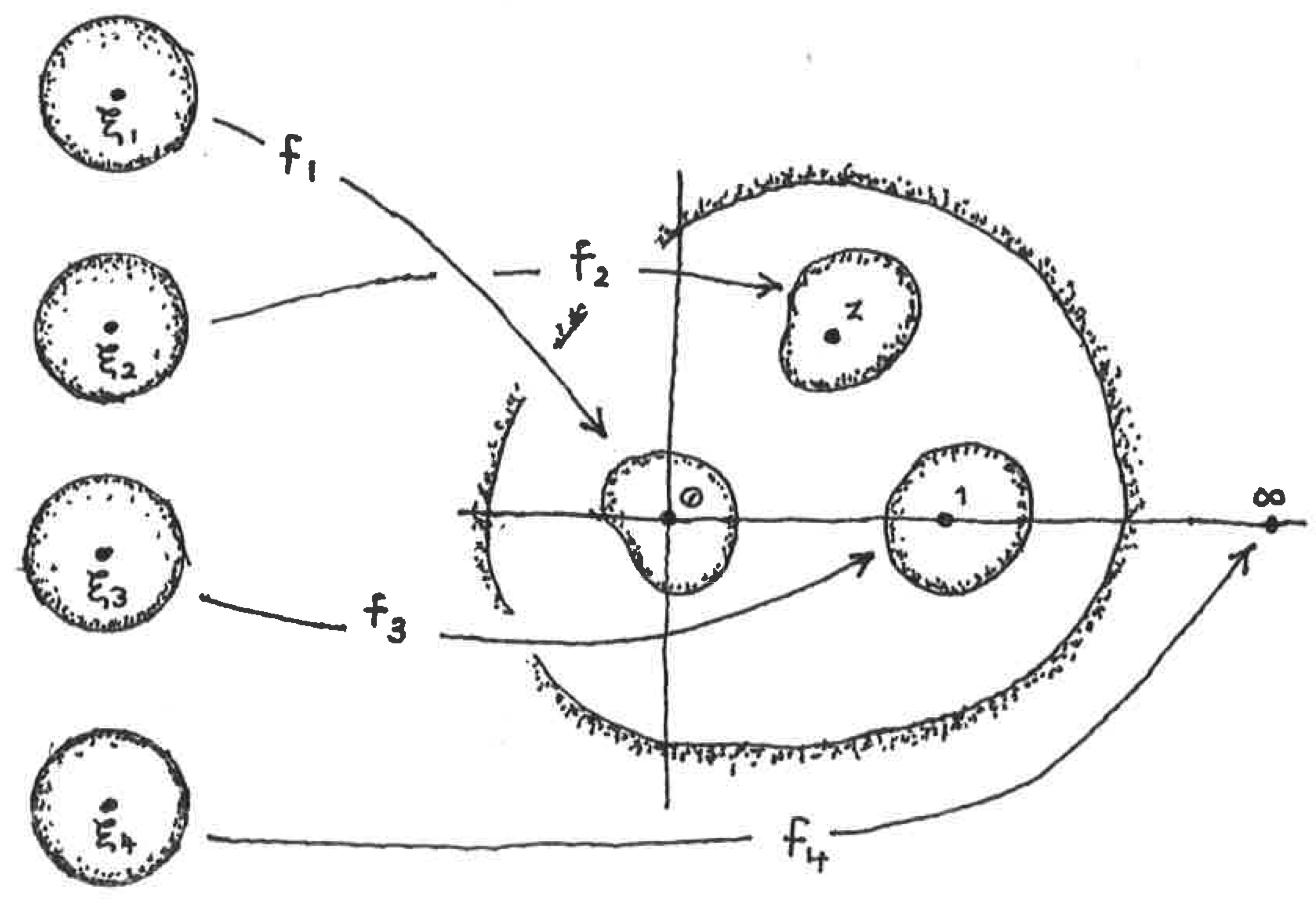}}
\end{wrapfigure}morphic maps of the disks into the Riemann sphere:
\begin{eqnarray}
z\lineup = f_1(\xi_1),\nonumber\\
z\lineup = f_2(\xi_2),\nonumber\\
z\lineup = f_3(\xi_3),\nonumber\\
z\lineup = f_4(\xi_4).
\end{eqnarray}
Some terminology: $\xi_1,...,\xi_4$ are called {\it local coordinates}, the maps $f_1,...,f_4$ are called {\it local coordinate}
{\it maps}, and the images of the unit disks on the Riemann sphere are called {\it local coordinate patches}. If we require that the local coordinate maps insert vertex operators at their standard positions in the global coordinate $z$ on the Riemann sphere, we have
\begin{equation}
f_1(0) = 0,\ \ \ f_2(0)=z,\ \ \ f_3(0)=1,\ \ \ f_4(0)=\infty.
\end{equation}
The proposal is that the off-shell amplitude should be defined by
\begin{equation}
\langle \mathcal{A}_{0,4}|\Phi_1\otimes\Phi_2\otimes\Phi_3\otimes\Phi_4 = \int_\mathbb{C}\Big\langle f_1\circ \Phi_1(0,0)f_2\circ\Phi_2(0,0)f_3\circ\Phi_3(0,0)f_4\circ\Phi_4(0,0)(b\ \mathrm{ghosts})\Big\rangle_\mathbb{C},
\end{equation}
where certain $b$-ghost insertions are needed to complete the definition of the measure; we turn to this shortly.

Therefore, the data of the off-shell amplitude is specified by a choice of local coordinate maps $f_1,...,f_4$ for each point $z$ in the moduli space. To avoid artificial singularities which could spoil BRST invariance, we require that the local coordinate maps vary continuously (though not necessarily analytically) as a function of $z$. To put this another way, an off-shell continuation is specified by a section of the fiber bundle $\P_{0,4}$ whose base is $\mathcal{M}_{0,4}$ and whose (infinite dimensional) fiber 
\begin{wrapfigure}{l}{.34\linewidth}
\centering
\resizebox{2.3in}{1.3in}{\includegraphics{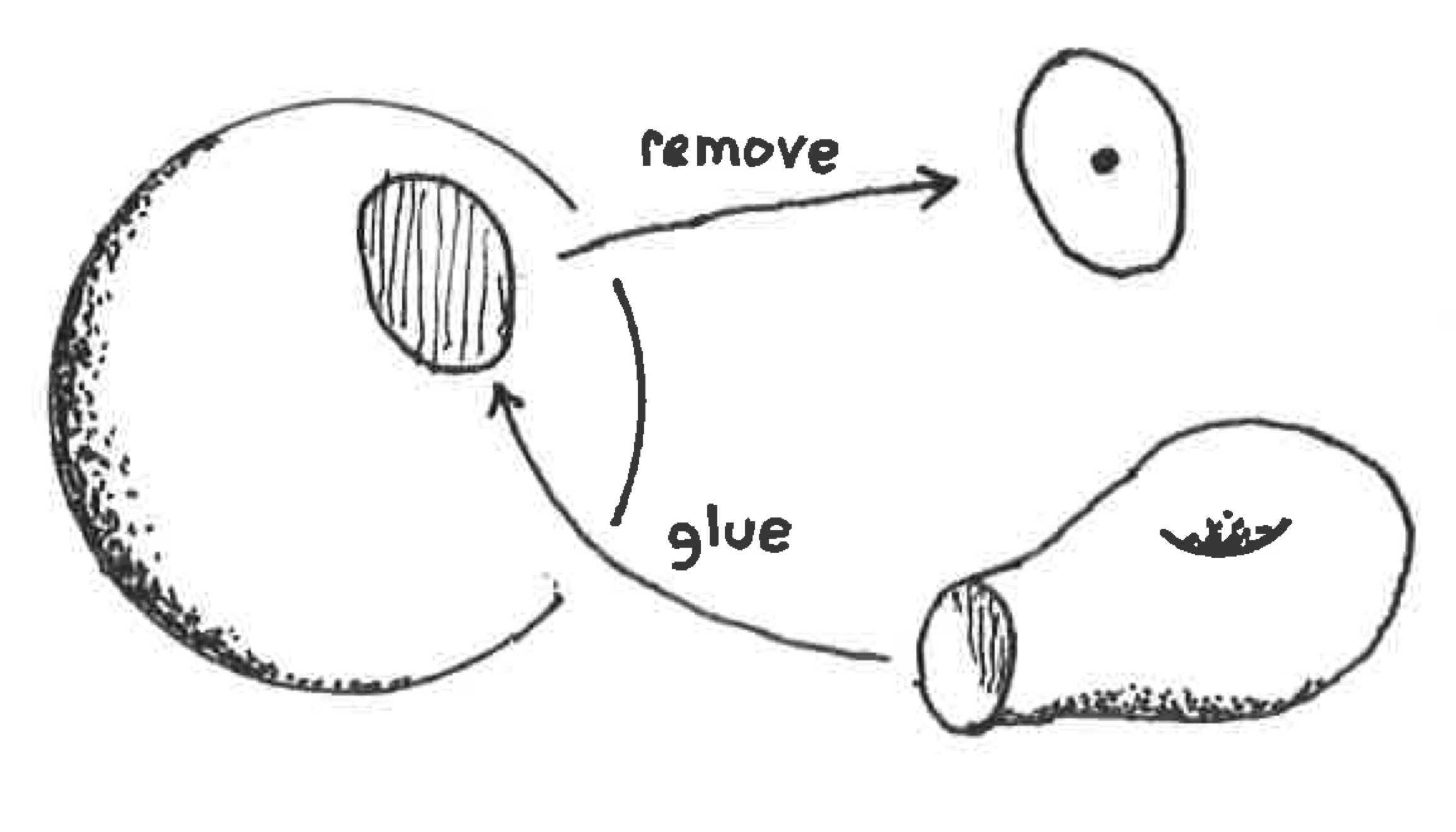}}
\end{wrapfigure} parameterizes possible local coordinate maps for a given $z\in\mathcal{M}_{0,4}$. There is an important condition 
the choice of section. We assume that the local coordinate patches on the Riemann sphere do not overlap. The reason is that we would like to be able to define the amplitude for arbitrary off-shell states, not only states defined by local vertex operators. Suppose for example that we wish to compute the amplitude with a state defined by a torus with hole removed. What we are supposed to do is cut out the local coordinate patch on the sphere, glue the hole of the torus to the hole on the sphere, and evaluate the path integral on the resulting surface. However, we can cut out a region of the sphere only once. If the local coordinate patches overlap, we cannot meaningfully cut them both out.

Since we need to integrate over the whole moduli space to get the off-shell amplitude, it appears we need a global section of $\P_{0,4}$. Unfortunately, there can be obstruction to finding a global section. This can be seen by considering a vector at the origin of the unit disk of $\Phi_2$:
\begin{equation} a \frac{\d}{\d\xi_2} + b\frac{\d}{\d \overline{\xi}_2}\ \ \ \mathrm{at}\ \xi_2=0.\end{equation}
Applying the local coordinate map $f_2$ defines a vector at a point $z$ on the Riemann sphere:
\begin{equation}a \frac{\d f_2}{\d\xi_2} \frac{\d}{\d z} + b \frac{\d\overline{f}_2}{\d\overline{\xi}_2}\frac{\d}{\d\zbar}\ \ \ \mathrm{at}\ z.\label{eq:vector}\end{equation}
This vector cannot vanish since $f_2$ must be one-to-one, which implies that its first derivative is nonvanishing. If we have a continuous global section of $\P_{0,4}$, it then appears that we have a continuous nonvanishing vector field on the sphere, which is impossible.\footnote{This argument is not quite correct, since it ignores punctures. In fact, since we assume that local coordinate patches do not overlap, the vector field \eq{vector} must vanish at the punctures. Then $\mathcal{P}_{0,4}$ {\it does} admit a continuous global section, but not of the kind that could be produced by an off-shell amplitude of closed SFT. A cleaner argument can be given in the context of the two loop tadpole amplitude, where analogous reasoning implies the existence of a continuous nonvanishing vector field on a genus 2 Riemann surface. Here the vector field must really be nonvanishing everywhere. I thank Ron Donagi for discussions of this point.} Therefore we have to work with sections that may have singularities. The established way to deal with this problem is to restrict the class of states to those which have no sense of ``direction" on the unit disk. That is, we will only attempt to define off-shell amplitudes for vertex operators satisfying
\begin{equation}e^{i\theta}\circ\Phi(0,0) = \Phi(0,0),\end{equation}
where the left hand side is a rotation of the unit disk. This is equivalent to the statement that states must be level matched. In fact, closed SFT requires an analogous condition involving the $b$-ghost. Without further motivation, we will state the conclusion, which is that the vector space $\Hhat$ on which off-shell amplitudes are defined is a linear subspace of $\H$ subject to the conditions 
\begin{equation} L_0^-\Phi = 0,\ \ \ b_0^-\Phi =0 \ \ \leftrightarrow\ \ \Phi\in \Hhat,\end{equation}
where $L_0^- = L_0 - \overline{L}_0$ and similarly for $b_0^-$. In a sense the $b_0^-$ constraint is the fundamental one, since the level matching condition follows from requiring a $b_0^-$ invariant subspace which is preserved by the action of $Q$. For states in $\Hhat$ there is an equivalence relation between the local coordinate maps defining the off-shell amplitude:
\begin{equation}f(\xi)\sim f(e^{i\theta}\xi).\end{equation}
Such states specify a boundary condition for the path integral along a simple, contractible closed curve on a Riemann surface containing a puncture. In this case there is no need to specify a marked point on the curve. This motivates the definition of the fiber bundle $\Phat_{0,4}$, which is $\P_{0,4}$ with the above equivalence relation imposed on the fiber. There are no obstructions to finding  a continuous global section of this bundle. In summary, the off-shell 4-point amplitude on the sphere can be defined on the subspace $\Hhat$ of states satisfying $b_0^-$ and level matching conditions, and is specified by an admissible global section of $\Phat_{0,4}$.

Now let us describe the $b$-ghost insertions needed to define a measure consistent with BRST invariance. For the on-shell amplitude, it is enough to integrate $b$-ghosts around $\Phi_2$; the reason is that the dependence of the integrand on $z$ only appears through the location of the $\Phi_2$ puncture. Off-shell, the story is more complicated since dependence on $z$ appears with all four punctures though the local coordinate maps. The local coordinate maps in fact must depend on $z$, since towards the boundaries of moduli space the local coordinate patches will need to adjust their shape to avoid overlapping. The key to defining the measure is understanding how the exterior derivative of the local coordinate maps can be computed through a contour integral of the energy momentum tensor around the puncture:
\begin{equation}d\Big(f_A\circ\Phi_A(0,0)\Big) = -dz f_A\circ\Big(T[v_z^A]\cdot\Phi_A(0,0)\Big) -d\zbar f_A\circ\Big(T[v_{\zbar}^A]\cdot\Phi_A(0,0)\Big),\ \ \ \ A=1,2,3,4,\ \ \ \ \label{eq:Schiff}\end{equation}
where
\begin{equation}T[v] \equiv \oint \frac{d\xi}{2\pi i}v(\xi) T(\xi) +\oint\frac{d\overline{\xi}}{2\pi i}\overline{v}(\overline{\xi})\overline{T}(\overline{\xi})\end{equation}
for a vector field $v(\xi)$ analytic in the vicinity of the unit circle. In the context of \eq{Schiff}, $v_z^A$ and $v_{\zbar}^A$ are called {\it Schiffer vector fields}. They are actually components of a 1-form on moduli space, but under conformal mapping of the unit disk they transform as vectors. Note that $v_z^A$ and $v_{\zbar}^A$ are independent vector fields, not related by complex conjugation. To see how the Schiffer vector fields are related to the local coordinate maps, we look at the $dz$ component of \eq{Schiff} (the $d\zbar$ component is similar) and write
\begin{equation}\frac{\d}{\d z}f_A\circ\Phi_A(0,0) = \frac{1}{\eps}\Big(f_A\circ\Phi_A(0,0)|_{z+\eps,\zbar} - f_A\circ\Phi_A(0,0)|_{z,\zbar}\Big)\end{equation}
for infinitesimal $\eps$. Then we have
\begin{equation}f_A\circ\Phi_A(0,0)|_{z+\eps,\zbar} = \left.f_A\circ\Big[(1-\eps T[v_z^A])\cdot\Phi_A(0,0)\Big]\right|_{z,\zbar}.\end{equation}
Since the energy momentum tensor generates infinitesimal conformal transformation, we have the relation 
\begin{equation}(1-\eps T[v_z^A])\cdot\Phi_A(0,0) = (1-\eps v_z^A)\circ\Phi_A(0,0),\end{equation}
where $1(\xi)=\xi$ is the identity conformal map.
\begin{exercise}Prove this\end{exercise}
\noindent Therefore
\begin{equation}f_A\circ\Phi_A(0,0)|_{z+\eps,\zbar} = \left.f_A\circ(1-\eps v_z^A)\circ\Phi_A(0,0)\right|_{z,\zbar}.\end{equation}
Explicitly indicating the dependence of the local coordinate maps on the moduli, this implies
\begin{equation}f_A(\xi_A,z+\eps,\zbar) = f_A(\xi_A-\eps v_z^A(\xi_A,z,\zbar),z,\zbar),\end{equation}
from which we learn
\begin{equation}v_z^A(\xi_A,z,\zbar) = -\left.\frac{\d f_A(\xi_A,z,\zbar)}{\d z}\right/\frac{\d f_A(\xi_A,z,\zbar)}{\d \xi_A}.\end{equation}
To describe the measure in a compact form it is helpful to introduce some notation. For each point in the moduli space we define a multilinear map
\begin{equation}\langle\Sigma_{0,4}|:\H^{\otimes 4}\to \H^{\otimes 0}\end{equation}
according to 
\begin{equation}\langle \Sigma_{0,4}| A_1\otimes A_2\otimes A_3\otimes A_4 = \Big\langle f_1\circ A_1(0,0) f_2\circ A_2(0,0)f_3\circ A_3(0,0)f_4\circ A_4(0,0)\Big\rangle_\mathbb{C}.\end{equation}
This is an example of a {\it surface state}. Generally, a surface state is defined by a correlation function on a Riemann surface with some number of states inserted into specified local coordinate patches. The surface state is BRST invariant
\begin{equation}\langle\Sigma_{0,4}|Q = 0.\end{equation}
The left hand side amounts to a BRST contour surrounding all four local coordinate patches on the Riemann sphere; we can then shrink the contour inside the sphere to a point, which gives zero. Next we introduce an operator valued 1-form on the moduli space:
\begin{equation}T^{(4)}:\H^{\otimes 4}\to \H^{\otimes 4}\end{equation}
defined by
\begin{eqnarray}
T^{(4)} \lineup = \Big(dz T[v^1_z]+ d\zbar T[v^1_{\zbar}]\Big)\otimes \mathbb{I}\otimes\mathbb{I}\otimes\mathbb{I} \, +\,  \mathbb{I}\otimes\Big(dz T[v^2_z]+ d\zbar T[v^2_{\zbar}]\Big)\otimes \mathbb{I}\otimes\mathbb{I}\nonumber\\
\lineup\ \ \ \ \,+\, \mathbb{I}\otimes\mathbb{I}\otimes\Big(dz T[v^3_z]+ d\zbar T[v^3_{\zbar}]\Big)\otimes \mathbb{I}\,+\, \mathbb{I}\otimes\mathbb{I}\otimes\mathbb{I}\otimes\Big(dz T[v^4_z]+ d\zbar T[v^4_{\zbar}]\Big).
\end{eqnarray}
To simplify signs, we assume that basis 1-forms on the moduli space are uniformly Grassmann odd objects. That is, they not only anticommute with each other, but anticommute with Grassmann odd worldsheet operators. See footnote \ref{footnote:5} for more explanation. Therefore $T$ is a Grassmann odd operator, even though the energy momentum tensor is Grassmann even. With the Schiffer vector fields defined as just described, it is easy to see that
\begin{equation}d\langle\Sigma_{0,4}| = -\langle \Sigma_{0,4}|T^{(4)}.\end{equation}
Using BRST invariance of the surface state, this can also be written as
\begin{equation}d\langle\Sigma_{0,4}| = -\langle \Sigma_{0,4}|b^{(4)} Q^{(4)},\end{equation}
where $b^{(4)}$ is defined as $T^{(4)}$ but with the $b$ ghost replacing the energy momentum tensor. Since $b^{(4)}$ involves the product of a Grassmann odd 1-form with a Grassmann odd $b$ ghost, in total it is Grassmann even. This last equation is the basic form we need for the measure; it relates the action of the BRST operator to the exterior derivative on the moduli space. However, multiplying the surface state with $b^{(4)}$ only gives a 1-form, and we need a 2-form to integrate over $\mathcal{M}_{0,4}$. The remedy, it turns out, is to simply multiply again by $b^{(4)}$. One can check that 
\begin{equation}d\Big(\langle \Sigma_{0,4}|b^{(4)}\Big) = -\langle\Sigma_{0,4}|\frac{1}{2!}(b^{(4)})^2 Q^{(4)}.\end{equation}
Note that $b^{(4)}$ is commuting and does not square to zero. This implies that 
\begin{equation}\langle\mathcal{A}_{0,4}| = \int_{\mathcal{M}_{0,4}}\langle \Sigma_{0,4}|\frac{1}{2!}(b^{(4)})^2\end{equation}
is an off-shell continuation of the 4-point amplitude on the sphere satisfying conditions (1)-(3) outlined at the beginning.

\begin{exercise}
Show that this reproduces the on-shell result when acting on states in $\mathcal{H}_Q$.
\end{exercise}

Let us describe how this story generalizes to arbitrary amplitudes. We consider a fiber bundle $\P_{g,n}$ whose base is the moduli space $\mathcal{M}_{g,n}$ of genus $g$ Riemann surfaces with $n$ punctures, and whose (infinite dimensional) fiber parameterizes possible ways of embedding unit disks around each puncture into the surface. $\Phat_{g,n}$ is the fiber bundle obtained after declaring that embeddings which differ by a rotation of the unit disks are equivalent. Given a point in $\P_{g,n}$ we define a  surface state
\begin{equation}\langle \Sigma_{g,n}|:\H^{\otimes n}\to\H^{\otimes 0}\end{equation}
which computes the correlation function on the Riemann surface at a point in $\mathcal{M}_{g,n}$ with vertex operators inserted at the punctures with the appropriate mapping of the unit disks $\xi^1,...,\xi^n$. The surface state is BRST invariant
\begin{equation}\langle\Sigma_{g,n}|Q^{(n)}=0.\end{equation}
Let $p^\alpha$ represent a coordinate system on $\P_{g,n}$; this includes $6g-6+2n$ coordinates on the moduli space together with an infinite number of other coordinates for the fiber. We have a list of Schiffer vector fields
\begin{equation}v_\alpha^A(\xi_A,p^\alpha),\ \ \ \ \ A=1,...,n\end{equation}
for each puncture, defined so that
\begin{equation}\frac{\d}{\d p^\alpha}\langle \Sigma_{g,n}|=-\langle \Sigma_{g,n}|\left(\sum_{A=1}^n\mathbb{I}^{\otimes A-1}\otimes T[v_\alpha^A]\otimes \mathbb{I}^{\otimes n- A}\right).\end{equation}
We introduce the exterior derivative on $\P_{g,n}$ and the Grassmann odd operator valued 1-form $T^{(n)}$:
\begin{equation}
d = dp^\alpha \frac{\d}{\d p^\alpha},\ \ \ \ T^{(n)}=dp^\alpha\sum_{A=1}^n\mathbb{I}^{\otimes A-1}\otimes T[v_\alpha^A]\otimes \mathbb{I}^{\otimes n- A},
\end{equation}
so that 
\begin{equation}
d\langle\Sigma_{g,n}| = -\langle \Sigma_{g,n}|T^{(n)}.
\end{equation}
Again we assume that the basis 1-forms $dp^\alpha$ are Grassmann odd.\footnote{\label{footnote:5}Operator valued forms on $\mathcal{P}_{g,n}$ come with two $\mathbb{Z}_2$ gradings: {\it Form degree} $|\mathcal{O}|_f$, which counts the rank of a form mod $\mathbb{Z}_2$, and {\it worldsheet Grassmann parity} $|\mathcal{O}|_w$, which counts the Grassmann parity of the worldsheet operators which multiply the basis forms. The total Grassmann parity is defined as
\begin{equation}|\mathcal{O}| = |\mathcal{O}|_f+|\mathcal{O}|_w.\end{equation} Other discussions such as \cite{Zwiebach} typically assume that basis forms commute through all worldsheet operators. The relation to our presentation can be understood as a difference in the definition of products of operator-valued forms. The relation is precisely
\begin{equation}(\mathcal{O}_1\mathcal{O}_2)_\text{here} = (-1)^{|\mathcal{O}_1|_w|\mathcal{O}_2|_f}(\mathcal{O}_1\mathcal{O}_2)_\text{there}.\end{equation} If an operator valued form is expressed so that all basis 1-forms are commuted to the left, the two definitions agree.} Note a slight difference in presentation from our discussion of the 4-point amplitude. We are discussing Schiffer vector fields and differential forms on $\P_{g,n}$ rather than on the moduli space. In the context of the 4-point amplitude we assumed that a section of $\P_{0,4}$ had been specified, so we could pull back forms on $\P_{0,4}$ into forms on the moduli space. Using $d^2=0$ we can show that
\begin{equation}\langle\Sigma_{g,n}|(dT^{(n)}-(T^{(n)})^2)=0.\label{eq:dT}\end{equation}
Next we introduce the operator valued 1-form $b^{(n)}$ defined by replacing the energy momentum tensor with $b$-ghost in $T^{(n)}$. It is Grassmann even and satisfies
\begin{equation}[b^{(n)},Q^{(n)}]=T^{(n)}.\end{equation}
Since the OPE of the $b$-ghost with the energy momentum tensor takes the same form as that of the energy momentum tensor with itself, \eq{dT} implies 
\begin{equation}\langle\Sigma_{g,n}|\left(db^{(n)}-\frac{1}{2}[T^{(n)},b^{(n)}]\right)=0.\end{equation}
With this we can define a measure on $\P_{g,n}$
\begin{equation}\langle \Omega_{g,n}|\equiv \left(-\frac{1}{2\pi i}\right)^{\mathrm{dim}_{\mathbb{C}}\mathcal{M}_{g,n}}\langle \Sigma_{g,n}|e^{b^{(n)}}.\end{equation}
The measure contains differential forms of every degree, and can be integrated over any submanifold of $\P_{g,n}$; ultimately, we want to integrate it over a global section of $\Phat_{g,n}$. The factors of $\frac{1}{2\pi i}$ are put there for later convenience in the context of closed SFT. The fundamental relation characterizing the measure is the so-called {\it BRST identity}:
\begin{equation}d\langle \Omega_{g,n}| = -\langle\Omega_{g,n}|Q^{(n)}.\end{equation}
\begin{exercise} Prove this\end{exercise}
\noindent Given a global section $\sigma(\Phat_{g,n})$ of $\Phat_{g,n}$, we can define an off-shell $n$-point amplitude at genus $g$ 
\begin{equation}\langle \mathcal{A}_{g,n}| = \int_{\sigma(\Phat_{g,n})}\langle \Omega_{g,n}|.\end{equation}
where the integrand is the pullback of the measure on the section. This definition achieves all desired properties (1)-(3) of an off-shell amplitude outlined at the beginning.

\pagebreak

\section{Lecture 2: Feynman Diagrams}

The off-shell amplitudes of closed SFT are of a special kind, since they all derive from a common set of vertices connected by propagators to form Feynman diagrams. Usually the Feynman graph expansion is deduced from the action; however, we don't know the form of the closed SFT action (yet), and presently it is actually easier to go the other way: construct a Feynman graph expansion 
\begin{wrapfigure}{l}{.28\linewidth}
\centering
\resizebox{1.9in}{1in}{\includegraphics{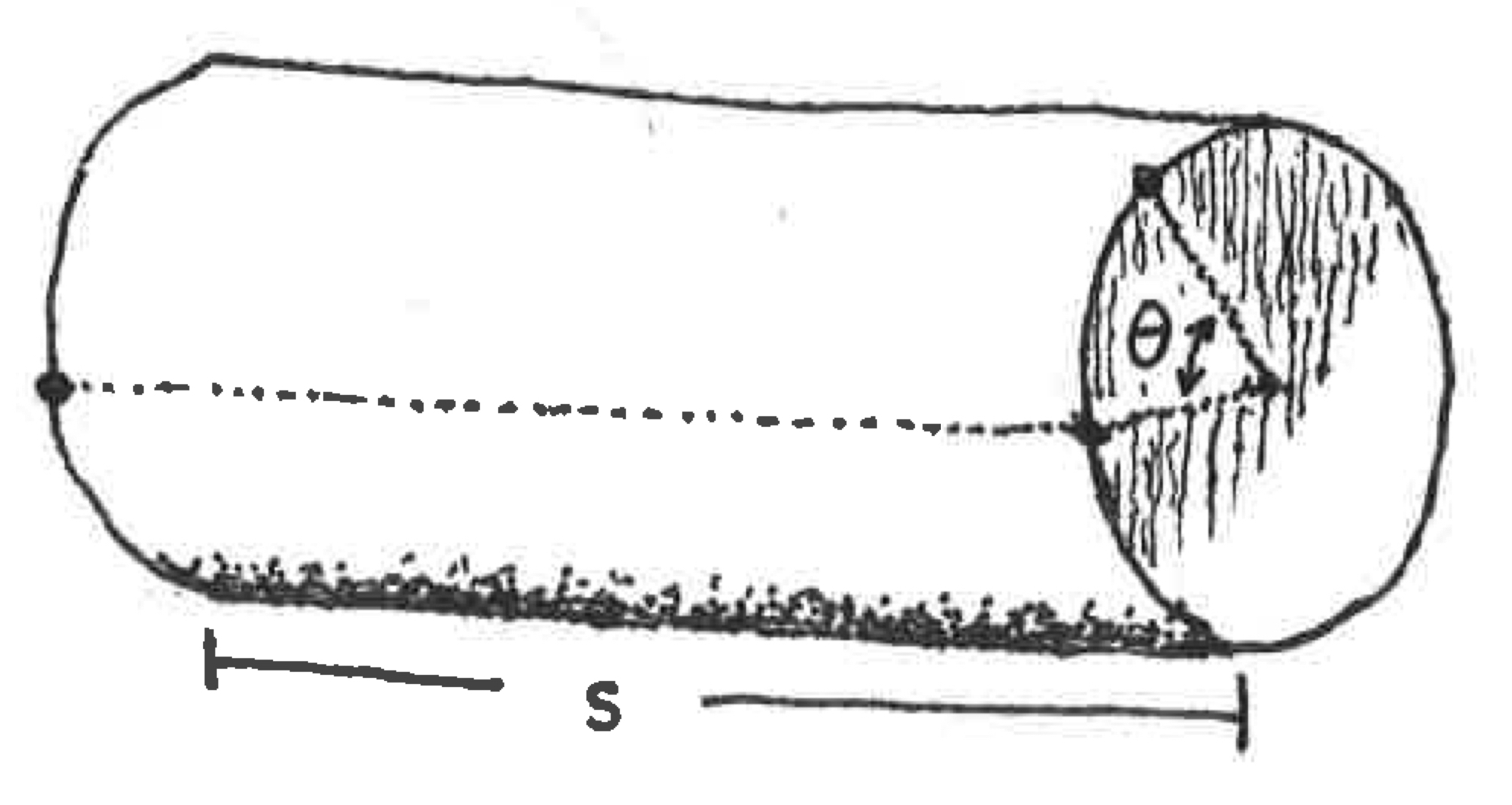}}
\end{wrapfigure}
of off-shell amplitudes, and use this to deduce the necessary form of the action. To construct Feynman diagrams we need at least a propagator and cubic vertex. 

The propagator can be visualized as a tube of closed string worldsheet of length $s$ and twist angle $\theta$. Specifically, $s,\theta$ will be coordinates on some portion of the moduli space of Riemann surfaces described by a Feynman diagram where the propagator appears, and we must  integrate over $s,\theta$ as part of the integration over the moduli 

\noindent space. A tube of length $s$ and twist angle $\theta$ can be attached to a state by applying the operator
\begin{wrapfigure}{l}{.18\linewidth}
\centering
\resizebox{1.3in}{1.2in}{\includegraphics{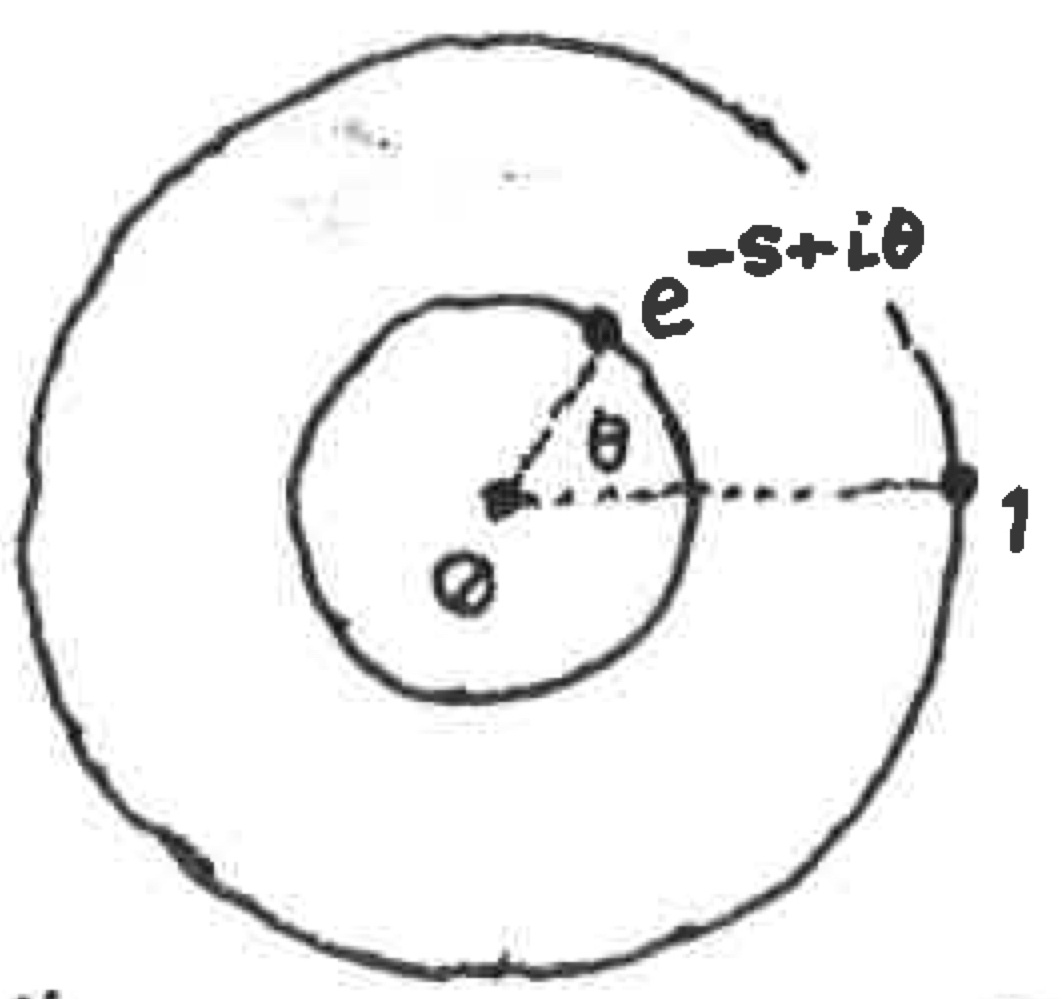}}
\end{wrapfigure}
\begin{equation}e^{-sL_0^+}e^{i\theta L_0^-},\ \ \ \ \ \ L_0^\pm = L_0\pm \overline{L}_0.\end{equation}
In the coordinate system of radial quantization, this operator implements a conformal transformation $\xi \to e^{-s+i\theta}\xi$ which shrinks and rotates the unit disk. The annular region $e^{-s}<|\xi|<1$ is the surface of the propagator in this frame. Integrating over $s,\theta$ and multiplying by the appropriate $b$-ghost insertions for the measure gives\footnote{The Schwinger integral representation of the propagator is only valid for states whose $L_0^+$ eigenvalue is positive. Unfortunately, intermediate states with negative eigenvalues appear generically in string amplitudes, so this representation of the propagator leads to unphysical divergences near the boundaries of moduli space. The simple resolution is to define the propagator as a sum over intermediate states divided by the eigenvalue of $L_0^+$. This option is not available in the standard worldsheet formulation. In loop amplitudes, the definition of the propagator must be refined with a stringy equivalent of the $i\eps$ prescription. Discussions of this appear in \cite{Wittenieps,Cutkosky,equivalence}.}
\begin{eqnarray}
\mathrm{propagator} \lineup = b_0^+b_0^- \int_0^\infty ds\int_0^{2\pi} \frac{d\theta}{2\pi}e^{-s L_0^+ +i\theta L_0^-}= \frac{b_0^+}{L_0^+}b_0^-\delta(L_0^-).
\end{eqnarray}
The $b$-ghost insertions ensure that the propagator is BRST invariant, in the sense that the commutator with $Q$ only gives contributions from the boundaries of the portion of moduli space represented by the propagator:
\begin{equation}[Q,\mathrm{propagator}] = b_0^- \delta(L_0^-)(1-e^{-\infty L_0^+}).\end{equation}
The $e^{-\infty L_0^+}$ contribution represents a true boundary of moduli space, where the closed string travels over an infinite distance $s\to\infty$. The $s\to0$ limit is not a boundary of moduli space; it is only a boundary of a portion of moduli space covered by the propagator. This will be canceled against other diagrams, so the BRST transformation of the full amplitude will only give contributions from the true boundaries of moduli space. The factor $1/L_0^+$ is analogous to $1/(p^2+m^2)$ familiar from the propagator of QFT. The factor 
\begin{equation}\delta(L_0^-) = \int_0^{2\pi} \frac{d\theta}{2\pi}e^{i \theta L_0^-}\end{equation}
is different; it is the projector onto level matched states. Its presence reflects the fact that a closed string, unlike a point particle, can ``twist." This twisting motion must be accommodated by an additional factor in the propagator. Note that the operator
\begin{equation}b_0^-\delta(L_0^-)\end{equation}
is BRST invariant; it will play an important role in the following.

Next we need a cubic vertex
\begin{equation}\langle \mathcal{V}_{0,3}|:\Hhat^{\otimes 3}\to\Hhat^{\otimes 0},\end{equation}
which can be defined by an off-shell 3-point amplitude at genus zero:
\begin{equation}\langle\mathcal{V}_{0,3}| =\langle\mathcal{A}_{0,3}| = \langle\Omega_{0,3}| = \langle \Sigma_{0,3}|.\end{equation}
There is some degeneracy of notation here. $\langle \mathcal{V}_{g,n}|$ will denote the $n$-point vertex in the closed SFT action at genus $g$; $\langle\mathcal{A}_{g,n}|$ as before is an off-shell $n$-point amplitude at genus $g$; $\langle \Omega_{g,n}|$ is the measure for this amplitude at some point in $\P_{g,n}$, and $\langle\Sigma_{g,n}|$ is the surface state at some point in $\P_{g,n}$. For $n=3$ and $g=0$ these notions are identical since the moduli space of the 3-punctured sphere is zero dimensional. The cubic vertex is specified by three local coordinate maps
\begin{equation}
\langle \mathcal{V}_{0,3}| A_1\otimes A_2\otimes A_3 = \Big\langle f_1\circ A_1(0,0)f_2\circ A_2(0,0) f_3\circ A_3(0,0)\Big\rangle_\mathbb{C}
.\label{eq:cubic}\end{equation}
Actually---as a matter of definition---a string field theory vertex should be symmetric under interchange of any pair of states. This is because the contribution to the action appears as $\langle \mathcal{V}_{g,n}|\Phi^{\otimes n}$, and since all $\Phi$s are identical, any asymmetric part will drop out.\footnote{One can of course add an asymmetric part, but in practice only the symmetric part of $\langle \mathcal{V}_{0,3}|$ will appear in computations. However, for the open string it is often useful (but not necessary) to define vertices which are {\it cyclically} symmetric, not totally symmetric. This is related to the notion of color ordering in gauge theory amplitudes, which does not have a clear gravitational counterpart.}  So really we should define $\langle \mathcal{V}_{0,3}|$ to be the totally symmetric part of the right hand side of \eq{cubic}. But then $\langle \mathcal{V}_{0,3}|$ would be defined by an ``average" of six distinct sections of $\Phat_{0,3}$. It is straightforward to generalize our previous construction of off-shell amplitudes accounting for the possibility of averages of sections. 
\begin{wrapfigure}{l}{.37\linewidth}
\centering
\resizebox{2.8in}{2.3in}{\includegraphics{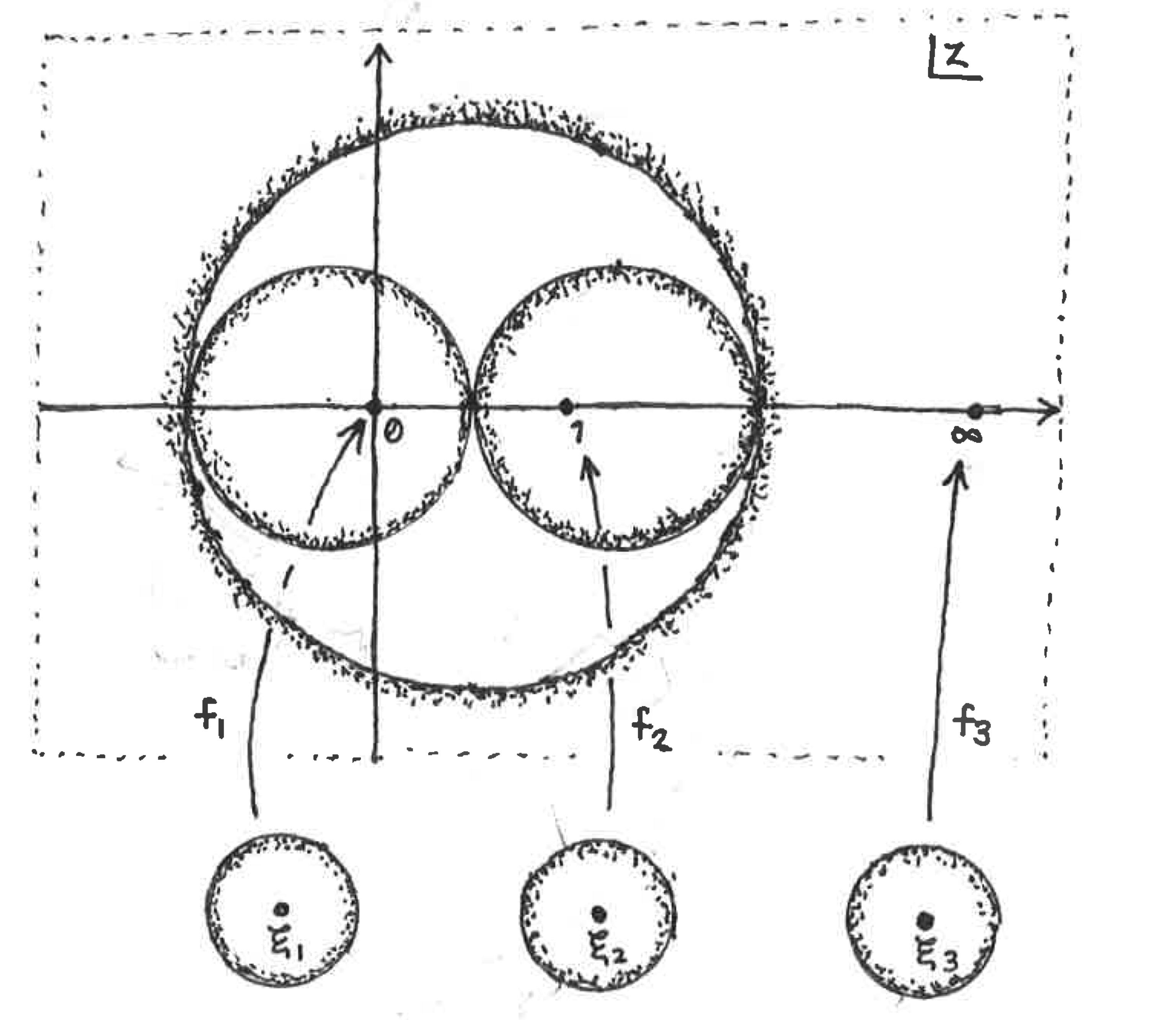}}
\end{wrapfigure}
But traditionally this is not considered ideal, since it factorially multiplies the number of off-shell amplitudes you need to keep track of. Therefore, we will require that the six sections of $\Phat_{0,3}$ obtained 
by permutations are identical; then \eq{cubic} is admissible without having to take the symmetric part. This amounts to a condition on the choice of local coordinate maps.

One simple choice of cubic vertex is defined by $SL(2,\mathbb{C})$ maps
\begin{equation}f_1(\xi_1) = \frac{2\xi_1}{\xi_1+3},\ \ \ \ f_2(\xi_2)=\frac{3-\xi_1}{3+\xi_1},\ \ \ \ f_3(\xi_3) = \frac{\xi_3+3}{2\xi_3}.\end{equation}
$f_1$ maps the first puncture to $0$, $f_2$ maps the second puncture to $1$, and $f_3$ maps the third puncture to $\infty$. Since these are fractional linear transformations, they map circles into circles. A picture of the local coordinate patches in the complex plane is shown above. A more famous choice of cubic vertex is the Witten vertex, defined by local coordinate maps
\begin{eqnarray}
f_1(\xi_1) = \lineup e^{2\pi i/3}\left(\frac{1+i \xi_1}{1- i\xi_1}\right)^{2/3},\ \ \ \ f_2(\xi_2) = \left(\frac{1+i \xi_2}{1- i\xi_2}\right)^{2/3}\ \ \ \ f_3(\xi_3) = e^{-2\pi i/3}\left(\frac{1+i \xi_3}{1- i\xi_3}\right)^{2/3}.
\end{eqnarray}
 The maps place punctures at $e^{2\pi i/ 3},1$ and $e^{-2\pi i/3}$. By $SL(2,\mathbb{C})$ transformation, we can alternatively place the punctures at $0,1$ and $\infty$. The resulting local coordinate patches are shown below:\\

\begin{wrapfigure}{l}{1\linewidth}
\centering
\resizebox{4.3in}{1.7in}{\includegraphics{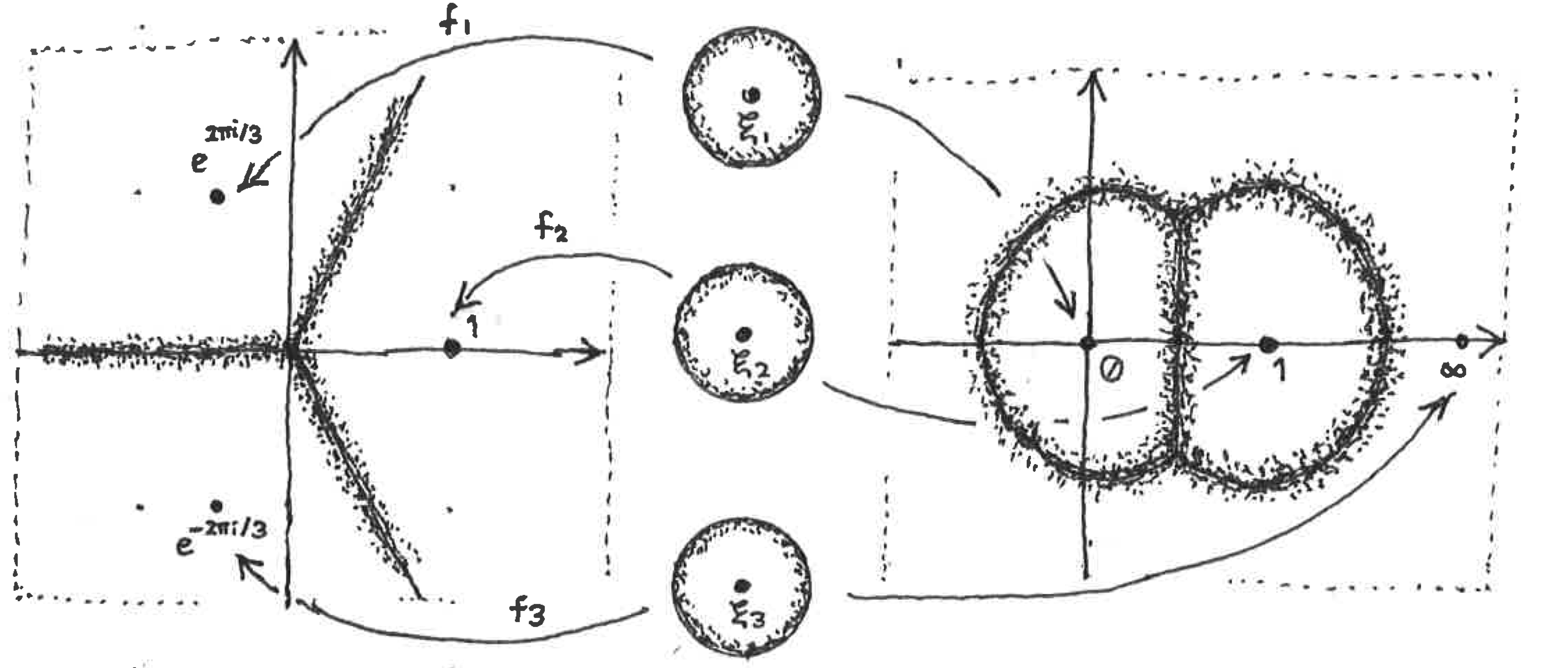}}
\end{wrapfigure}

\noindent \\ \\ \\ \\ \\ \\ \\ \\ \\ \\ \\ Unlike the $SL(2,\mathbb{C})$ vertex, the local coordinate patches of the Witten vertex fill the entire sphere. In closed SFT, the Witten vertex is notable as it is defined by the metric of minimal area on the 3-punctured sphere subject to the condition that nontrivial closed curves have length $2\pi$ or greater. This is an example of Zwiebach's generalized minimal area problem \cite{min_area}, which specifies a global section of $\Phat_{g,n}$ corresponding to a unique closed SFT. At present, the minimal area problem gives what is probably the closest thing to a canonical choice of field variable for closed SFT. However, calculations with vertices defined by minimal area metrics are extraordinarily difficult. The  $SL(2,\mathbb{C})$ vertex is simpler in this respect, at least for questions concerning low order amplitudes~at

\begin{wrapfigure}{l}{.4\linewidth}
\centering
\resizebox{2.5in}{2.5in}{\includegraphics{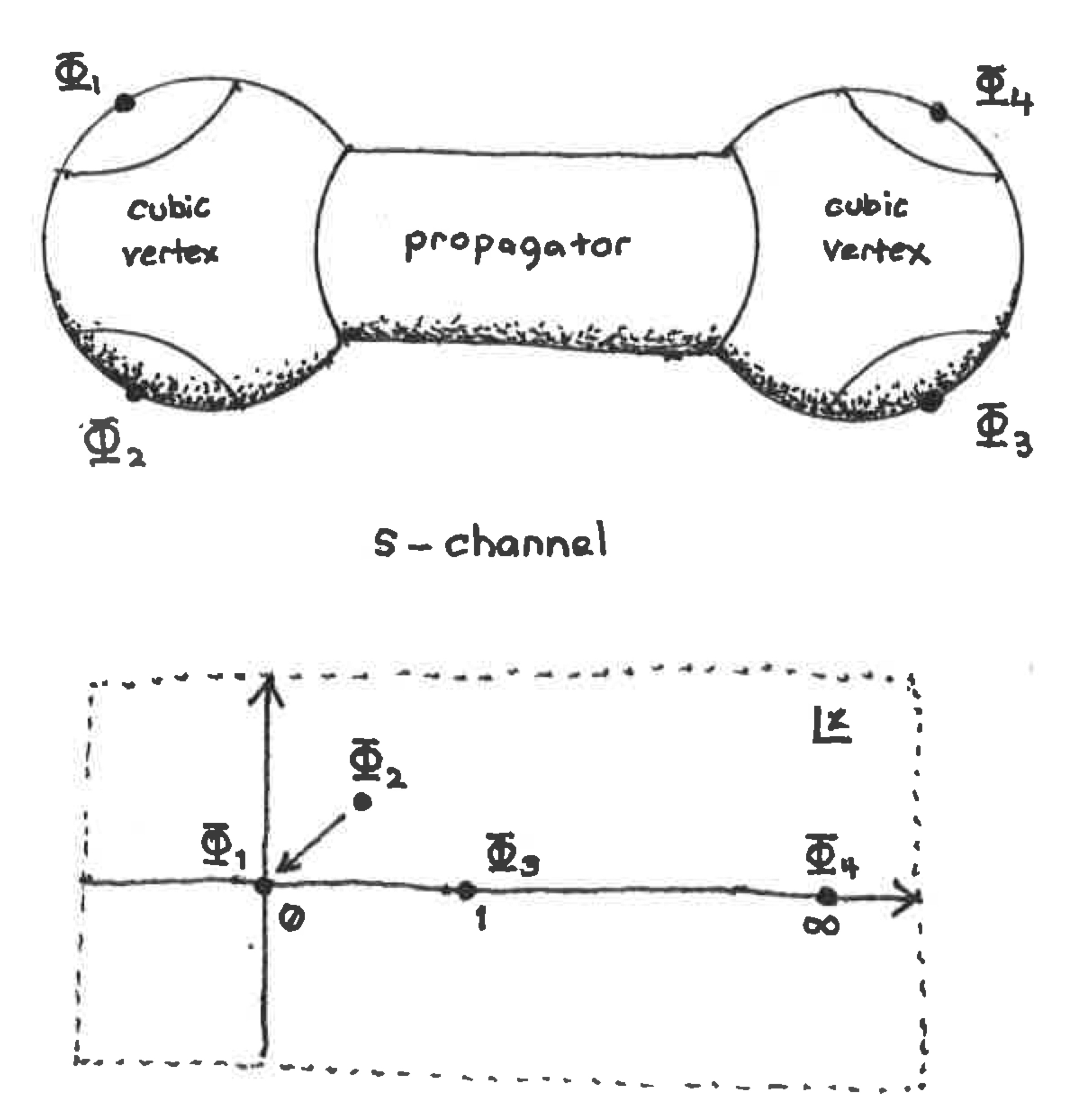}}
\end{wrapfigure}
\noindent genus zero. A more recent proposal is to define vertices using metrics of constant negative  curvature  \cite{Pius1}, which  is  proposed  to  lead  to some  simplification  in  performing moduli space integrals. The maps of the cubic vertex in this case are defined by the  modular $\lambda$ function, and the local coordinate patches resemble those of the $SL(2,\mathbb{C})$ vertex.

\begin{exercise}
Show that the $SL(2,\mathbb{C})$ vertex and the Witten vertex are invariant under permutations of the states.
\end{exercise}

Now we can attach cubic vertices and propagators to form Feynman diagrams. Consider the $s$-channel contribution to the 4-point amplitude on the sphere, shown left. This will define a section of $\Phat_{0,4}$ at least in the vicinity of the corner of the moduli space where the $\Phi_2$ puncture approaches the $\Phi_1$ puncture. One might guess that $s$-$t$-$u$ 

\begin{wrapfigure}{l}{.35\linewidth}
\centering
\resizebox{2.4in}{4in}{\includegraphics{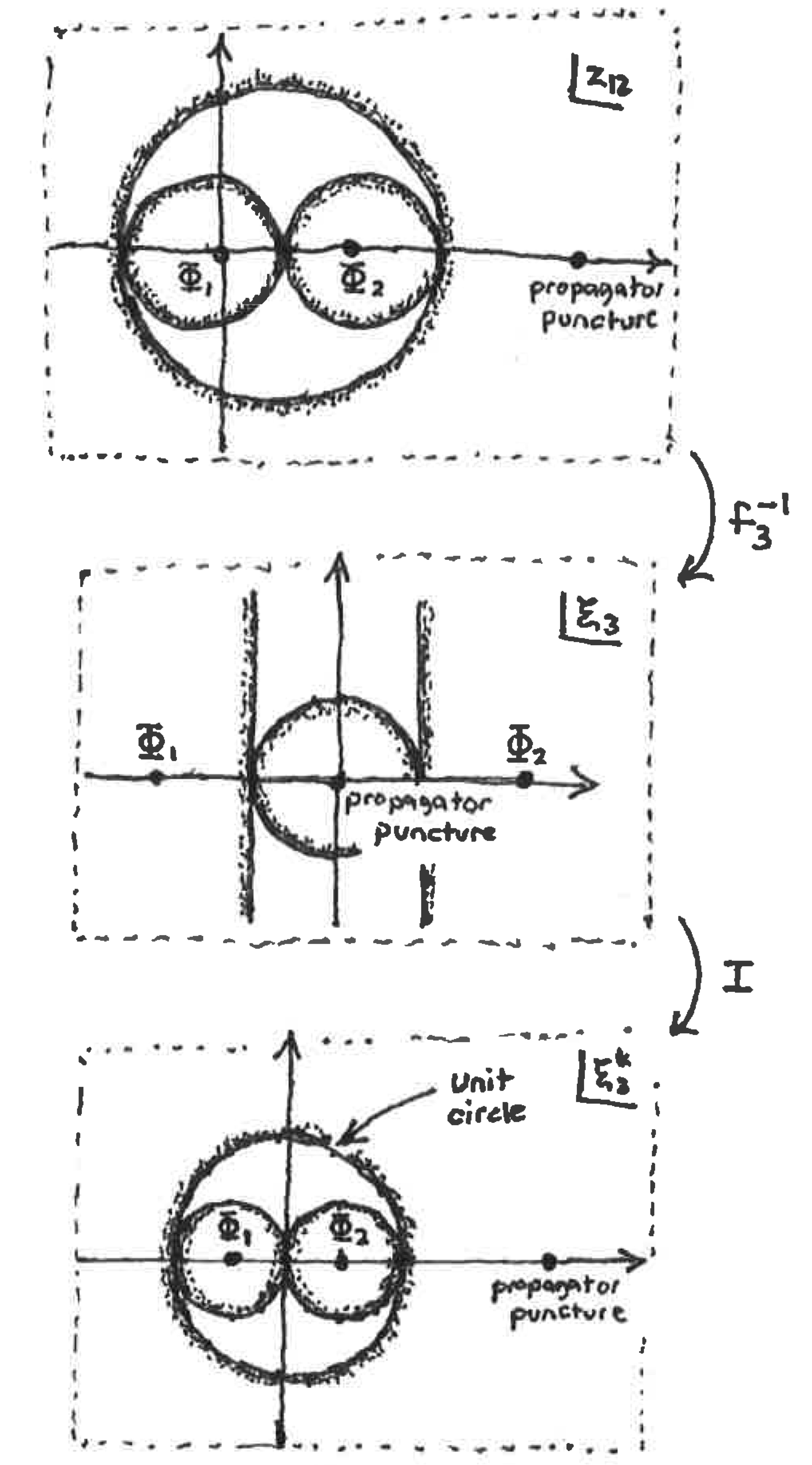}}
\resizebox{2.2in}{1.4in}{\includegraphics{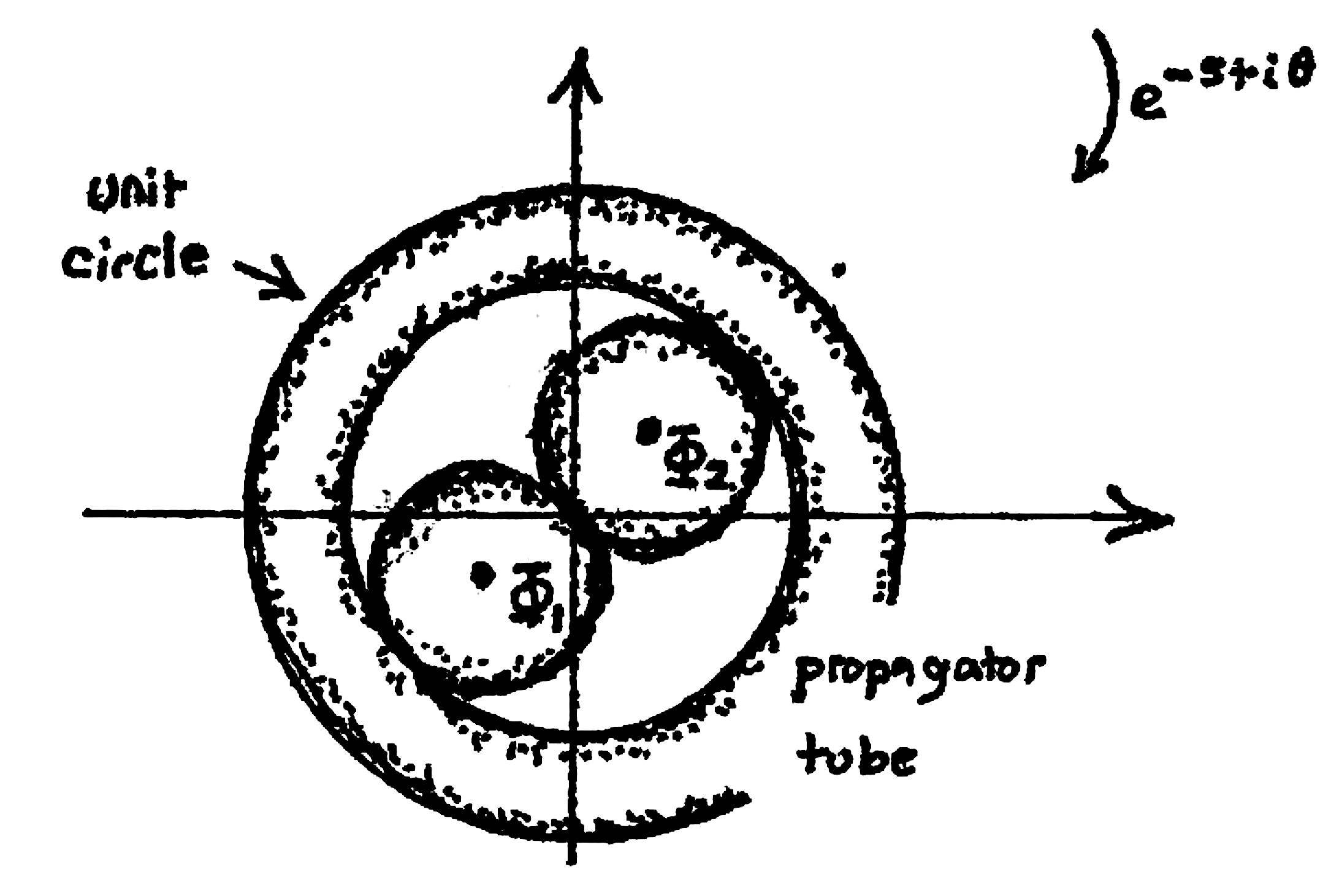}}
\end{wrapfigure}
\noindent channel duality will imply that this diagram extends to give a global section of $\Phat_{0,4}$. But, as we will see, this does not happen. The picture of a tube connecting spheres~is an intuitive way to visualize the $s$-channel process, but is not ideal for explicitly characterizing the resulting section~of 
$\Phat_{0,4}$. For this we need to express the process in terms of the global coordinate $z$ on the Riemann sphere. If we assume the $SL(2,\mathbb{C})$ cubic vertex, this can be done quite explicitly. It requires a number of steps. We start with the cubic vertex coupling $\Phi_1$ and $\Phi_2$, represented in the complex plane with the $\Phi_1$ puncture at the~origin,  the $\Phi_2$ puncture at $1$, and the puncture representing the propagator at  infinity. We~then perform a conformal transformation 
\begin{equation}\xi_3 = f_3^{-1}(z_{12}),\end{equation}
which transforms the local coordinate patch of the propagator into the unit disk. The local coordinate patches of $\Phi_1$ and $\Phi_2$ become semi-infinite planes in this coordinate. Next we perform  an inversion
\begin{equation}\xi_3^* =  I(\xi_3)=\frac{1}{\xi_3},\end{equation}
which interchanges the interior and exterior of the unit disk, mapping the puncture of the propagator back to infinity. What has been accomplished  by these transformations is we have mapped the cubic vertex, minus the local coordinate patch of the propagator, into the unit disk. The unit disk can then be viewed as a state representing the ``product" of $\Phi_1$ and $\Phi_2$. Attaching a propagator tube to this state shrinks 

\begin{wrapfigure}{l}{.45\linewidth}
\centering
\resizebox{3.2in}{2.9in}{\includegraphics{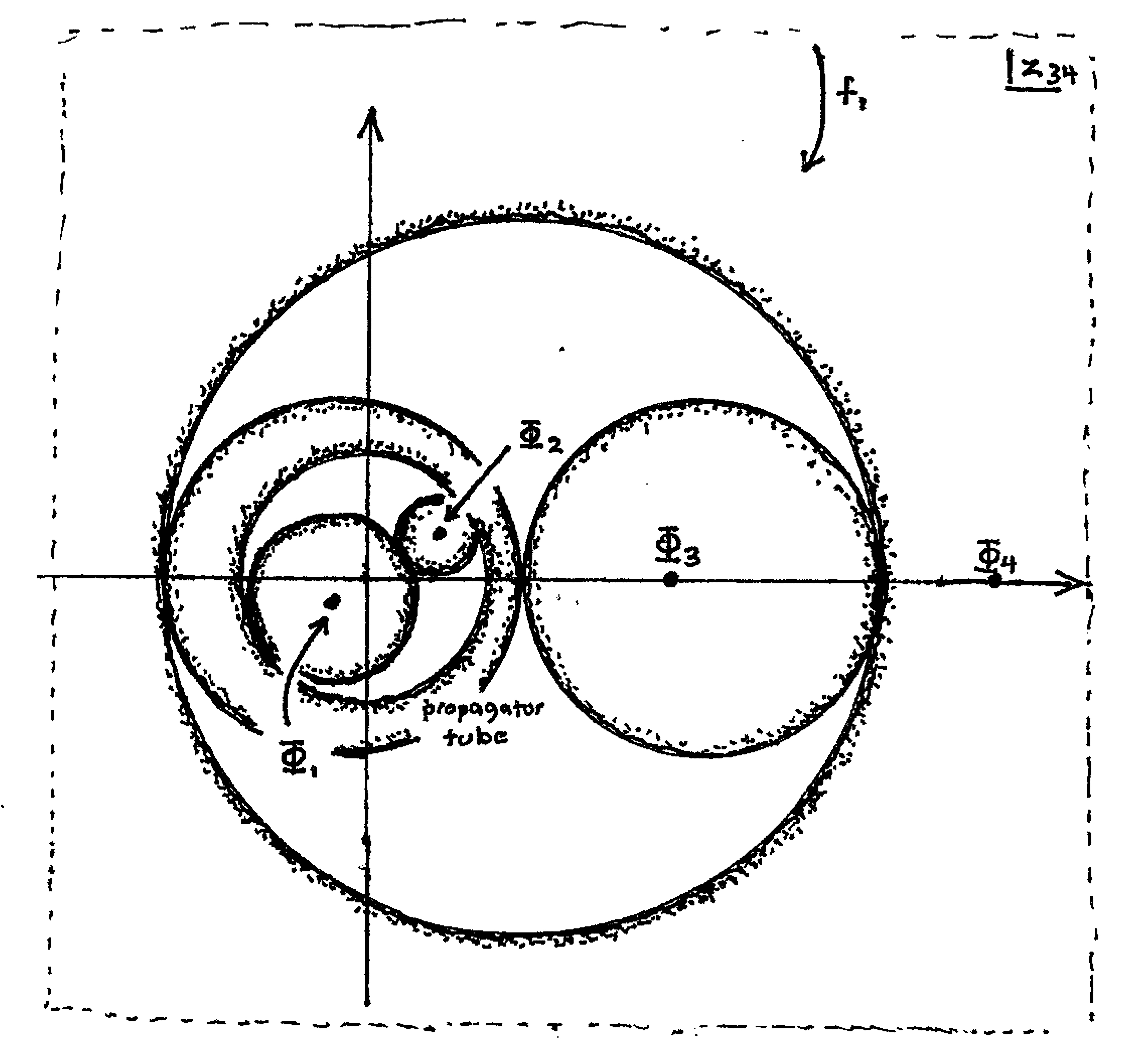}}
\end{wrapfigure}
\noindent and rotates the disk by a factor of $e^{-s+i\theta}$:
\begin{equation}\xi_\mathrm{propagator} = e^{-s+i\theta}\xi_3^*.\end{equation}
Now we can use 
\begin{equation}z_{34} = f_1(\xi_\mathrm{propagator})\end{equation}
to map the propagator tube attached to the vertex coupling $\Phi_1$ and $\Phi_2$ into the local coordinate patch around the origin of the vertex coupling $\Phi_3$ and $\Phi_4$. A further $SL(2,\mathbb{C})$ transformation can map the $\Phi_1$ puncture to the origin, so that the position of $\Phi_2$  is a coordinate on the moduli space $\mathcal{M}_{0,4}$. Since all the conformal 
transformations are known, we have explicit formulas for the local coordinate maps specifying a section of $\Phat_{0,4}$ in the neighborhood of the $s$-channel degeneration. Let us make a few comments:

\begin{itemize}
\item It is clear that in the $s$-channel diagram the puncture of $\Phi_2$ can never wander very far from the puncture of $\Phi_1$. In the coordinate $z_{34}$, the $\Phi_2$ puncture cannot leave the local coordinate patch around the origin. Therefore, the $s$-channel leaves much of the moduli space unaccounted for.
\item The quantity $\lambda = e^{-s+i\theta}$ is a holomorphic coordinate on the part of the moduli space covered by the $s$-channel diagram. It is clear that $\overline{\lambda}$ appears nowhere in the above sequence of conformal transformations, so the $s$-channel diagram produces a holomorphic local section of $\P_{0,4}$. Therefore we can analytically continue the section outside the domain of the $s$-channel diagram. However, this analytic continuation cannot produce an admissible global section of $\Phat_{0,4}$. This is because, in the $z_{34}$ coordinate, the local coordinate patches of $\Phi_3$ and $\Phi_4$ are independent of $\lambda$ and will remain independent of $\lambda$ upon analytic continuation; but at some stage the $\Phi_2$ puncture must approach the $\Phi_3$ and $\Phi_4$ punctures, and the respective local coordinate patches will have to adjust to avoid overlapping. This further demonstrates that $s$-$t$-$u$ channel duality is really lost when we go off-shell; there is no sense the $s$-channel diagram ``includes" contributions from other channels.
\end{itemize}

In any case, it is natural (and apparently necessary) to include the contributions from the $t$ and $u$ channel diagrams. If you do this you will find holomorphic local sections of $\P_{0,4}$ defined in the neighborhood of $0$, $1$ and $\infty$ on the moduli space. However, it turns out that we {\it still} do not cover the full moduli space. There is a missing region $\pi\cdot\mathcal{V}_{0,4}\subset \mathcal{M}_{0,4}$ which is unaccounted for.
\begin{exercise}
Assuming the $SL(2,\mathbb{C})$ cubic vertex, determine the region $\pi\cdot\mathcal{V}_{0,4}$.
\end{exercise} 

\begin{wrapfigure}{l}{.4\linewidth}
\centering
\resizebox{2.8in}{2.3in}{\includegraphics{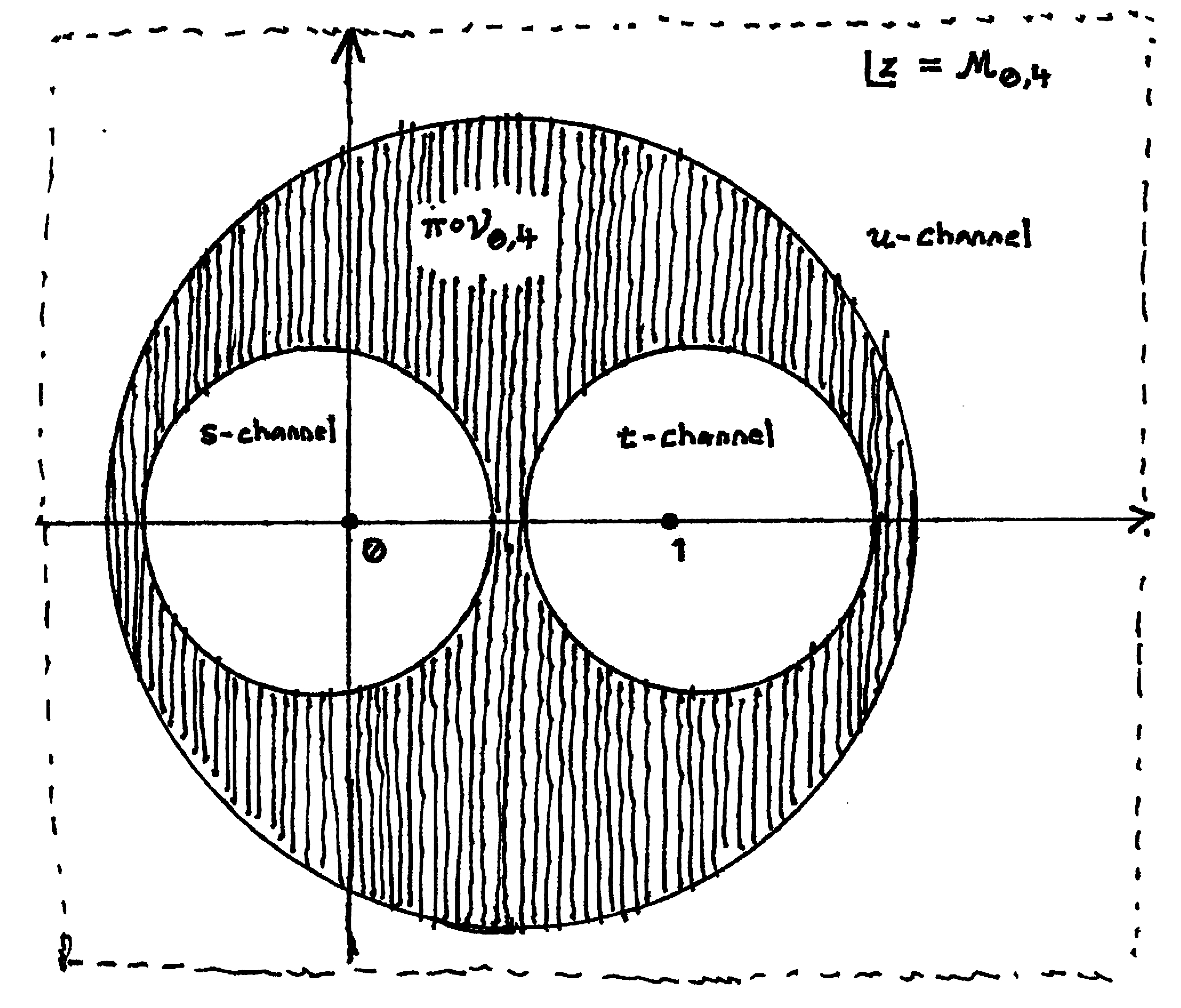}}
\end{wrapfigure}
\noindent The natural conclusion is that we are missing a contribution from an elementary quartic vertex. The 4-vertex would be defined by a local section of $\Phat_{0,4}$ defined on $\pi\cdot\mathcal{V}_{0,4}$. We denote this local section simply as $\mathcal{V}_{0,4}$. The quartic vertex is then 
\begin{equation}\langle \mathcal{V}_{0,4}|=\int_{\mathcal{V}_{0,4}}\langle \Omega_{0,4}|.\end{equation}
As anticipated earlier, the quartic vertex is simply an off-shell amplitude with integration towards the boundary of moduli space excluded. Since the 4-vertex should be symmetric, we assume that the 24 sections obtained by permuting the external states are identical. To ensure BRST invariance of the 4-point amplitude, $\mathcal{V}_{0,4}$ must be chosen so that the $s$, $t$, $u$ and quartic vertex diagrams patch together into a continuous global section of $\Phat_{0,4}$. This, in particular, requires that the local coordinate maps on the boundary of $\mathcal{V}_{0,4}$ match those of the $s$ $t$ and $u$ channel diagrams when the propagator tube collapses to zero length. This can be understood as part of a hierarchy of conditions on the local sections $\mathcal{V}_{g,n}\subset\Phat_{g,n}$ defining a vertex of the closed SFT action:
\begin{eqnarray}
\lineup \d^2=0,\nonumber\\
\lineup \d\mathcal{V}_{0,3}=0,\nonumber\\
\lineup \d\mathcal{V}_{0,4}+\mathcal{V}_{0,3}\cdot\mathcal{V}_{0,3} = 0, \nonumber\\
\lineup\ \ \ \ \ \ \ \ \ \ \ \vdots\ \ ,
\end{eqnarray}
where $\mathcal{V}_{0,3}\cdot\mathcal{V}_{0,3}$ is a section defined at the interface of quartic vertex and propagator regions defined by gluing cubic vertices with some twist angle $\theta$. This hierarchy of conditions are known collectively as the {\it geometrical BV equation}. It is a geometrical expression of the condition of nonlinear BRST invariance, and the existence of a consistent gauge-fixed path integral for closed SFT (specifically, we can obtain a solution to the BV master equation).

Proceeding to higher order amplitudes, we now have Feynman diagrams containing both cubic and quartic vertices. It should not come as a surprise that these diagrams will still fail to cover the moduli space, and for each off-shell amplitude $\langle\mathcal{A}_{g,n}|$ we need to introduce a new vertex
\begin{equation}\langle \mathcal{V}_{g,n}| = \int_{\mathcal{V}_{g,n}}\langle\Omega_{g,n}|\end{equation}
to fill in missing regions.\footnote{With a light-cone style vertex, the closed SFT action is cubic \cite{KugoZwiebach}. This action however will encounter difficulties at zero momentum, in particular in backgrounds with tadpoles. With covariant vertices of the kind we have been discussing, it is known that the action cannot be cubic \cite{notcubic}. It is strongly suspected that the action must be nonpolynomial, but this has not been proven.} The local section $\mathcal{V}_{g,n}$ is assumed to be symmetric, and must be chosen so that all Feynman diagrams patch together to define a continuous global section of $\Phat_{g,n}$, or equivalently, to give a solution to the geometrical BV equation.

We have essentially completed the task of giving a Feynman diagram construction of the off-shell 4-point amplitude. However, we would like to make it more explicit by giving a formula for $\langle \mathcal{A}_{0,4}|$ in terms of the vertices $\langle\mathcal{V}_{0,3}|,\langle\mathcal{V}_{0,4}|$ and the propagator. We would also like to understand how the geometrical BV equation imposes a condition on the vertices themselves, rather than sections of $\Phat_{g,n}$. To make these things explicit we will make more extensive use of the tensor product notation, so it will be a good time to review appendix \ref{app:tensor}. The first issue we encounter in this endeavor is that the cubic vertex is a dual state---a ``bra vector." But to represent the $s$-channel diagram, we need to convert it into a state which, after applying the propagator, can be ``fed in" to the other cubic vertex. We have seen that a state can be visualized as a unit disk. A dual state 
\begin{equation}\langle A|: \mathcal{H}\to \mathcal{H}^{\otimes 0}\end{equation}
\begin{wrapfigure}{l}{.35\linewidth}
\centering
\resizebox{2.2in}{1.1in}{\includegraphics{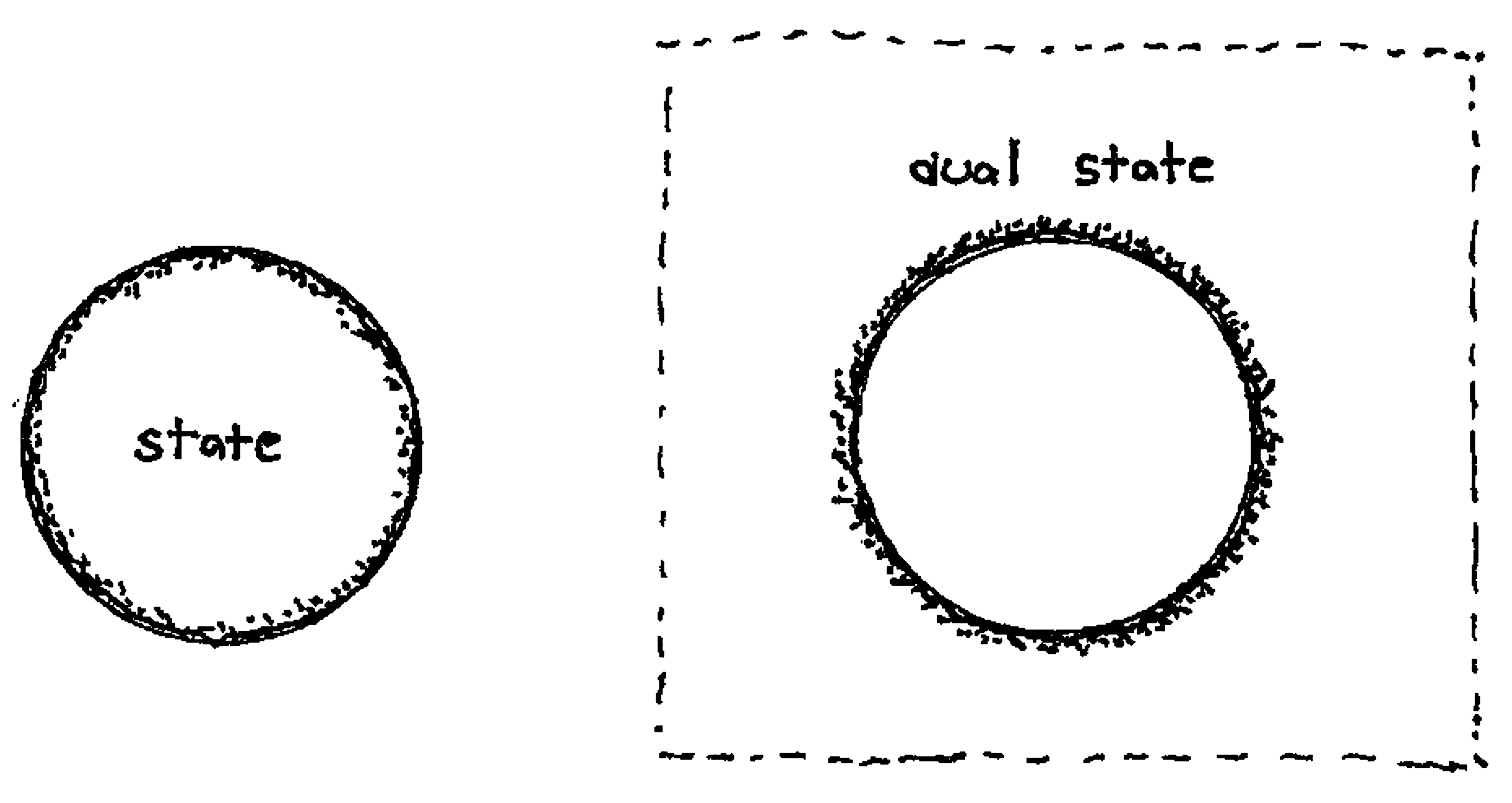}}
\end{wrapfigure}
can be visualized as the compliment of the unit disk in the complex plane. This way, we can patch a unit disk together with its compliment to produce a correlation function on the complex plane, which can be evaluated to give a number. For a unit disk centered at the origin, we can map a state into a dual state with an inversion
\begin{equation}I(\xi)=\frac{1}{\xi}.\end{equation}
This allows us to define a bilinear map
\begin{equation}\langle \mathrm{bpz}|:\mathcal{H}^{\otimes 2} \to\H^{\otimes 0}\end{equation}
through
\begin{equation}\langle \mathrm{bpz}|A_1\otimes A_2 = \Big\langle \big(I\circ A_1(0)\big)A_2(0)\Big\rangle_\mathbb{C}\equiv \langle A_1,A_2\rangle.\end{equation}
This is called the {\it BPZ inner product}. One can show that the inner product is graded symmetric:
\begin{equation}\langle A_1,A_2\rangle = (-1)^{|A_1||A_2|}\langle A_2,A_1\rangle.\end{equation}
Moreover, $L_0^\pm$ and $b_0^\pm$ are {\it BPZ even} in the sense that
\begin{eqnarray}\langle \mathrm{bpz}|L_0^\pm\otimes\mathbb{I} \lineup = \langle
\mathrm{bpz}|\mathbb{I}\otimes L_0^\pm,\\
\langle \mathrm{bpz}|b_0^\pm\otimes\mathbb{I}\lineup = \langle
\mathrm{bpz}|\mathbb{I}\otimes b_0^\pm,\end{eqnarray}
while the BRST operator is {\it BPZ odd}:
\begin{equation}\langle \mathrm{bpz}|(Q\otimes \mathbb{I}+\mathbb{I}\otimes Q) = \langle \mathrm{bpz}|Q^{(2)}=0.\end{equation}
The BPZ inner product is defined by a surface state, and the fact that surface states are BRST invariant implies that $Q$ is BPZ odd. 
\begin{exercise}
Prove these relations.
\end{exercise}
\noindent Given a state $|A\rangle$, the BPZ inner product defines a dual state through
\begin{equation}\langle A| = \langle \mathrm{bpz}|\Big(|A\rangle\otimes \mathbb{I}\Big).\end{equation}
On the other hand, inversion can map the exterior of the unit disk back into the interior, sending a dual state back into a state. This allows us to define the inverse of the BPZ inner product
\begin{equation}|\mathrm{bpz}^{-1}\rangle\in \mathcal{H}^{\otimes 2},\end{equation}
so that
\begin{equation}|A\rangle = \Big(\mathbb{I}\otimes \langle A| \Big)|\mathrm{bpz}^{-1}\rangle.\end{equation}
One can show
\begin{eqnarray} 
Q^{(2)}|\mathrm{bpz}^{-1}\rangle\lineup = 0,\\
 L_0^\pm\otimes\mathbb{I}|\mathrm{bpz}^{-1}\rangle\lineup = \mathbb{I}\otimes L_0^\pm|\mathrm{bpz}^{-1}\rangle,\\  b_0^\pm\otimes\mathbb{I}|\mathrm{bpz}^{-1}\rangle\lineup = \mathbb{I}\otimes b_0^\pm|\mathrm{bpz}^{-1}\rangle,
\end{eqnarray}
using the corresponding properties of $\langle\mathrm{bpz}|$. We have 
\begin{equation}
\Big(\langle\mathrm{bpz}|\otimes\mathbb{I}\Big)\Big(\mathbb{I}\otimes |\mathrm{bpz}^{-1}\rangle\Big) = \mathbb{I}.
\label{eq:bpzbpzinv}\end{equation}
In tensor product notation, this expresses that the BPZ inner product composed with its inverse gives the identity.

With these notational preliminaries, the $s$-channel contribution to the amplitude can be written
\begin{equation}\langle \mathcal{V}_{0,3}|\otimes\langle\mathcal{V}_{0,3}|\Phi_1\otimes\Phi_2\otimes\left(\left(\mathbb{I}\otimes \frac{b_0^+}{L_0^+}b_0^-\delta(L_0^-)\right)|\mathrm{bpz}^{-1}\rangle\right)\otimes \Phi_3\otimes\Phi_4.\end{equation}
The inverse BPZ inner product allows us to feed the propagator and cubic vertex into another cubic vertex. The combination
\begin{equation}|\omega^{-1}\rangle = \Big(\mathbb{I}\otimes b_0^-\delta(L_0^-)\Big)|\mathrm{bpz}^{-1}\rangle\end{equation}
is important, and is called the {\it Poisson bivector}. The Poisson bivector is Grassmann odd. It is annihilated by $b_0^-$ and $L_0^-$ when operating on either output, and so lives in a tensor product of two copies of $\Hhat$:
\begin{equation}|\omega^{-1}\rangle\in \Hhat^{\otimes 2}.\end{equation}
Moreover, since $b_0^-\delta(L_0^-)$ commutes with $Q$, the Poisson bivector is BRST invariant
\begin{equation}Q^{(2)}|\omega^{-1}\rangle = 0.\end{equation}
Including contributions from the $s,t,u$ and quartic vertex diagrams, the full off-shell 4-point amplitude is written
\begin{eqnarray}
\langle \mathcal{A}_{0,4}|\Phi_1\otimes\Phi_2\otimes\Phi_3\otimes\Phi_4 \lineup  = \langle\mathcal{V}_{0,4}|\Phi_1\otimes\Phi_2\otimes\Phi_3\otimes\Phi_4 \nonumber\\
\lineup\ \ \ -\langle \mathcal{V}_{0,3}|\otimes\langle\mathcal{V}_{0,3}|\left[\mathbb{I}\otimes\mathbb{I}\otimes\left(\left(\mathbb{I}\otimes \frac{b_0^+}{L_0^+}\right)|\omega^{-1}\rangle\right)\otimes\mathbb{I}\otimes\mathbb{I}\right]\times\nonumber\\
\lineup\ \ \ \  \Big(\Phi_1\otimes\Phi_2\otimes\Phi_3\otimes\Phi_4+\Phi_1\otimes\Phi_3\otimes\Phi_2\otimes\Phi_4 +\Phi_1\otimes\Phi_4\otimes\Phi_2\otimes\Phi_3\Big).\nonumber\\
\end{eqnarray}
To simplify signs, in this equation and the following we assume that states denoted $\Phi$ are Grassmann even. We can check BRST invariance by computing the amplitude with a sum of states of the form
\begin{equation}Q^{(4)}(\Lambda\otimes \Phi_1\otimes \Phi_2\otimes \Phi_3),\end{equation}
where $\Lambda$ is Grassmann odd. Using BRST invariance of $\langle\mathcal{V}_{0,3}|$ and $|\omega^{-1}\rangle$, and assuming\footnote{All commutators are graded: $[a,b]=-(-1)^{|a||b|}[b,a]$.}
\begin{equation}\left[Q,\frac{b_0^+}{L_0^+}\right]=\mathbb{I},\end{equation}
in particular, ignoring contributions from the boundary of moduli space, we obtain
\begin{eqnarray}
\langle \mathcal{A}_{0,4}| Q^{(4)} (\Lambda\otimes \Phi_1\otimes \Phi_2\otimes \Phi_3) \lineup = \langle \mathcal{V}_{0,4}| Q^{(4)} (\Lambda\otimes \Phi_1\otimes \Phi_2\otimes \Phi_3) \nonumber\\
\lineup\ \ \ + \langle \mathcal{V}_{0,3}|\otimes\langle\mathcal{V}_{0,3}|\Big(\mathbb{I}\otimes\mathbb{I}\otimes|\omega^{-1}\rangle\otimes\mathbb{I}\otimes\mathbb{I}\Big)\Big(\Lambda\otimes\Phi_1\otimes\Phi_2\otimes\Phi_3\nonumber\\
\lineup\ \ \ \ \ \ \ \ \ \ \ \ \ \ \ \ \ \ \ \ \ \ \ \ \ \ \ \ \ \ \ \ \ \ +\Lambda\otimes\Phi_2\otimes\Phi_1\otimes\Phi_3+\Lambda\otimes\Phi_3\otimes\Phi_1\otimes\Phi_2\Big),\nonumber\\
\end{eqnarray}
which should vanish. To simplify further we introduce {\it multi-string products}
\begin{equation}L_{0,2}:\Hhat^{\otimes 2} \to \Hhat,\ \ \ \ L_{0,3}:\Hhat^{\otimes 3}\to\Hhat,\end{equation}
defined by
\begin{eqnarray}
L_{0,2} \lineup = \Big(\mathbb{I}\otimes \langle\mathcal{V}_{0,3}|\Big)\Big(|\omega^{-1}\rangle\otimes\mathbb{I}\otimes\mathbb{I}\Big),\nonumber\\
L_{0,3}\lineup = \Big(\mathbb{I}\otimes \langle\mathcal{V}_{0,4}|\Big)\Big(|\omega^{-1}\rangle\otimes\mathbb{I}\otimes\mathbb{I}\otimes\mathbb{I}\Big),
\end{eqnarray}
We often write $L_{0,2}(\Phi_1\otimes\Phi_2) = L_{0,2}(\Phi_1,\Phi_2)$, and similarly for $L_{0,3}$. Since the vertices are symmetric, the products are symmetric:
\begin{eqnarray}
L_{0,2}(\Phi_1,\Phi_2) \lineup = L_{0,2}(\Phi_2,\Phi_1),\nonumber\\
L_{0,3}(\Phi_1,\Phi_2,\Phi_3)\lineup = L_{0,3}(\Phi_2,\Phi_1,\Phi_3),\ \ \ \mathrm{etc.}.\label{eq:prod}
\end{eqnarray}
This is similar to the antisymmetry of the Lie bracket; in fact, the difference is a matter of sign convention, so the products can be understood as defining some generalized Lie algebra. If we replace $\Lambda$ with $|\omega^{-1}\rangle$, BRST invariance of the 4-point amplitude translates into a condition on the products:
\begin{eqnarray}
QL_{0,3}(\Phi_1,\Phi_2,\Phi_3)+L_{0,3}(Q\Phi_1,\Phi_2,\Phi_3)+L_{0,3}(Q\Phi_2,\Phi_1,\Phi_3)+L_{0,3}(Q\Phi_3,\Phi_2,\Phi_1)\ \ \ \ \lineup \nonumber\\
+L_{0,2}(L_{0,2}(\Phi_1,\Phi_2),\Phi_3)+L_{0,2}(L_{0,2}(\Phi_1,\Phi_3),\Phi_2))+L_{0,2}(L_{0,2}(\Phi_3,\Phi_2),\Phi_1)\lineup = 0.\ \ \ \ 
\label{eq:L3}\end{eqnarray}
\noindent The second line of this equation can be interpreted as a Jacobiator. The first line defines the BRST variation of $L_{0,3}$. So this equation is effectively saying that the Jacobi identity holds up to BRST exact terms. Meanwhile, BRST invariance of the 3-point amplitude implies
\begin{equation}QL_{0,2}(\Phi_1,\Phi_2)+L_{0,2}(Q\Phi_1,\Phi_2)+L_{0,2}(Q\Phi_2,\Phi_1)=0,\label{eq:L2}\end{equation}
and obviously
\begin{equation}Q^2\Phi_1 = 0.\label{eq:L1}\end{equation}
We should be reminded of the geometrical BV equation. These relations are an echo of the geometrical BV equation at the level of an algebraic structure on $\Hhat$ directly related to the vertices. This is an {\it $L_\infty$ algebra}. Including higher genus vertices extends this further into a {\it quantum $L_\infty$ algebra}.

\pagebreak

\section{Lecture 3: The Action}

We are now ready to formulate the action of closed SFT. We start with the classical action (no loop vertices) where the dynamical string field is a Grassmann even state $\Phi\in\Hhat$ of ghost number~$2$. The action can be expressed
\begin{equation}
S_{g=0}  = \frac{1}{2!}\langle \mathcal{V}_{0,2}|\Phi\otimes\Phi +\frac{1}{3!}\langle \mathcal{V}_{0,3}|\Phi\otimes\Phi\otimes\Phi + \frac{1}{4!}\langle \mathcal{V}_{0,4}|\Phi\otimes\Phi\otimes\Phi\otimes\Phi+...\ \ .
\end{equation}
The subscript $g=0$ indicates that this is the classical action. In the last lecture we explained the procedure for constructing vertices $\langle \mathcal{V}_{g,n}|$ by filling in ``gaps" so that Feynman diagrams produce a continuous global section of $\Phat_{g,n}$. The quadratic vertex plays a special role, since it should imply the form of the propagator. It is natural to guess that the ``1-string product" of the theory will be the BRST operator
\begin{equation}L_{0,1}=Q,\end{equation}
which in analogy to \eq{prod} implies
\begin{equation}\Big(\mathbb{I}\otimes \langle\mathcal{V}_{0,2}|\Big)\Big(|\omega^{-1}\rangle\otimes\mathbb{I}\Big)=Q.\end{equation}
To derive $\langle \mathcal{V}_{0,2}|$ we have to invert the Poisson bivector. This defines a symplectic form
\begin{equation}\langle\omega|:\Hhat^{\otimes 2}\to\Hhat^{\otimes 0}.\end{equation}
Often we write $\langle\omega|A\otimes B = \omega(A,B)$. The symplectic form must somehow involve the BPZ inner product to cancel the $|\mathrm{bpz}^{-1}\rangle$ in the Poisson bivector. It also requires a $c$-ghost to cancel the $b_0^-$ of the Poisson bivector. The precise form of the $c$ ghost doesn't matter so much as long as it commutes with $b_0^-$ to give one. The conventional choice is 
\begin{equation}c_0^-\equiv \frac{1}{2}(c_0-\overline{c}_0),\end{equation}
which satisfies $[b_0^-,c_0^-]=\mathbb{I}$ and is BPZ odd
\begin{equation}
\langle \mathrm{bpz}|(c_0^-\otimes \mathbb{I}+\mathbb{I}\otimes c_0^-) = 0,\ \ \ \ (c_0^-\otimes\mathbb{I}+\mathbb{I}\otimes c_0^-)|\mathrm{bpz}^{-1}\rangle = 0.
\end{equation}
The symplectic form is 
\begin{equation}\langle\omega| = -\langle \mathrm{bpz}|c_0^-\otimes\mathbb{I}.\end{equation}
The easiest way to justify this is to see that it works:
\begin{eqnarray}
\Big(\langle\omega|\otimes\mathbb{I}\Big)\Big(\mathbb{I}\otimes|\omega^{-1}\rangle\Big)\lineup = -\Big(\langle\mathrm{bpz}|\otimes\mathbb{I}\Big)\Big(c_0^-\otimes\mathbb{I}\otimes\mathbb{I}\Big)\Big(\mathbb{I}\otimes\mathbb{I}\otimes b_0^-\delta(L_0^-)\Big)\Big(\mathbb{I}\otimes|\mathrm{bpz}^{-1}\rangle\Big)\nonumber\\
\lineup = \Big(\langle\mathrm{bpz}|\otimes\mathbb{I}\Big)\Big(\mathbb{I}\otimes c_0^-\otimes b_0^-\delta(L_0^-)\Big)\Big(\mathbb{I}\otimes|\mathrm{bpz}^{-1}\rangle\Big)\nonumber\\
\lineup = \Big(\langle\mathrm{bpz}|\otimes\mathbb{I}\Big)\Big(\mathbb{I}\otimes \mathbb{I}\otimes b_0^-\delta(L_0^-)c_0^-\Big)\Big(\mathbb{I}\otimes|\mathrm{bpz}^{-1}\rangle\Big)\nonumber\\
\lineup =  b_0^-\delta(L_0^-)c_0^-\Big(\langle\mathrm{bpz}|\otimes\mathbb{I}\Big)\Big(\mathbb{I}\otimes|\mathrm{bpz}^{-1}\rangle\Big)\nonumber\\
\lineup =  b_0^-\delta(L_0^-)c_0^-.
\end{eqnarray}
Here we used the BPZ odd property of $c_0^-$ to bring it together with $b_0^-\delta(L_0^-)$ and then canceled $\langle \mathrm{bpz}|$ against $|\mathrm{bpz}^{-1}\rangle$. The final result does not look like the identity operator. But it is equivalent to the identity operator when acting on states in $\Hhat$:
\begin{equation}b_0^-\delta(L_0^-)c_0^- A = A\ \ \ \mathrm{iff}\ \ \ A\in\Hhat.\end{equation}
Actually, this is a convenient way to characterize $\Hhat$: it is the subspace of states obtained after applying the projection operator $b_0^-\delta(L_0^-)c_0^-$. Graded symmetry of the BPZ inner product implies graded antisymmetry of $\omega$:
\begin{equation}\omega(A,B) = -(-1)^{|A||B|}\omega(B,A).\end{equation}
This is is why $\omega$ is referred to as a symplectic form. From this we learn that the 2-string vertex can be written
\begin{equation}\langle \mathcal{V}_{0,2}| = \langle\omega|\mathbb{I}\otimes Q,\end{equation}
and similarly
\begin{equation}\langle \mathcal{V}_{g,n}| = \langle\omega|\mathbb{I}\otimes L_{g,n},\end{equation}
where $L_{g,n}:\Hhat^{\otimes n}\to\Hhat$ are multi-string products. The classical action then takes the form
\begin{equation}S_{g=0}  = \frac{1}{2!}\omega(\Phi,Q\Phi)+\frac{1}{3!}\omega(\Phi,L_{0,2}(\Phi,\Phi)) + \frac{1}{4!}\omega(\Phi,L_{0,3}(\Phi,\Phi,\Phi))+...,\end{equation}
as written in the introduction.

For most computations it is more convenient to work with the products and symplectic form, rather than the vertices. One thing, however, that this language does not make manifest is the symmetry of the vertices, especially under permutations involving the first state. Consider for example
\begin{equation}\langle\mathcal{V}_{g,n}|A_1\otimes...\otimes A_{n-1}\otimes A_n =(-1)^{|A_n|(|A_1|+...+|A_{n-1}|)}\langle \mathcal{V}_{g,n}|A_n\otimes A_1\otimes...\otimes A_{n-1}.\end{equation}
We note that the products are uniformly Grassmann odd; the vertices contain $6g-6+2n$ $b$-ghost insertions, and the products contain one additional $b_0^-$ insertion from the Poisson bivector. So in total the products contain an odd number of anti-commuting objects. Therefore the above symmetry relation translates to
\begin{equation}(-1)^{|A_1|}\omega(A_1,L_{g,n}(A_2,...,A_{n-1},A_n))=(-1)^{|A_n|}(-1)^{|A_n|(|A_1|+...+|A_{n-1}|)}\omega(A_n,L_{g,n}(A_1,A_2,...,A_{n-1})),\end{equation}
where the additional sign appears from commuting $A_1$ and $A_n$ past $L_{g,n}$. Next we use antisymmetry of the symplectic form to rewrite the right hand side as
\begin{equation}-(-1)^{|A_n|(|A_1|+...+|A_{n-1}|+1)} (-1)^{|A_n|}(-1)^{|A_n|(|A_1|+...+|A_{n-1}|)}\omega(L_{g,n}(A_1,A_2,...,A_{n-1}),A_n).\end{equation}
Signs cancel, giving
\begin{equation}(-1)^{|A_1|}\omega(A_1,L_{g,n}(A_2,...,A_{n-1},A_n))=-\omega(L_{g,n}(A_1,A_2,...,A_{n-1}),A_n),\end{equation}
which implies 
\begin{equation}
\langle\omega|(L_{g,n}\otimes\mathbb{I}+\mathbb{I}\otimes L_{g,n})=0.
\end{equation}
This condition is conventionally referred to as {\it cyclicity}. Symmetry of the products together with cyclicity is equivalent to symmetry of vertices. Cyclicity of the BRST operator requires special consideration, unlike for higher products where it follows from symmetry of the corresponding vertex which is imposed by definition. We need
\begin{equation}\langle \mathrm{bpz}|c_0^- Q\otimes \mathbb{I} = -\langle\mathrm{bpz}|c_0^-\otimes Q.\end{equation}
This equality does not hold acting on arbitrary states in $\H$ since $c_0^-$ does not commute with $Q$. But the symplectic form is only intended to be defined on $\Hhat$. The trick is to insert the identity operator in the form of the projection onto $\Hhat$:
\begin{eqnarray}\langle \mathrm{bpz}|c_0^- Q\otimes \mathbb{I}\lineup = \langle \mathrm{bpz}|c_0^- Q b_0^-\delta(L_0^-)c_0^-\otimes \mathbb{I}\nonumber\\
\lineup = -\langle \mathrm{bpz}|c_0^- b_0^-\delta(L_0^-)Qc_0^-\otimes \mathbb{I}\nonumber\\
\lineup = -\langle \mathrm{bpz}|Q c_0^-\otimes  b_0^-\delta(L_0^-)c_0^-\nonumber\\
\lineup = -\langle \mathrm{bpz}|Q c_0^-\otimes  \mathbb{I}\nonumber\\
\lineup = -\langle \mathrm{bpz}| c_0^-\otimes Q.
\end{eqnarray}
where we used the BPZ even/odd properties of the operators involved. Therefore
\begin{equation}\langle \omega|(Q\otimes\mathbb{I} + \mathbb{I}\otimes Q)=0,\end{equation}
and the BRST operator is cyclic.

Cyclicity together with $Q^2=0$ implies that the free action
\begin{equation}S_\mathrm{free} = \frac{1}{2!}\omega(\Phi,Q\Phi)\end{equation}
is invariant under the linearized gauge transformation
\begin{equation}\Phi' = \Phi +Q\Lambda,\end{equation}
where $\Lambda\in \Hhat$ is Grassmann odd and ghost number 1. The linearized equations of motion are 
\begin{equation}Q\Phi=0,\end{equation}
which is the physical state condition. Since the free action has gauge invariance, to find the propagator we must fix a gauge. The conventional choice is {\it Siegel gauge}
\begin{equation}b_0^+\Phi=0.\end{equation}
The propagator is defined by solution to the equation
\begin{equation}Q\Phi = J\end{equation}
in Siegel gauge, where $J\in\Hhat$ is a ``source" serving as a stand-in for the nonlinear terms in the action. Multiplying by $b_0^+$ leads to
\begin{equation}\Phi = \frac{b_0^+}{L_0^+}J.\end{equation}
So the propagator is $b_0^+/L_0^+$. One might wonder what happened to the $b_0^-\delta(L_0^-)$ factor of the propagator discussed earlier. It is still there, but has been absorbed into other objects in our setup, notably the Poisson bivector and the outputs of the multi-string products. At a deeper level, the absence of $b_0^-\delta(L_0^-)$ reflects the fact that the dynamical variable of closed SFT captures the center of mass propagation of the closed string, but not its twisting motion due to the $b_0^-$ and level matching constraints. It is almost as though the fields which represent the twisting of the closed string have been integrated out. Typically integrating out degrees of freedom makes a theory more complicated. It is sometimes speculated that if the $b_0^-$ and level matching constraints could be lifted, closed SFT would take a fundamentally different and much simpler form. But no one seems to know how to do this.

The interacting theory has a gauge invariance which follows from a nonlinear generalization of $Q^2=0$, namely, the multi-string products form an $L_\infty$ algebra (at genus 0). Another name for this is homotopy Lie algebra; this is almost the same as a differential graded Lie algebra, except that the Jacobi identity only holds up to ``homotopy"---in the context of closed SFT, the Jacobiator is only BRST exact, instead of being strictly zero. The fact that the closed string products are symmetric, while the Lie bracket is antisymmetric, requires some explanation. Given $L_{0,2}$ we can define a ``bracket"
\begin{equation}[A,B] = (-1)^{|A|} L_{0,2}(A,B).\end{equation}
Symmetry of $L_{0,2}$ implies
\begin{equation}[A,B] = -(-1)^{(|A|+1)(|B|+1)}[B,A].\end{equation}
This says that the bracket is graded antisymmetric, but the relevant grading is shifted by $1$ relative to the conventional Grassmann grading. This shift in grading is called a {\it suspension}. In fact, with the appropriate signs for the higher products, we can consistently formulate the closed SFT action using the suspended grading, effectively treating Grassmann odd objects as commuting and Grassmann even objects as anticommuting. In practice this is not convenient. However, for open strings the dynamical field is Grassmann odd, and a formulation analogous to what we describe here requires the suspension. 

An $L_\infty$ algebra is characterized by an infinite hierarchy of Jacobi-like identities for multilinear products. A convenient and economical expression for these relations can be found using the {\it coalgebra formalism}. The products we consider are symmetric, which means that, when viewed as linear operators on tensor products of states, the ordering of states in the tensor product does not matter. Thus, for example, the tensor products
\begin{equation}A\otimes B,\ \ \ \ (-1)^{|A||B|}B\otimes A\end{equation}
should be seen as equivalent. The equivalence class will be denoted $A\wedge B$, where the ``wedge" is the symmetrized tensor product which satisfies
\begin{equation}A\wedge B = (-1)^{|A||B|}B\wedge A.\end{equation}
A symmetric $n$-string product $b_n$ can be viewed as a linear map
\begin{equation}b_n:\Hhat^{\wedge n}\to\Hhat,\end{equation}
where $\Hhat^{\wedge n}$ is given by wedge products of $n$ states in $\Hhat$. Now there is a useful way to extend the definition of $b_n$ so that it can operate on other symmetrized powers of $\Hhat$:
\begin{equation}
b_n:\Hhat^{\wedge m}\to\Hhat^{\wedge m-n+1},
\end{equation}
given by
\begin{eqnarray}
b_n(A_1\wedge ... \wedge A_m)\lineup  = 0,\ \ \ \ \ \ \ \ \ \ \ \ \ \ \ \ \ \ \ \ \ \ \ \ \ \ \ \ \ \ \ \ \ \ \ \ \ \ \ \ \ \ \ \ \ \ \ \ \ \ \ \ \ \ \ \ \ \ \ \ \ \ \ m<n,\nonumber\\
b_n(A_1\wedge ... \wedge A_n) \lineup =b_n(A_1,..., A_n)\ \ \ \ \ \ \ \ \ \ \ \ \ \ \ \ \ \ \ \ \ \ \ \ \ \ \ \ \ \ \ \ \ \ \ \ \ \ \ \ \ \ \ \ \ \ \ \ \, m=n,\nonumber\\
b_n(A_1\wedge ... \wedge A_m)\lineup = \sum_{\sigma}(-1)^{\sigma}b_n(A_{\sigma_1},...,A_{\sigma_n})\wedge A_{\sigma_{n+1}}\wedge...\wedge A_{\sigma_m}\ \ \ \ \ \ \ \ \ m>n.
\end{eqnarray}
The sum is over all distinct ways to partition integers $1,...,m$ into (unordered) sets 
\begin{equation}\sigma = \{\{\sigma_1,...,\sigma_n\},\{\sigma_{n+1},...,\sigma_m\}\}.\end{equation} The sign $(-1)^\sigma$ is defined so that
\begin{equation}A_1\wedge...\wedge A_n = (-1)^\sigma A_{\sigma_1}\wedge...\wedge A_{\sigma_n}.\end{equation}
Since $b_n$ can now act on any number of states wedged together, it can be viewed as an operator on the symmetrized tensor algebra
\begin{equation}b_n: S\Hhat\to S\Hhat,\end{equation}
where
\begin{equation}S\Hhat = \Hhat^{\wedge 0}\, \oplus\, \Hhat\, \oplus\, \Hhat^{\wedge 2}\, \oplus\, \Hhat^{\wedge 3}\, \oplus\, ...\ \ .\end{equation}
When a product is extended to an operator on the symmetrized tensor algebra in this way, it is called a {\it coderivation}.  The terminology connects with the fact that the symmetrized tensor algebra has a coalgebra structure. For present purposes we do not need to develop these definitions.

Therefore we can view the genus 0 products $L_{0,n}$ as coderivations on the symmetrized tensor algebra of $\Hhat$. Since in this context all products act on the same vector space, it is meaningful to add them:
\begin{equation}L_{g=0}  = Q+L_{0,2}+L_{0,3}+...\ \ .\end{equation}
The identities of an $L_\infty$ algebra are equivalent to the statement that this object is nilpotent:
\begin{equation}(L_{g=0})^2=0.\end{equation}
\begin{exercise}
Acting on $\Phi_1\wedge\Phi_2\wedge\Phi_3\wedge\Phi_4$ show that this equation implies the first three $L_\infty$ relations, as given in \eq{L3}-\eq{L1}.
\end{exercise}
\noindent This language makes it clear that the vertices of closed string field theory define a nonlinear generalization of the BRST operator. At the level of the local sections $\mathcal{V}_{0,n}$, the fact that $L_{g=0} $ squares to zero is equivalent to the tree level geometrical BV equation.

To describe the nonlinear gauge invariance it is helpful to introduce a few additional ingredients. Given a Grassmann even state $\Phi$ we can define an object in $S\Hhat$ called a {\it group-like element}:
\begin{equation}e^{\Phi} = 1_{S\Hhat}+\Phi + \frac{1}{2!}\Phi\wedge\Phi + \frac{1}{3!}\Phi\wedge\Phi\wedge\Phi+...\ \ .\end{equation}
Let us understand how an $n$-string product $b_n$, interpreted as a coderivation, acts on a group-like element:
\begin{eqnarray}
b_ne^\Phi \lineup = \sum_{m=0}^\infty \frac{1}{m!} b_n \Phi^{\wedge m}\nonumber\\
\lineup = \sum_{m=0}^\infty \frac{1}{m!}\left({m \atop n}\right) b_n(\Phi,...,\Phi)\wedge \Phi^{\wedge m-n}.
\end{eqnarray}
The binomial coefficient comes from the sum over partitions of $m$ $\Phi$s into sets with $n$ $\Phi$s and $m-n$ $\Phi$s. Since the $\Phi$s are identical, all of these partitions give the same result, and the sum produces a binomial factor. Continuing,
\begin{eqnarray}
b_n e^\Phi \lineup = b_n(\Phi,...,\Phi)\wedge \sum_{m=0}^\infty \frac{1}{m!}\frac{m!}{n!(m-n)!} \Phi^{\wedge m-n}\nonumber\\
\lineup= \frac{1}{n!}b_n(\Phi,...,\Phi)\wedge e^\Phi\nonumber\\
\lineup=\Big(\pi_1 b_n e^\Phi\Big)\wedge e^{\Phi},
\end{eqnarray}
where the projection operator $\pi_1$ acts as the identity on the $\Hhat^{\wedge 1}=\Hhat$ component of the tensor algebra, and sets everything else to zero. 
\begin{exercise}
Given some $\Lambda\in \Hhat$, show that 
\begin{equation}b_n \Big(\Lambda\wedge e^\Phi\Big) =\Big(\pi_1 b_n\,\Lambda\wedge e^\Phi\Big)\wedge e^\Phi + \Big(\pi_1 b_n e^\Phi\Big)\wedge\Lambda\wedge e^\Phi.\end{equation}
\end{exercise}
\begin{exercise} If $b_n$ is a cyclic product show that
\begin{equation}\omega\Big(A_1,\pi_1 b_n\Big( \,A_2\wedge...\wedge A_k\wedge e^{\Phi}\Big)\Big) = -(-1)^{|A_1||b_n|}\omega\Big(\pi_1 b_n\Big( \,A_1\wedge...\wedge A_{k-1}e^\Phi\Big),A_k\Big).\end{equation}
Since the wedge product is symmetric, this implies that we can place any of the states $A_1,...,A_k$ in the first entry of the symplectic form (with the appropriate sign) and the expression is unchanged. 
\end{exercise}

Having developed these definitions it seems natural to express the action in terms of the coderivation $L_{g=0} $ and the group-like element $e^\Phi$. At first we might guess that the action can be written as 
\begin{equation}\omega(\Phi,\pi_1L_{g=0} e^\Phi)=\omega(\Phi,Q\Phi)+\frac{1}{2!}\omega(\Phi,L_{0,2}(\Phi,\Phi))+\frac{1}{3!}\omega(\Phi,L_{0,3}(\Phi,\Phi,\Phi))+...\ ,\end{equation} 
but the factors are not right; the $(n+1)$-string vertex comes with a factor of $1/n!$ rather than $1/(n+1)!$. To get the extra factor of $n+1$ in the denominator we employ the following trick. We introduce an auxiliary parameter $t\in[0,1]$ and introduce a family of string fields $\Phi(t)\in \Hhat$ with boundary conditions
\begin{equation}\Phi(0)=0,\ \ \ \ \Phi(1)=\Phi = \mathrm{dynamical\ field}.\end{equation}
The classical action can then be written
\begin{equation}S_{g=0}  = \int_0^1 dt\,\omega\Big(\dot{\Phi}(t),\pi_1L_{g=0} e^{\Phi(t)}\Big),\end{equation}
where the dot denotes differentiation with respect to $t$. To see that this works, consider the contribution from the product $L_{0,n}$:
\begin{eqnarray}
\lineup \frac{1}{n!}\int_0^1 dt\,\omega\Big(\dot{\Phi}(t), L_{0,n}(\Phi(t),...,\Phi(t))\Big)\nonumber\\
\lineup\ \ \ \ \ \ \ \ \ = \frac{1}{n!}\int_0^1 dt\frac{1}{n+1}\left[\omega\Big(\dot{\Phi}(t), L_{0,n}(\Phi(t),...,\Phi(t))\Big)+...+\omega\Big(\Phi(t),L_{0,n}(\Phi(t),...,\dot{\Phi}(t))\Big)\right].\ \ \ \ \ \ \ \ \ 
\end{eqnarray}
We used cyclicity of the vertex to distribute the $t$-derivative symmetrically on each entry. The integrand is now a total derivative
\begin{eqnarray}
\frac{1}{n!}\int_0^1 dt\,\omega\Big(\dot{\Phi}(t),L_{0,n}(\Phi(t),...,\Phi(t))\Big) \lineup= \frac{1}{(n+1)!}\int_0^1 dt\frac{d}{dt}\omega\Big(\Phi(t), L_{0,n}(\Phi(t),...,\Phi(t))\Big)\nonumber\\
\lineup = \frac{1}{(n+1)!}\omega(\Phi, L_{0,n}(\Phi,...,\Phi)),
\end{eqnarray}
which gives the needed factorial factor for the vertex. Now we can derive the classical equations of motion:
\begin{equation}
\delta S_{g=0} =\int_0^1 dt\, \Big[\omega\Big(\delta\dot{\Phi}(t),\pi_1L_{g=0} e^{\Phi(t)}\Big)+\omega\Big(\dot{\Phi}(t),\pi_1L_{g=0}  \Big(\delta\Phi(t)\wedge e^{\Phi(t)}\Big)\Big)\Big].\label{eq:step}
\end{equation}
The result of exercise 10 implies that the second term can be written
\begin{equation}
\omega\Big(\delta\Phi(t),\pi_1\Big(L_{g=0} \dot{\Phi}(t)\wedge e^{\Phi(t)}\Big)\Big).
\end{equation}
Combining with the first term gives a total derivative
\begin{equation}
\delta S_{g=0} =\int_0^1 dt\frac{d}{dt}\omega\Big(\delta\Phi(t),\pi_1L_{g=0} e^{\Phi(t)}\Big)=\omega\Big(\delta\Phi,\pi_1L_{g=0} e^{\Phi}\Big).
\end{equation}
Setting the variation to zero implies the equations of motion:
\begin{eqnarray}
0\lineup =\pi_1L_{g=0} e^\Phi\nonumber\\
\lineup = Q\Phi + \frac{1}{2!}L_{0,2}(\Phi,\Phi)+\frac{1}{3!}L_{0,3}(\Phi,\Phi,\Phi)+...\ \ .
\end{eqnarray}
The infinitesimal gauge transformation is
\begin{eqnarray}
\delta\Phi \lineup = \pi_1 L_{g=0} \Big(\Lambda\wedge e^\Phi\Big)\nonumber\\
\lineup = Q\Lambda + L_{0,2}(\Lambda,\Phi)+\frac{1}{2!}L_{0,3}(\Lambda,\Phi,\Phi)+...\ \ .
\end{eqnarray}
First we can check gauge invariance of the equations of motion
\begin{eqnarray}
\delta\Big(\pi_1 L_{g=0} e^\Phi\Big)\lineup  = \pi_1 L_{g=0} \Big(\Big(\pi_1 L_{g=0} \,\Lambda\wedge e^\Phi\Big)\wedge e^\Phi\Big)\nonumber\\
\lineup = \pi_1(L_{g=0} )^2 \Big(\Lambda\wedge e^\Phi\Big) -\pi_1L_{g=0} \Big(\Big(\pi_1L_{g=0}  e^\Phi\Big)\wedge \Lambda\wedge e^\Phi\Big),
\end{eqnarray}
where we used the result of exercise 9. The first term vanishes due to $L_\infty$ relations, and the second term vanishes by the equations of motion. Next we check gauge invariance of the action. For definiteness, we assume that $\Phi(t)$ transforms in the same way as $\Phi$ but with some $\Lambda(t)$ with $\Lambda(0)=0$ and $\Lambda(1)=\Lambda$. The gauge variation of the action is
\begin{eqnarray}
\delta S_{g=0}  \lineup = \int_0^1 dt\, \omega\Big(\pi_1L_{g=0} \Big(\dot{\Lambda}(t)\wedge e^{\Phi(t)}\Big)+\pi_1 L_{g=0} \Big( \Lambda(t)\wedge\dot{\Phi}(t)\wedge e^{\Phi(t)}\Big),\pi_1 L_{g=0}  e^{\Phi(t)}\Big)\nonumber\\
\lineup\ \ \ \ +\int_0^1 dt\, \omega\Big(\dot{\Phi}(t), \pi_1L_{g=0} \Big(\pi_1 L_{g=0} \Big(\Lambda(t)\wedge e^{\Phi(t)}\Big)\wedge e^{\Phi(t)}\Big)\Big).
\end{eqnarray}
The strategy is to bring both $L_{g=0} $s into the same entry of the symplectic form using the result of exercise 10, and hope for cancellation from $(L_{g=0} ^2)=0$:
\begin{eqnarray}
\delta S_{g=0} \lineup = -\int_0^1 dt\, \omega\Big(\dot{\Lambda}(t),\pi_1L_{g=0}  \Big(e^{\Phi(t)}\wedge \Big(\pi_1 L_{g=0} e^{\Phi(t)}\Big)\Big)\Big)\nonumber\\ 
\lineup\ \ \ - \int_0^1 dt\,\omega\Big(\dot{\Phi}(t),\pi_1 L_{g=0} \Big(\Lambda(t)\wedge\Big(\pi_1 L_{g=0} e^{\Phi(t)}\Big)\wedge e^{\Phi(t)}\Big)\Big)\nonumber\\
\lineup\ \ \ +\int_0^1 dt\, \omega\Big(\dot{\Phi}(t), \pi_1L_{g=0} \Big(\pi_1 L_{g=0} \Big(\Lambda(t)\wedge e^\Phi(t)\Big)\wedge e^{\Phi(t)}\Big)\Big).
\end{eqnarray}
From the result of exercise 9 we find
\begin{equation}
\delta S_{g=0}  = -\int_0^1 dt\, \omega\Big(\dot{\Lambda}(t),\pi_1(L_{g=0} )^2 e^{\Phi(t)}\Big)+\int_0^1 dt\, \omega\Big(\dot{\Phi}(t), \pi_1(L_{g=0} )^2\Big(\Lambda(t)\wedge e^\Phi(t)\Big)\Big),
\end{equation}
which vanishes. Therefore the classical action is gauge invariant.

Since we have a gauge invariant classical theory, passing to the quantum theory requires a gauge-fixed path integral, which introduces Faddeev-Popov ghosts and so on. One novelty of closed SFT relative to, say, Yang-Mills theory is that the gauge transformations are themselves redundant. This can already be seen in the free theory; the linearized gauge transformation is
\begin{equation}\Phi' =\Phi + Q\Lambda,\end{equation}
but gauge parameters implement the same gauge transformation if
\begin{equation}\Lambda' = \Lambda + Q\mu,\end{equation}
with $\mu\in \Hhat$ Grassmann even and ghost number $0$. Furthermore, two $\mu$s imply the same relation between gauge parameters if $\mu'=\mu+Q\nu$, with $\nu\in\Hhat$ Grassmann odd and ghost number $-1$, and so on. Therefore, we need not only introduce Fadeev-Popov ghosts for gauge symmetry, but additional ghosts for the gauge symmetry of the gauge symmetry, and further for the gauge symmetry of that, {\it ad infinitum}. In complicated gauge systems such as closed SFT, the most sophisticated and systematic approach to defining the gauge fixed path integral is through the {\it Batalin-Vilkovisky formalism} (BV formalism). The BV formalism is worth studying as a subject in its own right, but for most purposes in closed SFT it is enough to know the result of the BV analysis. So we will only give a summary. In the BV formalism, analysis of the gauge invariance of a classical action allows one to define a space of ``fields and antifields" with an odd symplectic structure defining the so-called ``anti-bracket" denoted $(,)$. The space of fields and antifields includes and extends the space of fields which appear in the classical action. The classical action is therefore only defined on a submanifold of the space of fields and antifields. We look for an extension, called the ``master action,'' which is defined over the entire space. The master action is required to satisfy the ``master equation"
\begin{equation}(S,S)=0.\end{equation}
The Hamiltonian vector field generated by the anti-bracket with $S$  defines BRST transformations. Therefore the master equation says two things: first, that the master action is BRST invariant, and second that the BRST transformation is nilpotent. Note that the BRST symmetry discussed here is associated to a generic classical field theory; the string worldsheet is just a specific example. The BRST symmetry implied by the master action of closed SFT is different from worldsheet BRST symmetry, but they are closely related. So far this is only the classical BV formalism. In the quantum theory we have to think about BRST invariance of the path integral measure. The measure defines an additional operation on the space of fields and antifields called the symplectic Laplacian $\Delta$. In the quantum theory, the master action must be further corrected to satisfy 
\begin{equation}(S,S)+2\Delta S = 0.\end{equation}
This is the quantum master equation. The gauge fixed path integral is defined by integrating over a Lagrangian submanifold in the space of fields and antifields. If appropriately chosen, the quantum master action restricted to the submanifold will have no gauge symmetry, and there will be no divergence from integrating over the gauge orbit. The choice of Lagrangian submanifold can be viewed as a choice of gauge. The quantum master equation ensures that the choice of Lagrangian submanifold does not effect correlation functions of observable quantities. In gauge fixing it is usually convenient to consider a non-minimal extension of the space of fields and antifields to include ``ghost" and ``antighost" fields; these are the BV analogue of the Faddeev-Popov ghosts. Applying the BV machinery to classical closed SFT gives the following result:
\begin{itemize}
\item The classical master action takes exactly the same form as the classical action, but the Grassmann even string field $\Phi\in\Hhat$ now contains components of all ghost numbers. The antifields are components of $\Phi$ with ghost number $>2$. The fields are components with ghost number $\leq 2$, including the classical dynamical field at ghost number 2. Since states at odd ghost numbers are Grassmann odd, the contributions to $\Phi$ from odd ghost numbers must be multiplied by anticommuting parameters to ensure that $\Phi$ is Grassmann even. Fixing Siegel gauge effectively converts the fields and antifields into ghosts and antighosts.
\item The quantum master equation requires that the quantum master action contains loop vertices to fill in missing pieces of a continuous global section of $\Phat_{g,n}$.
\end{itemize}

\noindent These are very simple results. For a typical field theory, a solution to the master equation usually cannot be obtained in closed form. Indeed, BV structures permeate deeply into the structure of string field theory, at the level of the $L_\infty$ structure of the string products and the geometric BV equation defining the local sections~$\mathcal{V}_{g,n}$. It is almost as though the BV formalism was ``created" for the quantization of string field theory.

The structure of the quantum master action is perhaps not surprising. We have already anticipated the necessity of loop vertices. The fact that ghost number of the string field is unrestricted could also have been anticipated by inspection of the flow of ghost number through Feynman diagrams. At tree level, the ghost number of intermediate states is fixed by ghost number con-
\begin{wrapfigure}{l}{.4\linewidth}
\centering
\resizebox{2.3in}{.9in}{\includegraphics{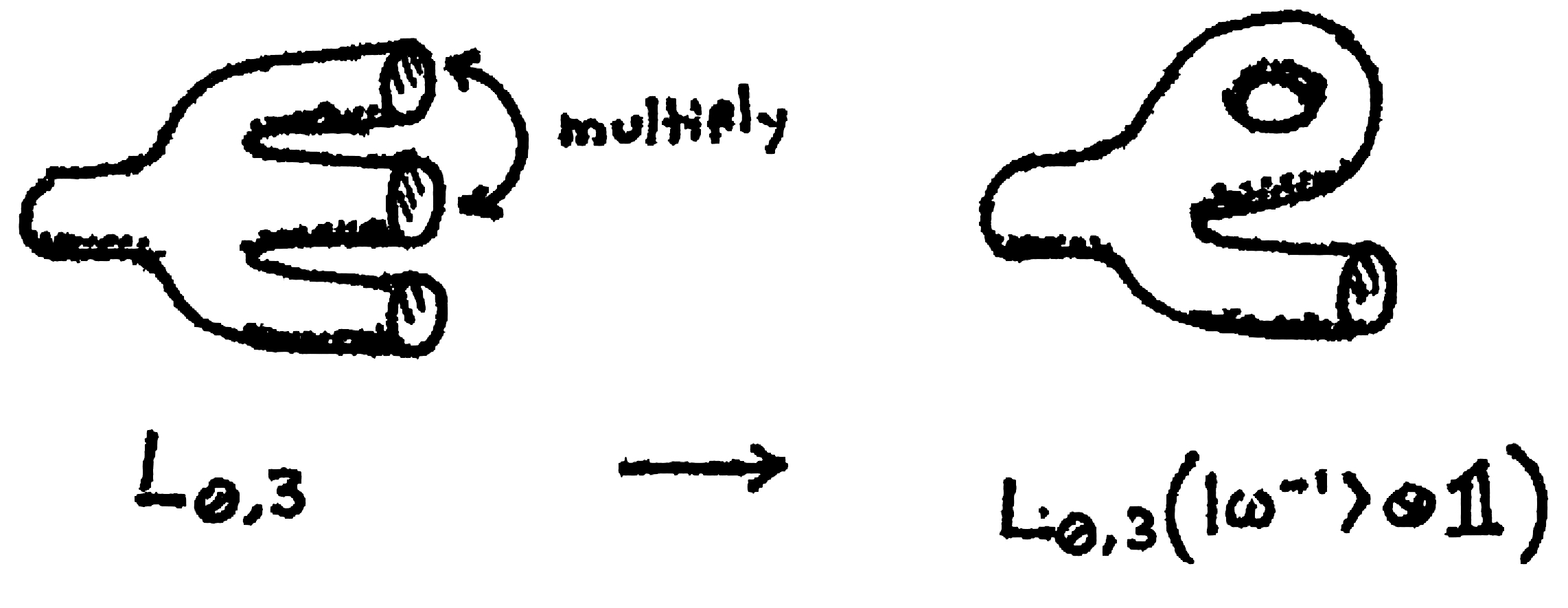}}
\end{wrapfigure} servation, but this is no longer the case in loops. The presence of the symplectic Laplacian in the quantum master equation implies that the higher genus products define a slightly more intricate algebraic structure than an ordinary $L_\infty$ algebra. This is reflected in the fact that a product can ``multiply itself" to form a product at higher genus with two fewer inputs. The resulting algebraic structure defines a quantum $L_\infty$ algebra. On the symmetrized tensor algebra, it can be characterized by summing the coderivations of all products at every genus together with the Poisson bivector
\begin{equation}L=\sum_{g,n=0}^\infty L_{g,n} + |\omega^{-1}\rangle,\end{equation}
where $|\omega^{-1}\rangle\in \Hhat^{\wedge 2}$ acts on $S\Hhat$ through the wedge product
\begin{equation}|\omega^{-1}\rangle(A_1\wedge ...\wedge A_n) = |\omega^{-1}\rangle\wedge A_1\wedge ... \wedge A_n.\end{equation}
Technically, since $|\omega^{-1}\rangle$ has two outputs it is not a coderivation; it is a so-called {\it second order coderivation} \cite{Markl}, an algebraic analogue of the symplectic Laplacian. A quantum $L_\infty$ algebra is characterized by the condition
\begin{equation}L^2=0.\end{equation}
\begin{exercise}
By expanding this equation, show that the 1-string product at genus 1 must satisfy
\begin{equation}QL_{1,1}(A) + L_{1,1}(QA)+L_{0,2}(L_{1,0},A)+L_{0,3}(|\omega^{-1}\rangle, A)= 0.\end{equation}
By inspection of Feynman diagrams, give an algebraic expression for the off-shell, 1-loop 2-point amplitude. Using the above and analogous relations for the 0-sting product at genus 1 $L_{1,0}$ as well as for $L_{0,2}$ and $L_{0,3}$, show that this amplitude is BRST invariant.
\end{exercise}

\pagebreak

\section{Lecture 4: Closed Superstring Field Theory}

Closed super SFT is the field theory of fluctuations of a closed string background in superstring theory---the heterotic or one of the Type II string theories. Closed super SFT has been characterized in the RNS formalism, based on worldsheet theories with $\mathcal{N}=(1,0)$ (heterotic) or $\mathcal{N}=(1,1)$ (Type II) supersymmetry. The theories are structurally similar to closed bosonic SFT: The dynamical string field $\Phi$ is Grassmann even and is subject to $b_0^-$ and level matching conditions; at the classical level it carries ghost number 2, and at the quantum level it contains components at all ghost numbers. There is a symplectic form and a hierarchy of products satisfying quantum $L_\infty$ relations. The main novelty in closed super SFT is that the products and symplectic form must contain additional operator insertions to soak up zero modes of the $\beta\gamma$ system. There are important differences here from the $bc$ system. If the $bc$ path integral on a Riemann surface has a $b$-ghost zero mode, it will vanish
\begin{equation}\int db = 0.\end{equation}
To get a nonvanishing result we therefore have to place a $b$-ghost operator inside the path integral to soak up the zero mode
\begin{equation}\int db \, b=1.\end{equation}
By contrast, if the $\beta\gamma$ path integral has a $\beta$ ghost zero mode, it will be divergent
\begin{equation}\int d\beta = \infty,\end{equation}
since $\beta$ is a Grassmann even variable. Inserting $\beta$ in the path integral only makes the problem worse. Instead we need to insert a delta function of $\beta$:
\begin{equation}\int d\beta \delta(\beta) = 1.\end{equation}
On second thought this is quite analogous to the $bc$ system, it just happens that for an odd variable $\delta(b)=b$. The delta function of $\beta$ however is a new kind of object which is not related in an elementary way to $\beta$. The number of delta functions of $\beta$ minus the number of delta functions of $\gamma$ defines a new grading on the vector space of states/operators of the superstring called {\it picture number}. 

The question is how objects such as $\delta(\beta)$ should be concretely defined. A remedy proposed long ago by Friedan, Martinec, and Shenker \cite{FMS} is to bosonize the $\beta\gamma$ ghosts
\begin{equation}\beta(z) = \d \xi e^{-\phi}(z),\ \ \ \ \gamma(z) = \eta e^\phi(z),\end{equation}
where $\phi$ is a holomorphic scalar with background charge and $\xi,\eta$ are analogous to $bc$ ghosts but with conformal weight $0$ and $1$ respectively. The bosonized fields come with quantum numbers
\begin{eqnarray}
\xi(z): \lineup\ \ \ \mathrm{Grassmann \ odd},\ \ \ \mathrm{gh}\#\ -1,\ \ \mathrm{picture}\ +1,\ \ \mathrm{weight}\ 0,\phantom{\Big)}\nonumber\\
\eta(z): \lineup\ \ \ \mathrm{Grassmann \ odd},\ \ \ \mathrm{gh}\#\ +1,\ \ \mathrm{picture}\ -1,\ \ \mathrm{weight}\ 1,\phantom{\Big)}\nonumber\\
e^{q\phi(z)}: \lineup\ \ \ \mathrm{Grassmann}\ q/\mathbb{Z}_2,\ \ \mathrm{gh}\#\ 0,\ \ \ \ \ \, \mathrm{picture}\ q,\ \ \ \ \ \ \mathrm{weight}\ -\frac{1}{2}q(q+2).
\end{eqnarray}
With these ingredients it is easy to describe operators with nonzero picture. For example
\begin{eqnarray}
\lineup \delta(\beta(z)) = e^{\phi(z)},\ \ \ \delta(\gamma(z)) = e^{-\phi(z)},\nonumber\\
\lineup \eta(z) = \d \gamma \delta(\gamma(z)),\ \ \ \d \xi = \d \beta\delta(\beta(z)).
\end{eqnarray}
Note that only $\d \xi$ appears in the bosonization formula, so the $\beta\gamma$ system knows nothing about the zero mode of the $\xi$ ghost. This means that the space of states/operators created by the bosonized ghosts, including the $\xi$ zero mode, is strictly larger than that of the $\beta\gamma$ system. This is called the {\it large Hilbert space}. The space of states/operators which are independent of the $\xi$ zero mode is the same as that of the $\beta\gamma$ system, and is called the {\it small Hilbert space}. The string field theories we discuss are based on the small Hilbert space. Large Hilbert space formulations also exist, but have a very different algebraic structure, and the solution of the BV master equation is not known in closed form. For applications to string perturbation theory we need to compute loop amplitudes, and the small Hilbert space approach is preferred. 

It is not necessary to bosonize the $\beta\gamma$ ghosts. Correlation functions of delta function $\beta\gamma$ operators can be understood with the proper definition of the $\beta\gamma$ path integral~\cite{revisited} (see also \cite{Ohmori}). However, closed super SFT has been largely developed using the bosonized $\eta,\xi,\phi$ system, due to the widespread use of this formalism. A related choice in the development of the theory is whether it should be understood in terms of ordinary Riemann surfaces with spin structure, or from the point of view of super-Riemann surfaces and supermoduli space. Presently the theory has been developed largely in the former perspective. However, at least conceptually a super-Riemann surface understanding may be useful. Some discussions in this direction can be found in \cite{Yeh,Jurco,Ohmori,Takezaki}. 

Understanding closed bosonic SFT gets you most of the way towards understanding closed super SFT. The superstring however presents two significant new hurdles: 
\begin{itemize}
\item Formulating a free action for Ramond sector string fields.
\item Defining a measure for off-shell amplitudes which avoids ``spurious singularities" which can appear in $\beta\gamma$ correlation functions at higher genus.
\end{itemize}
The issue with the Ramond action is a problem of picture number counting; at a deeper level, this has been thought to be related to the difficulty of formulating an action for the self-dual form of Type IIB supergravity. The issue with spurious singularities is believed to be related to subtle propertes of the supermoduli spaces of super-Riemann surfaces \cite{Donagi}. In the following we outline the current understanding of these issues. We discuss the heterotic string, so we only need to worry about $\beta\gamma$ ghosts in the holomorphic sector. For Type II we also need to keep track of them in the anti-holomorphic sector. At the present level of understanding, it appears that there is no obstruction to constructing an action for closed super SFT out to any finite order. What is missing is a concrete principle for defining the vertices at all orders, something analogous to Zweibach's minimal area problem in closed bosonic SFT.\footnote{There was an attempt to formulate an analogue of the minimal area problem for super-Riemann surfaces \cite{Yeh,Jurco}, but the proposed area functional is not positive or number-valued \cite{Jconv}.} For recent applications it has not been necessary to address this question.

There are two main ways to formulate a free action in the Ramond sector. For pedagogical reasons we start with the approach which is closest to the formulation of the free action for the closed bosonic string. This approach features prominently in recent work of Kunitomo and Okawa \cite{complete}, but goes back much further \cite{freeR,Yeh,Jurco}. We start with the claim that the dynamical field for the heterotic string should take the form
\begin{equation}\Phi = \Phi_\NS +\Phi_\R,\end{equation}
where $\Phi_\NS$ is the Neveu-Schwarz (NS) sector contribution, describing spacetime bosons, and $\Phi_\R$ is the Ramond (R) sector contribution, describing spacetime fermions. The field is subject to $b_0^-$ and level matching constraints, is Grassmann even, carries ghost number 2 classically and unrestricted ghost number quantum mechanically, and is GSO($+$) projected.  There are also picture number constraints: $\Phi_\NS$ has picture $-1$ and $\Phi_\R$ has picture $-1/2$. These are off-shell extensions of superconformal vertex operators in the $-1$ and $-1/2$ picture:
\begin{equation}c\overline{c}\mathcal{V}_\NS e^{-\phi},\ \ \ \ c\overline{c}\mathcal{V}_\R e^{-\phi/2},\end{equation}
where $\mathcal{V}_\NS,\mathcal{V}_\R$ are superconformal matter primaries of weight $(1/2,1)$ and $(5/8,1)$ respectively. It may be possible to formulate the theory with a Ramond field at picture $-3/2$, but other pictures are expected to be problematic since the spectrum of $L_0$ is unbounded from below, even at fixed momentum \cite{SenErbin}. Since vertices will typically come with a factor of $e^{-\lambda L_0^+}$, sums over intermediate states will cause divergence in loops. It is possible that negative weight states would be projected out with the proper definition of the propagator, but there are no string field theories at other pictures whose free actions are completely understood. 

The free action should take the form
\begin{equation}S_\mathrm{free} = \frac{1}{2!}\omega(\Phi,Q\Phi),\end{equation}
with the appropriately defined symplectic form. Correlation functions on a genus $g$ Riemann surface generically require operator insertions of net picture $2g-2$ to be well-defined. The free action is defined by a correlation function on the sphere, which needs $-2$ units of picture. Two NS states provide picture $-2$, and since the BRST operator and $c$ ghost carry vanishing picture, the symplectic form between NS states can be defined in exactly the same way as for the closed bosonic string. For the Ramond sector this does not work. Two Ramond states only provide picture $-1$, so we need an additional operator insertion in the symplectic form to remove one more unit of picture. An analogous problem already appeared in the bosonic string. Two closed string fields plus the BRST operator provide ghost number $5$, but we need ghost number $6$ to get a nonvanishing correlation function on the sphere. This is taken care of by $c_0^-$. But we know that $c_0^-$ is closely related to the $b_0^-$ and level matching constraints. This suggests that the Ramond string field must be subject to additional constraints related to picture changing. The geometrical origin of these constraints is somewhat obscure from the point of view of ordinary Riemann surfaces, but from the super-Riemann surface perspective they originate from postulating that local superconformal coordinates around a Ramond puncture should only be defined up to a shift along the odd direction.

To describe the required constraints, we introduce an operator $\mathscr{X}$ of picture $+1$ and an operator $\mathscr{Y}$ of picture $-1$ with the analogy
\begin{eqnarray}
b_0^- \delta(L_0^-)\lineup \ \longleftrightarrow\ \mathscr{X} = G_0\delta(\beta_0)+b_0\delta'(\beta_0),\nonumber\\
c_0^-\lineup \ \longleftrightarrow\ \mathscr{Y} = \delta'(\gamma_0)c_0.
\end{eqnarray}
$\mathscr{X}$ and $\mathscr{Y}$ are Grassmann even, BPZ even, and ghost number zero. $G_0$ is the zero mode of the superpartner of the energy momentum tensor and for example $\delta(\beta_0)$ is the delta function of the zero mode of the $\beta$ ghost. Note that this is not the same as the zero mode of the delta function of the $\beta$ ghost:
\begin{equation}\delta(\beta_0)\neq \oint \frac{dz}{2\pi i} \frac{1}{z^{5/2}}e^{\phi(z)}.\end{equation}
For example, we have $\beta_0\delta(\beta_0)=0$, which is clearly not satisfied on the right hand side. We have the relations
\begin{equation}\mathscr{X}\mathscr{Y}\mathscr{X} = \mathscr{X},\ \ \ [Q,\mathscr{X}]=0,\end{equation}
in analogy to $b_0^-\delta(L_0^-)$ and $c_0^-$. The operator $\mathscr{Y}$ is not BRST invariant and is not uniquely determined by these relations; our specific choice is conventional. The operator $\mathscr{X}$ is an example of a {\it picture changing operator} (PCO). For our purposes, a PCO is a representative of the cohomology class of the identity operator at picture $+1$. The Ramond string field is subject to the constraint
\begin{equation}\mathscr{X}\mathscr{Y}\Phi_\mathrm{R} = \Phi_\mathrm{R},\end{equation}
which can be shown to be equivalent to 
\begin{equation}\beta_0^2\Phi_\mathrm{R} = 0,\ \ \ \ (G_0\beta_0+\beta_0G_0)\Phi_\mathrm{R} = 0.\end{equation}
This is the Ramond sector analogue of the $b_0^-$ and level matching constraints. If $\mathcal{H}$ is the full GSO($+$) projected state space of the heterotic worldsheet theory, this leads to the definition of a restricted subspace
\begin{equation}\Hhat\subset\H\end{equation}
consisting of NS and R states at picture $-1,-1/2$ subject to $b_0^-$ and level matching constraints, and in addition satisfying the constraints on the Ramond sector outlined above. The symplectic form and Poisson bivector 
\begin{equation}\langle \omega|:\Hhat^{\otimes 2}\to\Hhat^{\otimes 0},\ \ \ \ |\omega^{-1}\rangle \in\Hhat^{\otimes 2}\end{equation}
are defined by
\begin{eqnarray}
\langle \omega| \lineup = -\langle \mathrm{bpz}|(c_0^- P_{-1}+\overline{c}_0\mathscr{Y}P_{-1/2})\otimes\mathbb{I},\\
|\omega^{-1}\rangle\lineup = \Big[b_0^-\delta(L_0^-)(P_{-1}+P_{-1/2}\mathscr{X})\Big]\otimes \mathbb{I}|\mathrm{bpz}^{-1}\rangle,\label{eq:PoissonOk}
\end{eqnarray}
where $P_p$ is the projection on to states of picture $p$. Picture number conservation implies
\begin{equation}\langle \mathrm{bpz}|P_p\otimes\mathbb{I} = \langle \mathrm{bpz}|\mathbb{I}\otimes P_{-2-p}.\end{equation}
\begin{exercise}
Using the identity $\delta(\beta_0)\delta(\gamma_0)\delta(\beta_0)=\delta(\beta_0)$, show that $\mathscr{X}\mathscr{Y}\mathscr{X}=\mathscr{X}$.
\end{exercise}
\begin{exercise}
Show that the symplectic form and Poisson bivector are inverses of each other, that the BRST operator is cyclic, and that the symplectic form is graded antisymmetric.
\end{exercise}

The vertices of the heterotic action\footnote{The notation $\bar{\mathcal{R}}_{g,n}$ follows Sen \cite{SenErbin}, who uses $\bar{\mathcal{R}}_{g,n_\mathrm{NS},n_\mathrm{R}}$ to denote averages of sections of the appropriate bundle defining a genus $g$ vertex with $n_\mathrm{NS}$ NS states and $n_\mathrm{R}$ Ramond states. $\mathcal{R}_{g,n_\mathrm{NS},n_\mathrm{R}}$ (without bar) denotes the corresponding sections for the vertices of the 1PI effective action. The notation $\mathcal{V}_{g,n}$ for the bosonic string follows Zwiebach \cite{Zwiebach}.} 
\begin{equation}\langle \bar{\mathcal{R}}_{g,n}|:\Hhat^{\otimes n}\to\Hhat^{\otimes 0}\end{equation}
are specified by appropriate local sections of $\Phat_{g,n}$ with some additional data specifying PCO locations. Specifically, if we split the string field into NS and R components, the $n$ string vertex decomposes into $\left\lfloor\frac{n}{2}\right\rfloor+1$ subvertices
\begin{equation}\langle \bar{\mathcal{R}}_{g,n_\mathrm{NS},n_\mathrm{R}}|,\ \ \ n_\mathrm{NS}+n_\mathrm{R}=n,\end{equation}
which couple $n_\mathrm{NS}$ states and $n_\mathrm{R}$ Ramond states. The amount of picture provided by the PCOs in each subvertex must be 
\begin{equation}N_{g,n_\mathrm{NS},n_\mathrm{R}} = 2g-2+n_\mathrm{NS}+\frac{1}{2}n_\mathrm{R}.\end{equation}
The heterotic string products are defined by
\begin{equation}L_{g,n}=\Big(\mathbb{I}\otimes\langle \bar{\mathcal{R}}_{g,n}|\Big)\Big(|\omega^{-1}\rangle\otimes\mathbb{I}^{\otimes n-1}\Big),\end{equation}
and should satisfy quantum $L_\infty$ relations. Note that products will always be proportional to $\mathscr{X}$ when producing a Ramond output. This ensures that string fields multiply consistently inside the subspace $\Hhat$, since the Ramond sector constraint is implied by $\mathscr{X}\mathscr{Y}\mathscr{X}=\mathscr{X}$. The Siegel gauge propagator is
\begin{equation}b_0^+/L_0^+.\end{equation}
As in lecture 2, it is natural to view the ``full propagator" as the Siegel gauge propagator multiplied by whatever operator insertions appear in the output of a product. If the propagator carries a Ramond state, this leads to 
\begin{equation}\frac{b_0^+b_0^- \delta(L_0^-)}{L_0^+}\mathscr{X}.\end{equation}
It is interesting to rework this expression a little bit:
\begin{eqnarray}
\frac{b_0^+b_0^- \delta(L_0^-)}{L_0^+}\mathscr{X}\lineup = \frac{(2b_0-b_0^-)b_0^-\delta(L_0^-)}{2L_0-L_0^-}(G_0\delta(\beta_0)+b_0\delta'(\beta_0))\nonumber\\
\lineup = \left(\frac{1}{G_0}\delta(\beta_0)b_0\right) b_0^-\delta(L_0^-).
\end{eqnarray}
$G_0$ is the stringy analogue of the Dirac operator, just as $L_0^+$ is analogous to the Klein-Gordon operator. Therefore the PCO from the Ramond output of a product turns the Siegel gauge propagator into a stringy analogue of the Dirac propagator, as is appropriate for fermions.

This formulation of the Ramond sector seems natural. But nonlocal delta function operators such as $\delta(\beta_0)$ are difficult to work with. Their correlation functions need to be derived from scratch, and (to my knowledge) they have not been studied from the point of view of the bosonized $\xi,\eta,\phi$ system. It is desireable to maintain a connection to bosonized language, since it provides a concrete description of Ramond vertex operators and allows the definition of the large Hilbert space, which is sometimes useful. In these respects, local delta function operators such as $\delta(\beta(z))$ are simpler objects to work with. Their bosonized description is straightforward, and general formulas for their correlation functions were derived on the 80's \cite{Verlinde,Lechtenfeld}. These formulas allow an explicit characterization of spurious singularities in $\beta\gamma$ correlation functions in terms of theta divisors, which is useful in the construction of superstring vertices. These practical considerations motivate a second approach to the Ramond sector, devised by Sen \cite{SenBV}, where only local delta function operators are needed.  The idea is to have a Feynman diagram expansion for string amplitudes almost the same as would be implied by the previous treatment of the Ramond sector, but without the additional constraint on the Ramond string field and with the PCO $\mathscr{X}$ replaced with the zero mode of the local PCO $X(z)=Q\cdot \xi(z)$:
\begin{equation}X_0 =\oint \frac{dz}{2\pi i}\frac{1}{z}X(z).\end{equation}
We will try to motivate the trick which achieves this. A first guess would be to take the Poisson bivector given in \eq{PoissonOk} and replace $\mathscr{X}$ with $X_0$:
\begin{equation}|\omega^{-1}\rangle\stackrel{?}{=} \Big[b_0^-\delta(L_0^-)(P_{-1}+P_{-1/2}X_0)\Big]\otimes \mathbb{I}|\mathrm{bpz}^{-1}\rangle.\end{equation}
The problem is that string field theory requires a symplectic form, and we want the symplectic form to be defined without further constraint on the Ramond sector.  This would be possible if the operator $P_{-1}+P_{-1/2}X_0$ had an inverse, but unfortunately it does not. So we need to modify it somehow.  We observe that, even though the physical Ramond string field should have picture $-1/2$, the operator $P_{-1}+P_{-1/2}X_0$ naturally acts on a vector space which also includes states at picture $-3/2$. Without picture $-3/2$, the second term would vanish. This suggests that we can modify the Poisson bivector by replacing $P_{-1}$ with the identity operator in the form
\begin{equation}P_{-1}+P_{-1/2}+P_{-3/2}.\end{equation}
Then 
\begin{equation}(P_{-1}+P_{-1/2}+P_{-3/2}+P_{-1/2}X_0)^{-1} = P_{-1}+P_{-1/2}+P_{-3/2}-X_0P_{-3/2}.\end{equation}
In this way we postulate a symplectic form and Poisson bivector:
\begin{eqnarray}
\langle \omega| \lineup = -\langle \mathrm{bpz}|\Big[c_0^-(\mathbb{I}-X_0 P_{-3/2})\Big]\otimes\mathbb{I},\\
|\omega^{-1}\rangle\lineup = \Big[b_0^-\delta(L_0^-)(P_{-1}+P_{-1/2}+P_{-3/2}+P_{-1/2}X_0)\Big]\otimes \mathbb{I}|\mathrm{bpz}^{-1}\rangle.
\end{eqnarray}
In this setup the subspace $\Hhat\subset\H$ should be identified with states at pictures $-3/2,-1,-1/2$ subject to $b_0^-$ and level matching constraints. No further constraints are needed in the Ramond sector at either picture. The heterotic string field 
\begin{equation}\Phi = \Phi_{-1}+\Phi_{-1/2}+\Phi_{-3/2}\in\Hhat\end{equation}
is Grassmann even, carries ghost number 2 classically and unrestricted ghost number quantum mechanically. The free action takes the form
\begin{eqnarray}
S_\mathrm{free} \lineup = \frac{1}{2!}\omega(\Phi,Q\Phi)\nonumber\\
\lineup = \frac{1}{2!}\langle \Phi_{-1},c_0^-Q\Phi_{-1}\rangle +\langle\Phi_{-3/2},c_0^-Q\Phi_{-1/2}\rangle -\frac{1}{2!}\langle \Phi_{-3/2},c_0^-X_0 Q\Phi_{-3/2}\rangle,
\end{eqnarray}
which leads to the linearized equations of motion
\begin{equation}Q\Phi = 0.\end{equation}
This is a consistent free action, but the string spectrum is wrong. We have two identical copies of the Ramond cohomology at picture $-1/2$ and $-3/2$. This deficiency however is rendered physically irrelevant by postulating that the interactions only couple Ramond states at picture $-1/2$. We define string vertices 
\begin{equation}\langle \bar{\mathcal{R}}_{g,n}|:\Hhat^{\otimes n}\to\Hhat^{\otimes 0}\end{equation}
so that the products 
\begin{equation}L_{g,n} = \Big(\mathbb{I}\otimes\langle \bar{\mathcal{R}}_{g,n}|\Big)\Big(|\omega^{-1}\rangle\otimes\mathbb{I}^{\otimes n-1}\Big)\end{equation}
satisfy quantum $L_\infty$ relations, and further we require that all vertices except $\langle \bar{\mathcal{R}}_{0,2}|$ vanish when applied to states at picture $-3/2$. In Feynman diagrams, vertices are always connected by a Siegel gauge propagator acting on the Poisson bivector
\begin{equation}\frac{b_0^+}{L_0^+}\otimes \mathbb{I}|\omega^{-1}\rangle.\end{equation}
Since states at picture $-3/2$ do not couple, we can project them out without changing anything: 
\begin{eqnarray}\lineup \Big[(P_{-1}+P_{-1/2})\otimes(P_{-1}+P_{-1/2})\Big]\frac{b_0^+}{L_0^+}\otimes \mathbb{I}|\omega^{-1}\rangle\nonumber\\\lineup\ \ \ = \left[\frac{b_0^+}{L_0^+}b_0^-\delta(L_0^-)(P_{-1}+P_{-1/2})(P_{-1}+P_{-1/2}+P_{-3/2}+P_{-1/2}X_0)(P_{-1}+P_{-3/2})\right]\otimes\mathbb{I}|\mathrm{bpz}^{-1}\rangle\ \ \ \ \ \ \ \ \ \nonumber\\
\lineup\ \ \ = \left[\frac{b_0^+}{L_0^+}b_0^-\delta(L_0^-)(P_{-1}+P_{-1/2}X_0)\right]\otimes\mathbb{I}|\mathrm{bpz}^{-1}\rangle.\end{eqnarray}
This is exactly what we wanted. The propagator is the same as with the previous treatment of the Ramond sector, but $\mathscr{X}$ is replaced with $X_0$ and we do not have to worry about extra constraints. The additional degrees of freedom at picture $-3/2$ do not appear as intermediate states or external states in Feynman diagrams, and can be effectively ignored. 

We have now completed the discussion of the free theory and can proceed to interactions. Let us describe the cubic vertex. The cubic vertex can couple either three NS states or one NS state with two R states:
\begin{equation}\langle \bar{\mathcal{R}}_{0,3}|N_1\otimes N_2\otimes N_3,\ \ \ \ \langle \bar{\mathcal{R}}_{0,3}|N_1\otimes R_2\otimes R_3.\end{equation}
In the later case the three states give picture $-2$, as needed for a well-defined correlation function on the sphere. For the purely NS coupling the three states give picture $-3$, so we need an additional insertion of picture $+1$ on the sphere. The insertion must be BRST invariant otherwise the on-shell 3-point amplitude would not be BRST invariant. Furthermore, the insertion cannot be BRST exact, otherwise the on-shell 3-point amplitude would vanish. Therefore the insertion needs to be a nontrivial element of the BRST cohomology at ghost number $0$ and picture $+1$; in other words, it needs to be a PCO. In the following we assume Sen's formulation of the Ramond sector, where it is consistent to assume that vertices contain only the local PCO insertion $X(z)$. The constrained formulation requires more general types of PCOs such as $\mathscr{X}$, though only in the quartic vertex and beyond.  The PCO $X(z)$ can be distributed as a sum or integral over different points on the sphere minus the three local coordinate patches. We do not insert PCOs in the local coordinate patches, since we need to be able to remove patches when gluing general off-shell states. Symmetry of the cubic vertex implies that $X(z)$ cannot be inserted at only one point. At minimum the PCO can be distributed as an average over two points, and in fact these two points are uniquely determined. In the coordinate on the Riemann sphere where the punctures are located at $0,1$ and $\infty$, the points are
\begin{equation}z=e^{\pm i\pi/3}.\end{equation}
Assuming symmetric local coordinate maps $f_1,f_2,f_3$ with punctures at these positions, the cubic vertex then can be defined by
\begin{eqnarray}
\langle \bar{\mathcal{R}}_{0,3}|N_1\otimes N_2\otimes N_3\lineup  = \left\langle \frac{X(e^{i\pi/3})+X(e^{-i\pi/3})}{2}f_1\circ N_1(0)f_2\circ N_2(0) f_3\circ N_3(0)\right\rangle_\mathbb{C},\nonumber\\
\langle \bar{\mathcal{R}}_{0,3}|N_1\otimes R_2\otimes R_3\lineup  = \Big\langle f_1\circ N_1(0)f_2\circ R_2(0) f_3\circ R_3(0)\Big\rangle_\mathbb{C}.
\end{eqnarray}
It is possible to distribute the PCOs in other ways; the various choices are related by field redefinition. 

The quartic vertex requires four local coordinate maps so that the $s,t,u$ and quartic vertex Feynman diagrams patch together to define a symmetric and continuous global section of $\Phat_{0,4}$. The quartic vertex also requires information about PCO positions. If all four states are NS, we need two PCOs, and they must be arranged to match the PCO positions in the $s,t,u$ channel diagrams at the boundary of the quartic vertex region of moduli space.  To describe all of this data it is helpful to introduce the fiber bundle $\widetilde{\mathcal{P}}_{g,n_\mathrm{NS},n_\mathrm{R}}$. The base of the fiber bundle is the moduli space of genus $g$ Riemann surfaces with spin structure and $n_\mathrm{NS},n_{\mathrm{R}}$ Neveu-Schwarz and Ramond punctures.\footnote{By moduli space with spin structure we mean a $4^g$-fold covering of the ordinary moduli space representing the possible spin structures on the Riemann surface. If the surface has a Ramond puncture, the spin structure can be described by doubling the range of the Ramond puncture around each of the $2g$ cycles. If there are only NS punctures, the moduli space with spin structure comes in two disconnected components, representing even and odd spin structures. Each component is given by adding the appropriate number of copies of the fundamental domain of the ordinary moduli space.} The (infinite dimensional) fiber parameterizes possible choices of $n_\mathrm{NS}+n_\mathrm{R}$ local coordinate maps, defined up to rotations of the unit disk, and positions of $N_{g,n_\mathrm{NS},n_\mathrm{R}}$ PCOs for a given point on the base.
In some circumstances, an off-shell amplitude can be defined by a admissible global section of this bundle. In general it is unclear whether it is possible to find an admissible global section. We will return to this issue shortly. In string field theory it is necessary to consider discrete and continuous averages of sections to ensure symmetry of vertices and to accommodate operators such as $X_0$ from the Ramond sector. 

We will say that a local section of $\widetilde{\mathcal{P}}_{g,n_\mathrm{NS},n_\mathrm{R}}$ is {\it admissible} if three conditions are satisfied: 
\begin{description} \item{1)} It implies an admissible local section of $\Phat_{g,n_\mathrm{NS}+n_\mathrm{R}}$;
\item{2)} The PCO locations vary continuously with the moduli and do not enter local coordinate patches; 
\item{3)} The PCO locations avoid {\it spurious singularities}. 
\end{description}
Spurious singularities are divergences in $\beta\gamma$ correlation functions that can appear for specific PCO configurations on a Riemann surface for given moduli, and are a manifestation of the appearance of additional zero modes of the $\gamma$ ghost. At genus zero, spurious singularities appear if PCOs collide, which can be understood as a more-or-less standard OPE divergence. More surprisingly, spurious singularities can appear at higher genus even for configurations where PCOs do not collide. There is no local mechanism for understanding the origin of this divergence. In any case, an admissible local section should avoid spurious singularities.

Let us describe the measure for integration on $\widetilde{\mathcal{P}}_{g,n_\mathrm{NS},n_\mathrm{R}}$. For a given point in this bundle, we introduce a surface state
\begin{equation}\langle \Sigma_{g,n+1}|:\H^{\otimes n+1}\to\H^{\otimes 0}.\end{equation}
The surface state is partially specified by the $n=n_\mathrm{NS}+n_{\mathrm{R}}$ local coordinate patches of the NS and R states. In addition we introduce one more local coordinate patch, whose coordinate we denote as $y$ with $|y|<1$. This will give a coordinate system on the portion of the Riemann surface where we wish to insert PCOs. The choice of coordinate system for PCOs on the surface is arbitrary, and is not part of the data given to us by the fiber bundle. But for present discussion it is convenient to fix a choice of coordinate system for each point in $\widetilde{\mathcal{P}}_{g,n_\mathrm{NS},n_\mathrm{R}}$. So as not to redundantly parameterize the PCO positions, we assume that the coordinate system can depend on the moduli and the other local coordinate patches but not on the PCO positions themselves. Suppose we wish to insert $N$ operators $\mathcal{O}^1,...,\mathcal{O}^N$ at points $y^1,...,y^N$ in this coordinate system. Eventually these operators will be related to PCOs, but for later discussion it is helpful to be more general. The surface state with operator insertions is given by
\begin{equation}\langle \Sigma_{g,n+1}|\mathbb{I}^{\otimes n}\otimes \Big(\mathcal{O}^1(y^1)...\mathcal{O}^N(y^N)|0\rangle\Big).\end{equation}
We would like to turn this into a differential form which can be integrated over $\widetilde{\mathcal{P}}_{g,n_\mathrm{NS},n_\mathrm{R}}$. We introduce coordinates on $\widetilde{\mathcal{P}}_{g,n_\mathrm{NS},n_\mathrm{R}}$
\begin{equation}p^\alpha, \  y^\mu,\ \ \ \ \mu=1,...,N_{g,n_\mathrm{NS},n_\mathrm{R}} .\end{equation}
The coordinates $p^\alpha$ parameterize the moduli space and the possible choices of $n$ local coordinate patches for the NS and R states. The coordinates $y^\mu$ specify PCO locations. We have Schiffer vector fields 
\begin{eqnarray}\lineup v_\alpha^A(\xi_A,p^\alpha),\ \ \ \ A=1,...,n,\\
\lineup v_\alpha^{n+1}(y,p^\alpha).
\end{eqnarray}
The first $n$ vector fields represent deformations of the local coordinate patches of the NS and R states, and the final vector field represents a deformation of the $y$ coordinate patch as implied by our choice of coordinate system for the PCOs at each point in $\widetilde{\mathcal{P}}_{g,n_\mathrm{NS},n_\mathrm{R}}$. We introduce the exterior derivative and operator-valued 1-form
\begin{eqnarray}
d\lineup = dp^\alpha\frac{\d}{\d p^\alpha} + dy^\mu \frac{\d}{\d y^\mu}+d\overline{y}^\mu\frac{\d}{\d \overline{y}^\mu},\\
b \lineup = dp^\alpha\left(\sum_{A=1}^N \mathbb{I}^{\otimes A-1}\otimes b[v_\alpha^A]\otimes \mathbb{I}^{\otimes N-A+1} + \mathbb{I}^{\otimes n}\otimes b[v_\alpha^{n+1}]\right),
\end{eqnarray}
in the same way as in lecture 1. Note that the local coordinate patches of the NS and R states, and (by assumption) of $y$ are independent of PCO locations, so that the 1-form $b$ does not have any $dy^\mu$ coordinate. With this we can define a measure 
\begin{equation}\langle \mathcal{O}^1...\mathcal{O}^N| = \langle \Sigma_{g,n+1}|e^b \,\mathbb{I}^{\otimes n}\otimes \Big(\mathcal{O}^1(y^1)...\mathcal{O}^N(y^N)|0\rangle\Big),\end{equation}
which satisfies the BRST identity
\begin{equation}
d \langle \mathcal{O}^1...\mathcal{O}^N|+\big\langle (Q-d)(\mathcal{O}^1...\mathcal{O}^N)\big| = -(-1)^{|\mathcal{O}^1|+...+|\mathcal{O}^N|}\langle\mathcal{O}^1...\mathcal{O}^N| Q.\label{eq:BRSTid}
\end{equation}
This is not yet the measure for off-shell amplitudes of the heterotic string, since we have not specified the operator insertions, and the action of $Q$ produces a total derivative only if the second term above vanishes. One might initially guess that the operator insertions should be PCOs, but this is not quite right since PCOs are not annihilated by $Q-d$, and the second term will not vanish. However the combination $X(y^\mu)-d\xi(y^\mu)$ is invariant under $Q-d$. Setting $N = N_{g,n_\mathrm{NS},n_\mathrm{R}}$ we therefore choose
\begin{equation}
\mathcal{O}^1(y^1)...\mathcal{O}^N(y^N) = \Big(X(y^1)-d\xi(y^1)\Big)\,...\,\Big(X(y^{N_{g,n_\mathrm{NS},n_\mathrm{R}}} )-d\xi(y^{N_{g,n_\mathrm{NS},n_\mathrm{R}}} )\Big).
\end{equation}
Writing $X^1 = X(y^1)$ etc. for short, the measure is therefore
\begin{equation}
\langle \Omega_{g,n_\mathrm{NS},n_\mathrm{R}}| = \left(-\frac{1}{2\pi i}\right)^{\mathrm{dim}_\mathbb{C}\mathcal{M}_{g,n_\mathrm{NS},n_\mathrm{R}}}\Big\langle \big(X^1-d\xi^1\big)\,...\,\big(X^{N_{g,n_\mathrm{NS},n_\mathrm{R}} }-d\xi^{N_{g,n_\mathrm{NS},n_\mathrm{R}} }\big)\Big|,
\end{equation}
with a convenient normalization. This satisfies the expected BRST identity
\begin{equation}d\langle \Omega_{g,n_\mathrm{NS},n_\mathrm{R}}| = -\langle \Omega_{g,n_\mathrm{NS},n_\mathrm{R}}|Q.\end{equation}
Given an admissible local section $\sigma(\widetilde{\mathcal{P}}_{g,n_\mathrm{NS},n_\mathrm{R}})$ we can evaluate the integral
\begin{equation}\int_{\sigma(\widetilde{\mathcal{P}}_{g,n_\mathrm{NS},n_\mathrm{R}})}\langle \Omega_{g,n_\mathrm{NS},n_\mathrm{R}}|, \end{equation}
where the integrand is the pullback of the measure. This will be BRST invariant up to contributions from the boundary of the local section. 

\begin{wrapfigure}{l}{.3\linewidth}
\centering
\resizebox{2in}{3in}{\includegraphics{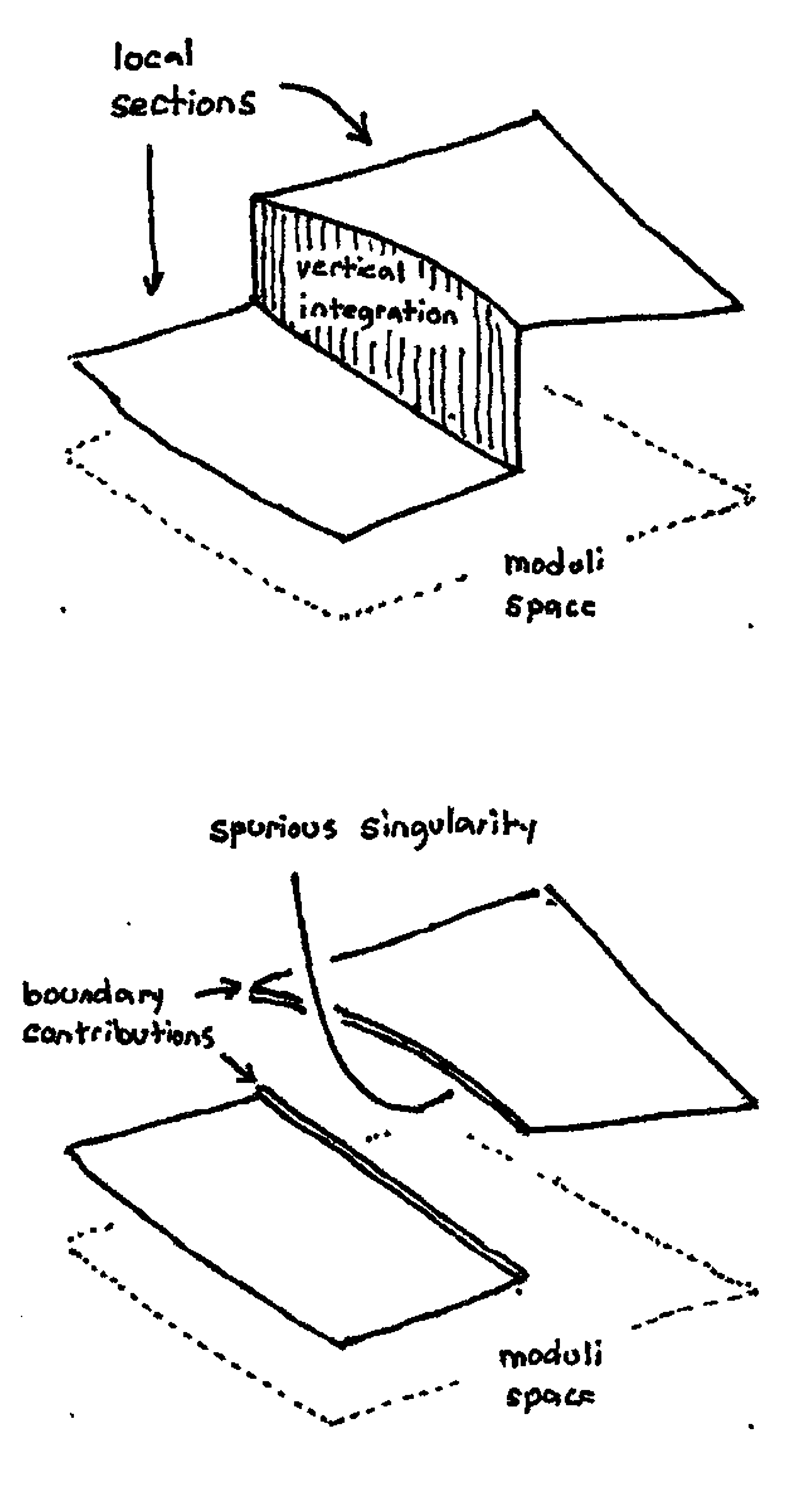}}
\end{wrapfigure}
To define an off-shell amplitude it seems we should extend the local section into an admissible global section. However, it is not clear whether this is possible. Spurious singularities could lead to a topological obstruction to the existence of an admissible global section, though I do not think this has been established. In any case, finding a global section which avoids spurious singularities everywhere may be inconvenient. A remedy proposed by Sen \cite{SenOffShell}, and later fully articulated by Sen and Witten \cite{SenWitten} is called {\it vertical integration}. The idea is to cover the moduli space by a patchwork of admissible local sections. At the interface of local sections there will be discontinuities. The idea is to form a closed integration cycle in $\widetilde{\mathcal{P}}_{g,n_\mathrm{NS},n_\mathrm{R}}$ by integrating along the fiber (i.e. `vertically') at discontinuities  to join local sections. If the integration is suitably implemented, the measure will be a total derivative in the fiber coordinate, which allows us to jump across spurious singularities which might otherwise prevent smoothly deforming and joining neighboring local sections. The boundary contributions from integrating along the fiber are called  {\it vertical corrections}, and restore BRST invariance of the off-shell amplitude. 

We will describe vertical corrections from a somewhat different point of view, following~\cite{vertical}. Presently we work with fixed genus and number of NS and R punctures, so we will omit these labels. We also omit the normalization factor for the measure, to reduce clutter. Suppose for the moment that we can ignore spurious singularities, and that an off-shell amplitude can be specified by a global section. We can express the amplitude as an integral over the moduli space 
\begin{equation}\langle \mathcal{A}^N| = \int_{\mathcal{M}}\big\langle (X^1-d\xi^1)\,...\,(X^{N}-d\xi^{N})\big| ,\end{equation}
where the local coordinate patches and PCO locations have been specified as a function of the moduli by the choice of global section. It is interesting to observe that this expression can be formally derived following an iterative procedure which starts from an amplitude where explicit PCO insertions are absent. It requires working in the large Hilbert space at intermediate steps. The idea is to insert the $\xi$ ghost $\xi(y^1)$ into the measure, act with the BRST operator, insert $\xi(y^2)$ into the measure, act with the BRST operator and so on. Repeating this process for $N$ steps gives the amplitude with $N$ PCOs:
\begin{eqnarray}
\langle \mathcal{A}^0|  = \int_{\mathcal{M}} \big\langle 1\big| \ \ \ \ \ \ \ \ \ \ \ \ \ \ \ \ \ \ \ \lineup\underset{\mathrm{insert}\ \xi}{\longrightarrow} \ \ \ \ \ \ \ \ \ \ \ \ \ \ \ \ \ \ \ \ \  \int_{\mathcal{M}} \big\langle \xi^1\big|\ \ \ \ \ \ \ \ \ \ \ \ \ \ \ \ \ \ \  \ \ \ \underset{\mathrm{act}\ Q}{\longrightarrow}\phantom{\Bigg]}\nonumber\\
\langle\mathcal{A}^1| =  \int_{\mathcal{M}} \big\langle (X^1-d\xi^1)\big|\ \ \ \ \ \ \ \ \ \ \ \lineup\underset{\mathrm{insert}\ \xi}{\longrightarrow} \ \ \ \ \ \ \ \ \  \ \ \  \ \ \   \int_{\mathcal{M}} \big\langle (X^1-d\xi^1)\xi^2\big|\ \ \  \ \ \  \ \ \   \ \ \  \ \ \underset{\mathrm{act}\ Q}{\longrightarrow}\phantom{\Bigg]}\nonumber\\
\langle\mathcal{A}^2|  =  \int_{\mathcal{M}} \big\langle (X^1-d\xi^1)(X^2 -d\xi^2)\big| \ \ \ \ \ \lineup\underset{\mathrm{insert}\ \xi}{\longrightarrow} \ \  \ \ \  \ \ \  \ \ \  \ \ \  \ \ \  \ \ \  \ \ \  ... \phantom{\Bigg]}\nonumber\\
...\ \ \  \ \ \  \ \ \  \ \ \  \ \ \  \ \ \  \ \ \  \ \ \   \lineup\underset{\mathrm{insert}\ \xi}{\longrightarrow} \  \int_{\mathcal{M}}\big\langle (X^1-d\xi^1)\,...\,(X^{N-1}-d\xi^{N-1})\xi^N\big| \phantom{\Bigg]}\underset{\mathrm{act}\ Q}{\longrightarrow} \ \nonumber\\
 \langle \mathcal{A}^N|  = \int_{\mathcal{M}}\big\langle (X^1-d\xi^1)\,...\,(X^{N}-d\xi^{N})\big|.\ \lineup \phantom{\Bigg]}
\end{eqnarray}
When acting with $Q$ we used the BRST identity \eq{BRSTid} in the presence of operator insertions and discarded contributions from the boundaries of moduli space. Now the claim is that this procedure also works if the PCO positions vary discontinuously with the moduli as may be needed to jump across spurious poles, and vertical corrections will be generated automatically. Let us illustrate 
\begin{wrapfigure}{l}{.25\linewidth}
\centering
\resizebox{1.6in}{1.4in}{\includegraphics{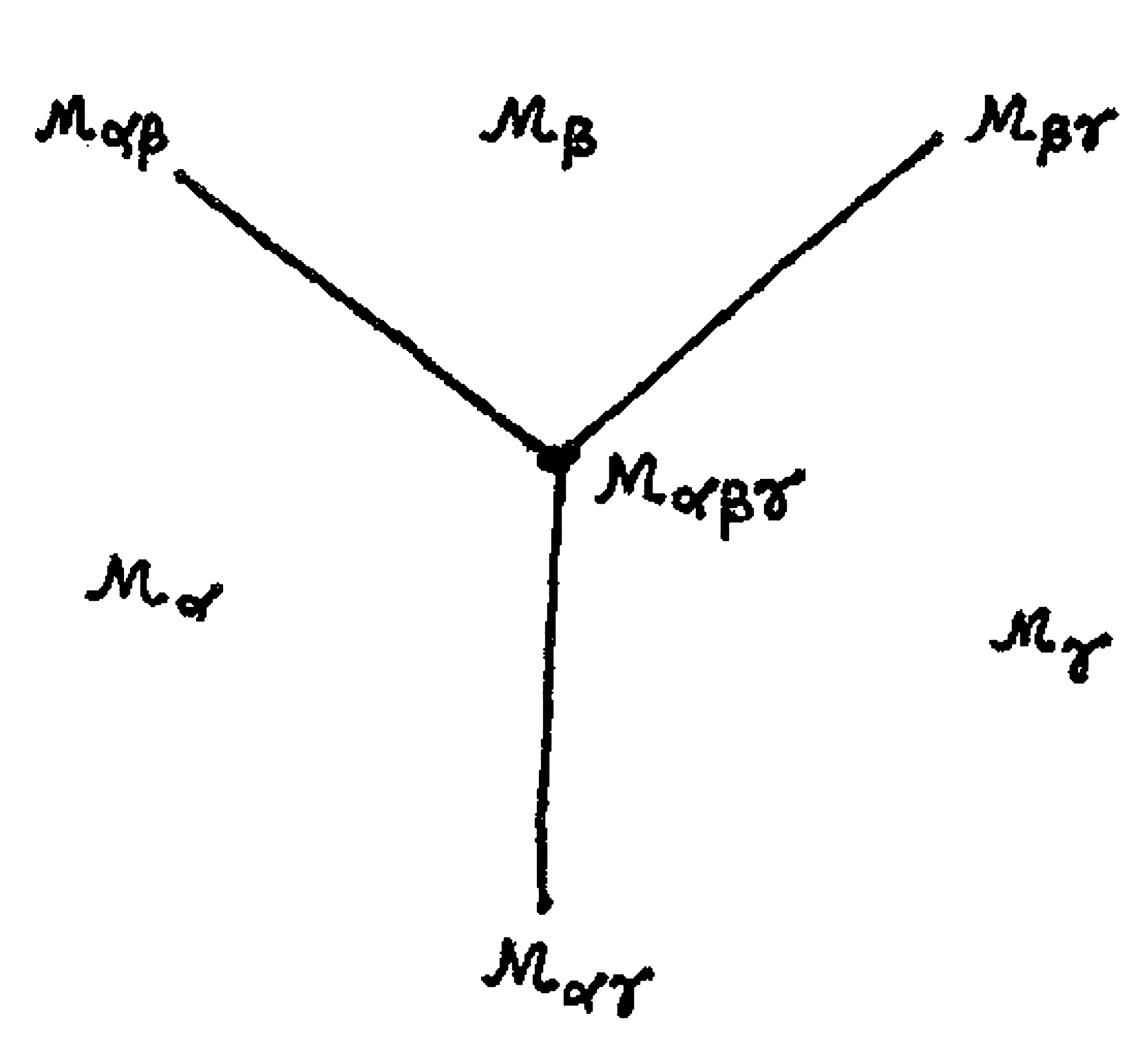}}
\end{wrapfigure}
this with an example. Suppose that we can cover the moduli space $\mathcal{M}$ with three admissible local sections defined on non-overlapping regions $\mathcal{M}_\alpha,\mathcal{M}_\beta$ and $\mathcal{M}_\gamma$. We have
\begin{equation}\mathcal{M} = \mathcal{M}_\alpha\cup \mathcal{M}_\beta\cup\mathcal{M}_\gamma.\end{equation}
At the intersection between these regions are submanifolds of codimension one and two:
\begin{eqnarray}
\lineup \mathcal{M}_{\alpha\beta} = \mathcal{M}_\alpha\cap\mathcal{M}_\beta,\ \ \ \mathcal{M}_{\beta\gamma} = \mathcal{M}_\beta\cap\mathcal{M}_\gamma,\ \ \ \mathcal{M}_{\alpha\gamma} = \mathcal{M}_\alpha\cap\mathcal{M}_\gamma;\nonumber\\
\lineup \mathcal{M}_{\alpha\beta\gamma} = \mathcal{M}_\alpha\cap\mathcal{M}_\beta\cap \mathcal{M}_\gamma.
\end{eqnarray}
We use the order of Greek indices to indicate orientation, so that index orderings which differ by an odd (even) permutation are the same manifolds with opposite (identical) orientation. In this sense the Greek indices can be regarded as totally antisymmetric, and we define for example $\mathcal{M}_{\alpha\alpha}$ to be the empty set. The orientation on the moduli space determines an orientation on $\mathcal{M}_\alpha$, $\mathcal{M}_\beta$ and $\mathcal{M}_\gamma$, and the orientation on the higher codimension intersections will be defined by
\begin{equation}\d \mathcal{M}_{\alpha_1,...,\alpha_n} = -\sum_{\alpha_{n+1}} \mathcal{M}_{\alpha_1...\alpha_n\alpha_{n+1}}.\end{equation}
Suppose at first that we only need one PCO. The local section on $\mathcal{M}_\alpha$ is specified by the PCO position $y_\alpha^1$ as a function of $\mathcal{M}_\alpha$, and likewise $y_\beta^1$ for $\mathcal{M}_\beta$ and  $y_\gamma^1$ for $\mathcal{M}_\gamma$. Inserting the $\xi$ ghost at these positions, the integration over the moduli space splits into three pieces:
\begin{equation}\int_{\mathcal{M}_\alpha} \big\langle \xi_\alpha^1\big|+\int_{\mathcal{M}_\beta}\big\langle \xi_\beta^1\big|+\int_{\mathcal{M}_\gamma}\big\langle \xi_\gamma^1\big|,\end{equation}
where we write $\xi_\alpha^1=\xi(y_\alpha^1)$ etc. Next we act with $Q$. Using the BRST identity and integrating the total derivative contribution  gives 
\begin{eqnarray}
\langle\mathcal{A}^1| \lineup = \int_{\mathcal{M}_\alpha} \big\langle( X^1_\alpha -d\xi_\alpha^1)\big|+\int_{\mathcal{M}_\beta}\big\langle (X^1_\beta-d\xi_\beta^1)\big|+\int_{\mathcal{M}_\gamma}\big\langle (X^1_\gamma-d\xi_\gamma^1)\big|
\nonumber\\
\lineup\ \ \  +\int_{\mathcal{M}_{\alpha\beta}}\big\langle(\xi_\beta^1 -\xi_\alpha^1)\big|+\int_{\mathcal{M}_{\beta\gamma}}\big\langle(\xi_\gamma^1 -\xi_\beta^1)\big|+\int_{\mathcal{M}_{\alpha\gamma}}\big\langle(\xi_\gamma^1 -\xi_\alpha^1)\big|.
\end{eqnarray}
The first three terms are contributions to the amplitude from the local sections on $\mathcal{M}_\alpha,\mathcal{M}_\beta$ and $\mathcal{M}_\gamma$; the second three terms are vertical corrections. Now suppose that we needed two PCOs to define the amplitude. The local sections on $\mathcal{M}_\alpha,\mathcal{M}_\beta$ and $\mathcal{M}_\gamma$ would specify the positions $y^2_\alpha,y^2_\beta$ and $y^2_\gamma$ of a second PCO. To determine the vertical corrections, in addition it will be necessary to specify a PCO position on the codimension 1 interfaces $\mathcal{M}_{\alpha\beta},\mathcal{M}_{\beta\gamma}$ and $\mathcal{M}_{\alpha\gamma}$, which we write as $y^2_{\alpha\beta},y^2_{\beta\gamma}$ and $y^2_{\alpha\gamma}$ respectively. Inserting the appropriate $\xi$ ghost in each of the six terms above gives
\begin{eqnarray}
\lineup \int_{\mathcal{M}_\alpha} \big\langle( X^1_\alpha -d\xi_\alpha^1)\xi^2_\alpha\big|+\int_{\mathcal{M}_\beta}\big\langle (X^1_\beta-d\xi_\beta^1)\xi^2_\beta\big|+\int_{\mathcal{M}_\gamma}\big\langle (X^1_\gamma-d\xi_\gamma^1)\xi^2_\gamma\big|
\nonumber\\
\lineup  +\int_{\mathcal{M}_{\alpha\beta}}\big\langle(\xi_\beta^1 -\xi_\alpha^1)\xi^2_{\alpha\beta}\big|+\int_{\mathcal{M}_{\beta\gamma}}\big\langle(\xi_\gamma^1 -\xi_\beta^1)\xi^2_{\beta\gamma}\big|+\int_{\mathcal{M}_{\alpha\gamma}}\big\langle(\xi_\gamma^1 -\xi_\alpha^1)\xi^2_{\alpha\gamma}\big|.
\end{eqnarray}
Next we act with $Q$ to find
\begin{eqnarray}
\langle\mathcal{A}^2| \lineup = \int_{\mathcal{M}_\alpha}\! \big\langle( X^1_\alpha -d\xi_\alpha^1)( X^2_\alpha -d\xi_\alpha^2)\big|\!+\!\int_{\mathcal{M}_\beta}\!\big\langle (X^1_\beta-d\xi_\beta^1)(X^2_\beta-d\xi_\beta^2)\big|\!+\!\int_{\mathcal{M}_\gamma}\!\big\langle (X^1_\gamma-d\xi_\gamma^1)(X^2_\gamma-d\xi_\gamma^2)\big|
\nonumber\\
\lineup\ \ \  +\int_{\mathcal{M}_{\alpha\beta}}\Big\langle\Big[(X^2_{\alpha\beta}-d\xi^2_{\alpha\beta})(\xi_\beta^1 -\xi_\alpha^1)+(\xi^2_{\alpha\beta}-\xi^2_\alpha)(X^1_\alpha-d\xi^1_\alpha) - (\xi^2_{\alpha\beta}-\xi^2_\beta)(X^1_\beta-d\xi^1_\beta)\Big]\Big|\nonumber\\
\lineup\ \ \ +\int_{\mathcal{M}_{\beta\gamma}}\Big\langle\Big[(X^2_{\beta\gamma}-d\xi^2_{\beta\gamma})(\xi_\gamma^1 -\xi_\beta^1)+(\xi^2_{\beta\gamma}-\xi^2_\beta)(X^1_\beta-d\xi^1_\beta) - (\xi^2_{\beta\gamma}-\xi^2_\gamma)(X^1_\gamma-d\xi^1_\gamma)\Big]\Big|\nonumber\\
\lineup\ \ \ +\int_{\mathcal{M}_{\alpha\gamma}}\Big\langle\Big[(X^2_{\alpha\gamma}-d\xi^2_{\alpha\gamma})(\xi_\gamma^1 -\xi_\alpha^1)+(\xi^2_{\alpha\gamma}-\xi^2_\alpha)(X^1_\alpha-d\xi^1_\alpha) - (\xi^2_{\alpha\gamma}-\xi^2_\beta)(X^1_\gamma-d\xi^1_\gamma)\Big]\Big|\nonumber\\
\lineup \ \ \ +\int_{\mathcal{M}_{\alpha\beta\gamma}}\Big\langle\Big[\xi^2_{\alpha\beta}(\xi^1_\beta-\xi^1_\alpha)+\xi^2_{\beta\gamma}(\xi^1_\gamma-\xi^1_\beta)-\xi^2_{\alpha\gamma}(\xi^1_\gamma-\xi^1_\alpha)\Big]\Big|.
\end{eqnarray}
Again, the first three terms are contributions to the amplitude from the local sections $\mathcal{M}_\alpha,\mathcal{M}_\beta$ and $\mathcal{M}_\gamma$; the remaining terms are vertical corrections. Note that the last term is a dedicated vertical correction from the codimension 2 interface. In more complicated examples we may have four or more local sections and interfaces of codimension 3 and higher. For an amplitude with $p$ PCOs there will be vertical corrections from interfaces up to codimension $p$.

A vertex in the heterotic SFT action is characterized by integration over a subspace of the full moduli space. Generally it may be necessary to cover this subspace by a patchwork of local sections of $\widetilde{\mathcal{P}}_{g,n_\mathrm{NS},n_\mathrm{R}}$ which avoid spurious singularities, and then add vertical corrections to compensate for discontinuities between local sections. In addition we will need to consider averages of sections to achieve symmetry and couple to Ramond states. While the configuration of PCOs inside a string vertex may be complicated, the important point is that spurious singularities do not present an obstruction to the existence of the vertex. One subtle issue, however, is that the configuration of spurious singularities depends on the moduli of the Riemann surface and the location of the punctures. In string field theory, we should be able to cut out a local coordinate patch around each puncture and glue a generic off-shell state to the vertex.  If the off-shell state contains a vertex operator whose position is displaced from the puncture, or more dramatically if the state itself is defined by path integral over a nontrivial surface, the location of spurious singularities will be modified. Gluing such states to a vertex is  exactly what is needed to form the Feynman graph expansion from the action. Therefore, if a vertex avoids spurious singularities when contracted with a basis of local vertex operators, it is far from obvious that spurious singularities will still be absent when computing the S-matrix. One way to address this difficulty is to assume that vertices have a tube of empty worldsheet attached to each external state: 
\begin{equation}\langle \bar{\mathcal{R}}_{g,n}| = \langle\bar{\mathcal{R}}_{g,n}' |e^{-\lambda L_0^+}\otimes ...\otimes e^{-\lambda L_0^+}, \end{equation}
where $\langle\bar{\mathcal{R}}_{g,n}' |$ contains all operator insertions. The tube of worldsheet is often referred to as a {\it stub}. In a vertex with stubs, the position of PCOs is not only excluded from the local coordinate patches, but from a finite and potentially large region around the local coordinate patches. When gluing vertices and propagators in Feynman diagrams, the stubs will exponentially suppress states of large conformal weight in the propagator. If the exponential suppression is strong enough---equivalently, if the stub length $\lambda$ is sufficiently long---the sum over intermediate states will converge. Therefore, while the locus of spurious singularities may adjust slightly when gluing vertices with propagators, for sufficiently large $\lambda$ it cannot change so drastically as to render a previously admissible PCO configuration divergent. Not much detail is known about how large $\lambda$ needs to be to avoid problems; it will depend in some way on how close the PCO configurations in the vertices are to spurious singularity. But it is presumed that $\lambda$ can be finite.

The presence of stubs raises an interesting point. In each internal line of a Feynman diagram, the propagator is accompanied by a factor of $e^{-2\lambda L_0^+}$, so effectively we are working with a propagator
\begin{equation}\frac{b_0^+e^{-2\lambda L_0^+}}{L_0^+}=b_0^+\int_{2\lambda}^\infty dt e^{-tL_0^+}.\end{equation}
From this point of view it seems that the propagator is missing tubes of worldsheet with length less than $2\lambda$. This part of the propagator gives the dominant contribution from states of large conformal weight. Therefore with long stubs it appears as though the physics of highly excited string states is being transferred from the propagators into the vertices; the high energy behavior is essentially being integrated out, in a manner similar to the Wilsonian effective action \cite{SenWilsonian}. Therefore it is generally felt that adding long stubs will make nonperturbative physics more difficult to access in closed SFT; perhaps, for example, nonperturbative vacua will be more difficult to find. On the other hand, in Zwiebach's minimal area prescription \cite{Zwiebach} vertices always come with stubs of length~$\pi$. In as far as the minimal area metric gives the optimal definition of closed string vertices, this suggests that closed string field theory should necessarily be viewed as an effective field theory. In this respect, open string field theory \cite{Witten} is on a very different footing, as integration over the moduli space is implemented by the propagator alone \cite{ZwiebachWitten}. 
However, there has been little concrete progress in making sense of open bosonic string field theory quantum mechanically, and for the open superstring spurious singularities create new complications.  While string field theory gives a rigorous definition of perturbative string theory, it remains to be seen if it can provide a definition of nonperturbative string theory.

\vspace{.25cm}

\noindent {\bf Acknowledgements}

\vspace{.25cm}

\noindent I would like to thank the organizers of the workshop ``String Theory from a worldsheet perspective" in spring 2019 for inviting me to give these lectures. I also thank the Galileo Galilei Institute for Theoretical Physics and INFN for hospitality and partial support during my stay at this workshop. I am grateful to R. Donagi for discussion and encouraging me to write up these notes, and S. Stieberger and O. Lechtenfeld for comments. This work was supported by ERDF and M\v{S}MT (Project CoGraDS -CZ.02.1.01/0.0/0.0/15\_ 003/0000437) and the GA{\v C}R project 18-07776S and RVO: 67985840.

\pagebreak

\begin{appendix}

\section{Tensor Products}
\label{app:tensor}

Consider a $\mathbb{Z}_2$ graded vector space $\H$. The grading refers to Grassmann parity, which will be denoted $|V|$ for $|V\rangle\in \H$. Modulo addition by an even integer, a Grassmann odd vector has $|V|=1$ and a Grassmann even vector has $|V|=0$.  We may introduce a  basis $|e_i\rangle$ for $\H$, where $i$ ranges over some set. Tensor products of elements of $\H$ generates a new vector space 
\begin{equation}
\H\otimes \H = \H^{\otimes 2},
\end{equation}
which consists of linear combinations of basis vectors $|e_i\rangle\otimes |e_j\rangle$. Similarly we may define $\H^{\otimes n}$ for any positive integer $n$. The tensor powers of $\H$ inherit a $\mathbb{Z}_2$ grading from $\H$, defined as the sum of the Grassmann parities of all vectors appearing in the tensor product:
\begin{equation}\big||V_1\rangle\otimes...\otimes |V_n\rangle\big| = | V_1|+...+ |V_n|.\end{equation}
 We also introduce $\H^{\otimes 0}$. If $\H$ is a complex vector space, $\H^{\otimes 0}$ is isomorphic to complex numbers~$\mathbb{C}$. However, we prefer to distinguish $\H^{\otimes 0}$ from $\mathbb{C}$ for two reasons. First, in quantum string field theory and in the Ramond sector of classical superstring field theory, basis elements appear in linear combinations with coefficients that can be anticommuting numbers. In this context $\H^{\otimes 0}$ should be extended to a complex Grassmann algebra. Second, we regard $\H^{\otimes 0}$ as a 1-dimensional vector space consisting of scalar multiples of the identity element of the tensor product. The identity element satisfies 
\begin{equation}1\otimes |V\rangle =|V\rangle\otimes 1 =|V\rangle,\ \ \ |V\rangle \in \H.\end{equation}
Since there is no {\it a priori} notion of the tensor product of a vector with a complex number, $\H^{\otimes 0}$ includes a little more structure than $\mathbb{C}$.

We may consider the dual space $\H^\star$. This consists of linear combinations of dual basis vectors $\langle e^i|$ satisfying $\langle e^i|e_j\rangle =\delta^i_j$. A dual vector can be viewed as a linear map from $\H$ into $\H^{\otimes 0}$:
\begin{equation}\langle V|:\H\to \H^{\otimes 0}.\end{equation}
Conversely, a vector can be viewed as a linear map from $\H^{\otimes 0}$ into $\H$:
\begin{equation}|V\rangle: \H^{\otimes 0}\to \H.\end{equation}
This can be understood as multiplication of a scalar by a vector from the left.

We may consider linear maps between various tensor powers of $\H$. The Grassmann parity of a map  $M:\H^{\otimes m}\to\H^{\otimes m'}$ is given by 
\begin{equation}|M| = \big| M(|V_1\rangle\otimes...\otimes |V_m\rangle) \big|-\big||V_1\rangle\otimes...\otimes |V_m\rangle\big|.\end{equation}
This is the Grassmann parity of the input states minus the Grassmann parity of the output states. A useful concept is the notion of tensor product of linear maps. Suppose 
\begin{eqnarray}
\lineup M:\H^{\otimes m}\to\H^{\otimes m'},\nonumber\\
\lineup N:\H^{\otimes n}\to\H^{\otimes n'}
\end{eqnarray}
are two maps between tensor powers of $\H$. The tensor product map 
\begin{equation}M\otimes N:\H^{\otimes m+n}\to\H^{\otimes m'+n'}\end{equation}
is defined by 
\begin{eqnarray}\lineup(M\otimes N)(|V_1\rangle\otimes...\otimes |V_{m+n}\rangle) = \nonumber\\
\lineup\ \ \ \ \ \ \ \ \ \ \ \ \ \ \ \ \ \ \ \ (-1)^{|N|(|V_1|+...+|V_m|)}\Big(M (|V_1\rangle\otimes...\otimes |V_m\rangle)\Big)\otimes\Big(N (|V_{m+1}\rangle \otimes...\otimes |V_{m+n}\rangle)\Big),\ \ \ \ \ \ \ \ \label{eq:tensorMN}\end{eqnarray}
where $|V_1\rangle,..,|V_{m+n}\rangle$ are arbitrary vectors in $\H$. If $M'$ and $N'$ are linear maps whose input space is identical to the output space of $M$ and $N$, respectively, the above definition implies
\begin{equation}(M'\otimes N')(M\otimes N) = (-1)^{|M||N'|}(M'M)\otimes (N'N).\end{equation}
To give an example, consider
\begin{equation}Q\otimes \mathbb{I}+\mathbb{I}\otimes Q,\end{equation}
where $Q$ is the BRST operator. This acts on the tensor product of two vectors as 
\begin{equation}\Big(Q\otimes \mathbb{I}+\mathbb{I}\otimes Q\Big)|V_1\rangle\otimes |V_2\rangle = \big(Q|V_1\rangle\big)\otimes |V_2\rangle  + (-1)^{|V_1|}|V_1\rangle \otimes\big(Q|V_2\rangle\big).\end{equation}
We can also take the tensor product of maps to and from $\H^{\otimes 0}$. Since $\H^{\otimes 0}$ is special among tensor powers of $\H$, this case requires special attention. We give two examples. The identity operator can be expressed in a few equivalent ways: 
\begin{eqnarray}
\mathbb{I} \lineup =|e_i\rangle\langle e^i| \\
\lineup = |e_i\rangle\otimes\langle e^i|,\label{eq:Idee}\\
\lineup = (-1)^{|e_i|}\langle e^i|\otimes |e_i\rangle.\label{eq:Idee2}
\end{eqnarray}
where the repeated index is summed. In the first expression, the product of the vector and dual vector is defined by composition of maps to and from $\H^{\otimes 0}$. The second expression can be viewed as representing the identity operator as a map from $\H^{\otimes 0}\otimes \H$ into $\H\otimes\H^{\otimes 0}$, and the third expression does this in reverse. To check that the second expression acts as the identity, we apply the definition:
\begin{eqnarray}
\mathbb{I}|e_j\rangle \lineup = \big(|e_i\rangle\otimes\langle e^i|\big)|e_j\rangle\nonumber\\
\lineup = \big(|e_i\rangle\otimes\langle e^i|\big)\big(1\otimes |e_j\rangle\big)\nonumber\\
\lineup = \big(|e_i\rangle\cdot 1\big)\otimes\big(\langle e^i|e_j\rangle\big)\nonumber\\
\lineup = |e_i\rangle\otimes\delta^i_j\nonumber\\
\lineup = |e_j\rangle\otimes 1\nonumber\\
\lineup = |e_j\rangle.
\end{eqnarray}
To apply the definition \eq{tensorMN} in the second step we pulled out a trivial tensor product with the identity. A second example is the inverse relation \eq{wwinv} between the symplectic form  and Poisson bivector mentioned in the introduction: 
\begin{equation}(\langle \omega|\otimes\mathbb{I})(\mathbb{I}\otimes|\omega^{-1}\rangle) = \mathbb{I}.\end{equation}
We can write
\begin{eqnarray}
\langle \omega| \lineup = \omega_{ij}\langle e^i|\otimes\langle e^j|,\\
|\omega^{-1}\rangle\lineup = (-1)^{|e_j|}\omega^{ij}|e_i\rangle\otimes |e_j\rangle.\label{eq:winvee}
\end{eqnarray}
The origin of the sign in the components of the Poisson bivector will be clear in a moment. Substituting
\begin{eqnarray}
(\langle \omega|\otimes\mathbb{I})(\mathbb{I}\otimes|\omega^{-1}\rangle)\lineup =\omega_{ij}\omega^{kl}(-1)^{|e_l|}(\langle e^i|\otimes\langle e^j|\otimes\mathbb{I})(\mathbb{I}\otimes |e_k\rangle\otimes |e_l\rangle)\nonumber\\
\lineup = \omega_{ij}\omega^{kl}(-1)^{|e_l|}\langle e^i|\otimes \langle e^j|e_k\rangle\otimes |e_l\rangle\nonumber\\
\lineup = \omega_{ij}\omega^{kl}(-1)^{|e_l|}\langle e^i|\otimes \delta^j_k\otimes |e_l\rangle\nonumber\\
\lineup = \omega_{ij}\omega^{jk}(-1)^{|e_k|}\langle e^i|\otimes 1 \otimes |e_k\rangle\nonumber\\
\lineup = \omega_{ij}\omega^{jk}(-1)^{|e_k|}\langle e^i|\otimes |e_k\rangle.
\end{eqnarray}
Comparing with the identity operator as expressed in \eq{Idee2} we learn that
\begin{equation}\omega_{ij}\omega^{jk}=\delta_i^k.\end{equation}
This is a more familiar expression of the statement that the Poisson bivector inverts the symplectic form.

Let us make a comment about the sign in \eq{winvee}. Graded antisymmetry of the symplectic form implies that
\begin{equation}\omega_{ij}=-(-1)^{|e_i||e_j|}\omega_{ji}.\end{equation}
In string field theory we are concerned with an {\it odd} symplectic form, in the sense that it produces a nonzero result only when acting on basis vectors whose total Grassmann parity is odd:
\begin{equation}\omega_{ij}\neq 0\text{  only if  }|e_i|+|e_j| = 1\text{ mod }\mathbb{Z}_2.\end{equation}
In particular, for the nonvanishing components of an odd symplectic form $|e_i||e_j|$ is an even integer. Therefore the components are strictly antisymmetric:
\begin{equation}\omega_{ij} = -\omega_{ji}.\end{equation}
Similarly the components of the inverse $\omega^{ij}$ are nonvanishing only if $|e_i|+|e_j|$ is odd, and $\omega^{ij}$ is strictly antisymmetric. The components of the Poisson bivector in \eq{winvee} however come with a sign. With the sign, the components are actually {\it symmetric}:
\begin{equation}(-1)^{|e_j|}\omega^{ij}=(-1)^{|e_i|+1}\omega^{ij} = (-1)^{|e_i|}\omega^{ji}.\end{equation}
Therefore the Poisson bivector lives in the symmetrized tensor product of two copies of $\H$. This is actually necessary, since otherwise the Poisson bivector would give a vanishing result when fed into the string products, which are symmetric. The result must be nonvanishing to correctly account for gauge invariance in loop amplitudes.

\end{appendix}

\end{document}